\documentclass[conference, letterpaper, nofonttune, 10pt]{IEEEtran}

\usepackage{cite}
\usepackage{xspace}
\usepackage{relsize}
\usepackage{amsmath}
\usepackage{amssymb}
\usepackage{amsfonts}
\usepackage{multicol}
\usepackage{flushend}
\usepackage{stfloats}

\usepackage{enumerate}
\usepackage[shortlabels]{enumitem}

\usepackage{algorithmic}
\usepackage{graphicx}
\usepackage{microtype}
\usepackage[normalem]{ulem}
\usepackage[hyphens]{url}
\usepackage{hyperref}
\usepackage{textcomp}
\usepackage{arydshln}

\usepackage{capt-of}
\usepackage{refcount}
\usepackage[table]{xcolor}
\usepackage{booktabs}
\usepackage{tikz}

\usepackage{times}

\newcommand{\circleed}[1]%
{
\raisebox{.5pt}{\textcircled{\raisebox{-.9pt} {#1}}}
}

\newcommand{\twodigits}[1]%
{%
    \ifnum #1<10 0#1%
    \else%
    \number#1%
    \fi%
}

\usepackage{listings}
\usepackage{lstlinebgrd}
\definecolor{dkgreen}{rgb}{0,0.6,0}
\definecolor{gray}{rgb}{0.5,0.5,0.5}
\definecolor{mauve}{rgb}{0.58,0,0.82}
\definecolor{dkgray}{rgb}{0.47,0,0}

\definecolor{highlight}{rgb}{0,0,0}

\hypersetup{
    colorlinks=true,
    linkcolor=blue,
    filecolor=blue,      
    citecolor=blue,      
    urlcolor=blue,
}

\lstdefinestyle{intext}{
  frame=trBL,
  language=Java,
  numbers=left,
  numbersep=3.5pt,
  xleftmargin=2.5em,
  xrightmargin=2.5em,
  framexleftmargin=0.9em,
  framexrightmargin=0em,
  aboveskip=2mm,
  belowskip=0mm,
  columns=flexible,
  showstringspaces=false,
  basicstyle={\scriptsize\ttfamily},
  numberstyle={\scriptsize\color{darkgray}},
  keywordstyle={\color{blue}},
  commentstyle={\bfseries\color{dkgreen}},
  stringstyle={\bfseries\color{mauve}},
  breaklines=true,
  breakatwhitespace=true,
}

\lstdefinestyle{check}{
  frame=none,
  numbers=none,
  escapeinside={(@}{@)},
  backgroundcolor = \color{white},
  xleftmargin=2.6em,
  xrightmargin=1em,
  framexleftmargin=0em,
  framexrightmargin=0em,
  aboveskip=3mm,
  belowskip=2mm,
  columns=flexible,
  showstringspaces=false,
  basicstyle={\small\ttfamily},
  numberstyle={\small\ttfamily},
  keywordstyle={\small\ttfamily},
  commentstyle={\scritpsize\ttfamily},
  stringstyle={\small\ttfamily},
  breaklines=true,
  breakatwhitespace=true,
}

\newcommand{\our}[0]{{\textsl{\small{CRYLOGGER}}}\xspace}
\newcommand{\ourff}[0]{{\textsl{\footnotesize{CRYLOGGER}}}\xspace}
\newcommand{\ourabs}[0]{{{\footnotesize{CRYLOGGER}}}\xspace}
\newcommand{\cryrule}[1]{\textit{R-\twodigits{\getrefnumber{#1}}}}

\newcommand{\crysl}[0]{kruger_ecoop2019}
\newcommand{\cryslname}[0]{{CrySL}\xspace}
\newcommand{\cryslnref}[0]{{\cryslname}~\cite{\crysl}\xspace}

\newcommand{\cryptolint}[0]{egele_ccs2013}
\newcommand{\cryptolintname}[0]{{CryptoLint}\xspace}
\newcommand{\cryptolintnref}[0]{{\cryptolintname}~\cite{\cryptolint}\xspace}

\newcommand{\cryptoguard}[0]{rehaman_ccs2019}
\newcommand{\cryptoguardname}[0]{{CryptoGuard}\xspace}
\newcommand{\cryptoguardnref}[0]{{\cryptoguardname}~\cite{\cryptoguard}\xspace}

\newcommand{\mallodroid}[0]{fahl_ccs2012}
\newcommand{\mallodroidname}[0]{{MalloDroid}\xspace}
\newcommand{\mallodroidnref}[0]{{\mallodroidname}~\cite{\mallodroid}\xspace}

\newcommand{\cognicrypt}[0]{kruger_ase2017}
\newcommand{\cognicryptname}[0]{{CogniCrypt}\xspace}
\newcommand{\cognicryptnref}[0]{{\cognicryptname}~\cite{\cognicrypt}\xspace}

\newcommand{\cryptoapi}[0]{afrose_secdev2019}
\newcommand{\cryptoapiname}[0]{{CryptoAPI-Bench}\xspace}
\newcommand{\cryptoapinref}[0]{{\cryptoapiname}~\cite{\cryptoapi}\xspace}

\newcommand{\cma}[0]{shuai_dasc2014}
\newcommand{\cmaname}[0]{{CMA}\xspace}
\newcommand{\cmanref}[0]{{\cmaname}~\cite{\cma}\xspace}

\newcommand{\smvhunter}[0]{sounth_ndss2014}
\newcommand{\smvhuntername}[0]{{SMV-Hunter}\xspace}
\newcommand{\smvhunternref}[0]{{\smvhuntername}~\cite{\smvhunter}\xspace}

\newcommand{\icryptotracer}[0]{li_ndss2014}
\newcommand{\icryptotracername}[0]{{iCryptoTracer}\xspace}
\newcommand{\icryptotracernref}[0]{{\icryptotracername}~\cite{\icryptotracer}\xspace}

\newcommand{\androssl}[0]{gagnon_fps2015}
\newcommand{\androsslname}[0]{{AndroSSL}\xspace}
\newcommand{\androsslnref}[0]{{\androsslname}~\cite{\androssl}\xspace}

\newcommand{\khunt}[0]{li_ccs2018}
\newcommand{\khuntname}[0]{{K-Hunt}\xspace}
\newcommand{\khuntnref}[0]{{\khuntname}~\cite{\khunt}\xspace}

\begin{document}

\title{CRYLOGGER: \\ Detecting Crypto Misuses Dynamically}
\thanks{}
\author{
Luca Piccolboni,
Giuseppe Di Guglielmo,
Luca P. Carloni, Simha Sethumadhavan \\
\{piccolboni, giuseppe, luca, simha\}@cs.columbia.edu \\
Columbia University, New York, NY, USA 
}

\maketitle

\renewcommand{\baselinestretch}{1.15}

\begin{abstract}
Cryptographic (crypto) algorithms are the essential ingredients of all secure
systems: crypto hash functions and encryption algorithms, for example, can
guarantee properties such as integrity and confidentiality.
Developers, however, can misuse the application programming interfaces (API) of
such algorithms by using constant keys and weak passwords.
This paper presents \ourabs, the first open-source tool to detect crypto
misuses dynamically.
\ourabs logs the parameters that are passed to the crypto APIs during the
execution and checks their legitimacy offline by using a list of crypto rules.
We compare \ourabs with CryptoGuard, one of the most effective static tools to
detect crypto misuses.
We show that our tool complements the results of CryptoGuard, making the case
for combining static and dynamic approaches.
We analyze 1780 popular Android apps downloaded from the Google Play Store to
show that \ourabs can detect crypto misuses on thousands of apps dynamically
and automatically.
We reverse-engineer 28 Android apps and confirm the issues flagged by
\ourabs. We also disclose the most critical vulnerabilities to app developers
and collect their feedback.

\end{abstract}

\smallskip

{
\small\textbf{\emph{Repository---}%
{\url{https://github.com/lucapiccolboni/crylogger}~\cite{repo}}}
}

\renewcommand{\baselinestretch}{0.965}

%
%

\section{Introduction}\label{sec:intro}

{
Cryptographic (crypto) algorithms are the key ingredients of all secure
systems~\cite{rivest_1990}. Crypto algorithms can guarantee that the
communication between two entities satisfies strong properties such as data
confidentiality (with encryption) and data integrity (with hashing).
While the crypto theory can formally guarantee that those properties are
satisfied, in practice poor implementations of the crypto
algorithms~\cite{zinzi_ccs2017} can jeopardize communication security.
For instance, Brumley et al.~\cite{brumley_rsa2012} showed how to obtain the
entire private key of an encryption algorithm, which is based on elliptic
curves, by exploiting an arithmetic bug in OpenSSL.
Unfortunately, ensuring that the actual implementation of the crypto algorithms
is correct as well as secure is not sufficient. The crypto algorithms can be,
in fact, \textit{misused}.
Egele et al.~\cite{egele_ccs2013} showed that $88\%$ of the Android apps they
downloaded from the Google Play Store had at least one crypto misuse. For
example, thousands of apps used hard-coded keys for encryption instead of
truly-random keys, thus compromising data confidentiality.
Similarly, Rahaman et al.~\cite{\cryptoguard} showed that $86\%$ of the Android
apps they analyzed used broken hash functions, e.g., {SHA$1$}, for which
collisions can be produced~\cite{stevens_2012}, threatening data integrity.
}

\smallskip

{
Recently, researchers analyzed the causes of crypto misuses in many contexts.
Fischer et al.~\cite{fischer_sp2017} found that many Android apps included
snippets of code taken from Stack Overflow and $98\%$ of these snippets
included several crypto issues.
Nadi et al.~\cite{nadi_icse2016} claimed that the complexity of application
programming interfaces (APIs) is the main origin of crypto misuses in Java.
Developers have to take low-level decisions, e.g., select the type of padding
of an encryption algorithm, instead of focusing on high-level tasks.
Acar et al.~\cite{acar_sp2017} compared $5$ crypto libraries for Python and
argued that poor documentation, lack of code examples and bad choices of
default values in the APIs are the main causes of crypto misuses.
Muslukhov et al.~\cite{muslukhov_asiaccs2018} showed that $90\%$ of the misuses
in Android originated from third-party libraries, a result that was later
confirmed by Rahaman et al.~\cite{\cryptoguard}.
}

\smallskip

{ 
At the same time, researchers started to implement tools to automatically
detect crypto misuses, e.g.,~\cite{\cryptoguard, \cryptolint}. The idea is to
define a set of \textit{crypto rules} and check if an application respects them
by verifying the parameters passed to the crypto APIs.
The rules usually come from (i) papers that show the vulnerabilities caused by
some crypto algorithms or their misconfigurations,
e.g.,~\cite{vaudenay_ecrypt2002}, and (ii) organizations and agencies, e.g.,
NIST and IETF, that define crypto-related standards to prevent attacks.
Examples of crypto rules are setting (i) a minimum key size for encryption,
e.g., $2048$ bits for RSA~\cite{barker_nist2018} or (ii) a minimum number of
iterations for key derivation, e.g., $1000$ for PKCS\#5~\cite{rfc8018}.
}

\smallskip

{
To check the crypto rules, researchers developed static as well as dynamic
solutions. Static approaches, e.g., \cryslnref, \cryptolintnref,
\cryptoguardnref, \mallodroidnref, \cognicryptnref and \cmanref, examine the
code with program slicing~\cite{weiser_icse1981} to check the values of the
parameters that are passed to the APIs of the crypto algorithms. 
Static analysis has the benefit that the code is analyzed entirely without the
need of executing it. Also, it can scale up to a large number of applications.
Static analysis produces, however, false positives, i.e., alarms can be raised
on legit calls to crypto algorithms.
Some static approaches, e.g. \cryptoguardname, suffer also from false
negatives, i.e., some misuses escape detection, because the exploration is
pruned prematurely to improve scalability on complex programs. It is also
possible that static analysis misses some crypto misuses in the code that is
loaded dynamically~\cite{poeplau_ndss2014}.
Most of the recent research efforts focused on static
approaches~\cite{braga_tres2019}, while little has been done to bring dynamic
approaches to the same level of completeness and effectiveness.
Few approaches have been proposed towards this direction, e.g., \smvhunternref,
\androsslnref, \khuntnref, and \icryptotracernref. Dynamic approaches are
usually more difficult to use since they require to trigger the crypto APIs at
runtime to expose the misuses, but they do not usually produce false positives.
Unfortunately, these dynamic approaches do not support as many crypto rules as
the current static approaches. \smvhuntername and \androsslname consider only
rules for SSL/TLS, and \khuntname focuses on crypto keys. 
\icryptotracername attacks the hard problem of detecting misuses in iOS apps.
\icryptotracername supports few rules as it needs to rely on API hooking
techniques.
}

\vspace{-0.1cm}
\subsection{Contributions}

{
In this paper, we present \our, an open-source tool to detect crypto misuses
dynamically.
It consists of (i) a \textit{logger} that monitors the APIs of the crypto
algorithms and stores the values of the relevant parameters in a log file, and
(ii) a \textit{checker} that analyzes the log file and reports the crypto rules
that have been violated.
The key insights of this work are:
(1) we log the relevant parameters of the crypto API calls by instrumenting few
classes that are used by a large number of applications;
(2) we log the values of the parameters of the crypto APIs at runtime, while we
check the rules offline to reduce the impact on the applications performance;
(3) we show that, for most Android apps, the calls to the crypto APIs can be
easily triggered at runtime, and thus a dynamic approach can be effective in
detecting misuses even if the code of the applications has not been explored
entirely;
(4) we show that, for Android apps, it is sufficient to execute an application
for a relatively short amount of time to find many of the crypto misuses that
are reported by the current static tools.
}

\smallskip

{
We envision two main uses of \our. (1) Developers can use it to find
crypto misuses in their applications as well as in the third-party libraries
they include. \our can exploit the input sequences that are defined by
developers for verification purposes to detect the misuses. \our can be used
alongside static tools as it complements their analysis
(Section~\ref{sec:res1}).  Using \our also helps to reduce the false positives
reported by static tools.
(2) \our can be used to check the apps submitted to app stores, e.g., the
Google Play Store. Using a dynamic tool on a large number of apps is hard, but
\our can refine the misuses identified with static analysis because, typically,
many of them are false positives that cannot be discarded manually on such a
large number of apps.
}

\smallskip

{
We make the following contributions:

\begin{enumerate}[1., leftmargin=3.6em]
\itemsep0.25em

\item 
we describe \our, the first open-source tool to detect crypto misuses
dynamically; the tool is available at:
{\textbf{\small\url{https://github.com/lucapiccolboni/crylogger}}}\cite{repo};

\item
we implement \our for Android and Java apps; we support $26$ crypto rules,
and we decouple the logging and the checking mechanisms so that new rules can
be easily added and checked with \our;

\item
we compare \our against \cryptoguardnref, one of the most effective static
tools to detect misuses: we use $150$ popular Android apps of the Google Play
Store for the comparison; we show that \our reports misuses that
\cryptoguardname misses, but we show that the opposite is also possible, thus
making the case for combining static and dynamic approaches;

\item
we reverse engineer $150$ Android apps to evaluate the false positives of
\cryptoguardname; we show that for some rules many false positives are
reported due to insecure, but untriggerable, code included in the apps;

\item
we compare \our against \cryptoguardname by using the \cryptoapinref, a set of
Java programs that include misuses; we also extend the \cryptoapiname with
tests cases suited for dynamic tools;

\item 
we use \our to analyze $1780$ Android apps downloaded from the Google Play
Store (the dataset was collected between September and October 2019). These are
the most popular apps of $33$ different categories. We confirm the results
previously reported with static tools~\cite{\cryptolint,\cryptoguard} and
report new misuses;

\item
we disclose the vulnerabilities we found to $306$ app and library developers and we
report the feedback we received from the $10$ who replied; we manually
reverse-engineer $28$ apps to determine if the vulnerabilities reported by
\our can actually be exploited.

\end{enumerate}

\begin{figure}[t!]
\centering
\includegraphics[width=0.88\linewidth]{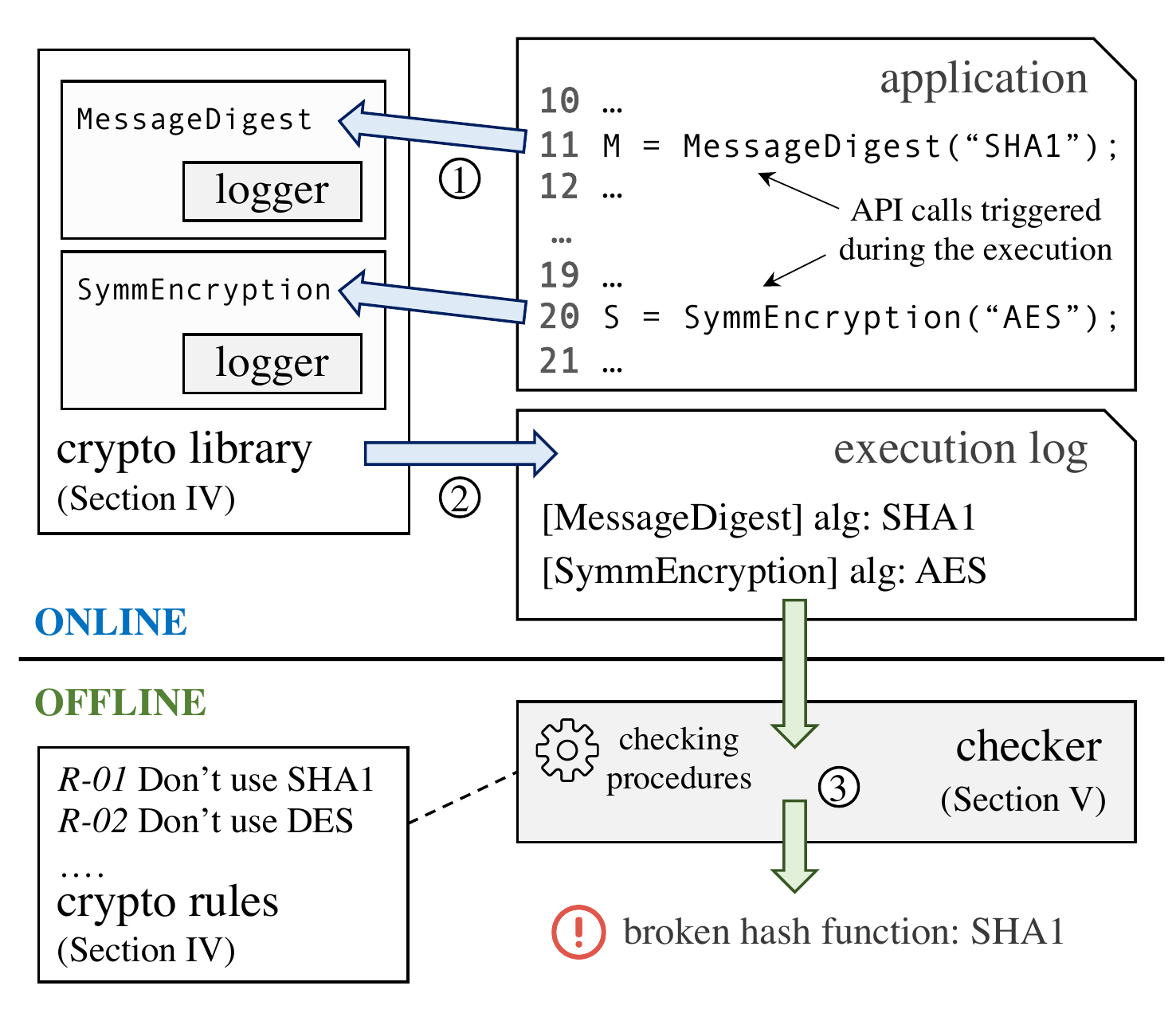}
\vspace{-0.3cm}
\caption{Overview of \ourff. \circleed{$1$} We run the application with an
instrumented crypto library. \circleed{$2$} We generate a log containing 
the parameters of the crypto API calls. \circleed{$3$} We check the crypto
rules and report all the violations.}\label{fig:cry}
\vspace{-0.4cm}
\end{figure}

\section{Overview}\label{sec:overv}

{
Fig.~\ref{fig:cry} provides an overview of \our. It consists of:
}

\begin{enumerate}[1., leftmargin=3.8em]
\itemsep0.3em
\item \textit{logger}:
the logger extends a crypto library, for example the Java crypto library, to
trace the API calls to crypto algorithms; for each of these calls, it logs the
relevant parameters that must be used to check the crypto rules; for example,
in Fig.~\ref{fig:cry}, the logger saves the names of the algorithms chosen by
the application for message digest (SHA$1$) and symmetric encryption (AES);
\item \textit{checker}:
the checker analyzes the log offline, after the application has been executed,
and it produces a list of all the crypto rules violated by the application. To
check the rules it uses a set of checking procedures, each of which covers many
crypto rules; for instance, in Fig.~\ref{fig:cry}, the checker finds that the
application uses the broken algorithm SHA$1$ as message digest algorithm.
\end{enumerate}
We decouple logging from checking for $4$ main reasons: (1) the parameters of
interest of the crypto library are more stable, i.e., it is unlikely that new
parameters are added; for example, the main parameters of an algorithm for key
derivation are the salt, the password and the number of iterations, (2) the
crypto rules are likely to change: for example, new rules can be added when new
vulnerabilities are found as well as current rules may need to be updated (for
example the minimum key size of RSA), (3) crypto rules are context-dependent:
some rules may be not relevant for certain applications or contexts, and (4)
checking rules offline does not affect the application performance, which is
important, for instance, when the application response is critical (Android).

\smallskip

Similarly to most of the current static solutions, we developed \our primarily
to check Java and Android applications. Our ideas, however, could be adapted to
other contexts.
In the next sections, we describe our tool in more detail.
In Section~\ref{sec:rel}, we discuss the related work.
In Section~\ref{sec:pre}, we describe a generic crypto library that we use to
define the crypto rules and \mbox{the API} parameters that must be logged.
In Section~\ref{sec:check}, we explain how \our checks the rules.
In Section~\ref{sec:impl}, we present an implementation of \our for Java and
Android~\cite{repo}, by explaining which APIs we instrumented and how we
analyzed a large number of Android apps.
In Section~\ref{sec:res0}, we describe the dataset of apps we use for
the evaluation.
In Section~\ref{sec:res1}, we perform a comparison of \our against
\cryptoguardname by using $150$ Android apps and the \cryptoapinref.
In Section~\ref{sec:res2}, we present an analysis of $1780$ apps from the
Google Play Store. We also report the feedback received for disclosing the
vulnerabilities and our reverse-engineering analysis of the vulnerabilities
found in $28$ apps.
In Section~\ref{sec:lim}, we discuss the limitations of our approach before
\mbox{concluding in Section~\ref{sec:conc}.}

%
%

\section{Related Work}\label{sec:rel}


\subsection{Detection of Crypto Misuses}\label{sec:rel:misuses}

{
Several tools exist to detect crypto misuses. Most of them are based on
\textit{static analysis}, e.g., \cryptolintnref, \cryptoguardnref, \cryslnref,
\mallodroidnref, \cognicryptnref and \cmanref.
These tools differ in the crypto rules that they support and in the slicing
algorithms~\cite{weiser_icse1981} that they adopt for analysis. Among them,
\cryptoguardname covers the highest number of crypto rules.
As discussed in~\cite{johnson_icse2013}, the main problem with static analysis
is the high number of false positives, which requires the users to manually
examine the results and determine the true positives.
Recent studies~\cite{afrose_secdev2019, rehaman_ccs2019} showed that
\cryptoguardname is one of the most effective tools in reducing the false
positives, thanks to rule-specific algorithms that refine the results of the
static analysis.  We show, however, that \cryptoguardname still produces many
false positives in practice by reporting crypto misuses that can never be
triggered at runtime (Section~\ref{sec:res1}).
To achieve scalability on complex apps, some tools ``cut off'' some branches of
the static explorations, e.g., \cryptoguardname clips orthogonal explorations.
This causes false negatives in addition to false positives. False negatives are
also caused by code that is loaded at runtime~\cite{poeplau_ndss2014}.
}

\smallskip

{
Other tools identify crypto misuses by employing \textit{dynamic analysis}.
\smvhunternref and \androsslnref, for example, detect misuses of the SSL/TLS
protocol. \khuntnref detects badly-generated keys, insecurely-negotiated keys
and recoverable keys by analyzing execution traces of Java programs.
\icryptotracernref detects misuses in iOS apps, which is a complex task that
must be implemented through API hooking techniques.
To the best of our knowledge, there are no approaches that are as exhaustive
and effective as static approaches and cover many crypto tasks, e.g.,
encryption, authentication, and SSL/TLS. This motivated us to develop \our, a
tool that supports more crypto rules than current static approaches and covers
several crypto tasks. 
The main disadvantage of all dynamic tools is the possibility of missing
vulnerabilities due to poor coverage~\cite{zheng_spsm2012}.  Some misuses can
remain undetected if the application are not explored thoroughly.
We show, however, that \our is capable of finding most of the crypto misuses
that \cryptoguardname reports even if the apps are not fully explored
(Section~\ref{sec:res1}).
}

\subsection{Other Related Research}\label{sec:rel:other}

{
The problem of crypto misuses has been studied from many different
perspectives. Fischer et al.~\cite{fischer_sp2017} analyzed security-related
code snippets taken from Stack Overflow. They found that $>$$15\%$ of the apps
of the Google Play Store contained snippets of code directly taken from Stack
Overflow and $\sim$$98\%$ of these had at least one misuse.
In a more recent work~\cite{fischer_usenix2019}, they showed that nudges
\cite{wang_www2013} significantly helped developers in making better decisions
when crypto tasks need to be implemented.
Nadi et al.~\cite{nadi_icse2016} showed that the main cause of misuses lies in
the complexity of the APIs rather than in the lack of security knowledge in
developers. Acar et al.~\cite{acar_sp2017} showed that poor documentation, lack
of code examples and bad choices of default values in the crypto APIs
contribute to many of the crypto misuses. Green et al.~\cite{green_ieeesp2016}
made the case for developing security-friendly APIs that help developers to
avoid common mistakes.
Many recent works, e.g.,~\cite{rehaman_ccs2019, muslukhov_asiaccs2018} showed
that third-party libraries cause most of the crypto misuses in Android, up to
$90\%$ in some cases.
To simplify the work for developers, several approaches display security tips or
warnings in an integrated development environment. For example, \cognicryptnref
generates code snippets in Eclipse, which can be used when crypto tasks need to
be implemented. Similarly, FixDroid~\cite{nguyen_ccs2017} provides suggestions
to developers on how to fix crypto-related issues in Android Studio.
To remove the burden of fixing misuses from developers, some approaches repair
problematic code snippets automatically~\cite{ma_ccs2016, ma_eso2017,
singleton_mob2019, krug_cgo2020}.
}

\subsection{Testing Android Apps}\label{sec:rel:dyn}

{
Analyzing Android apps dynamically and automatically is considered a hard
problem~\cite{choudhary_ase2015,zheng_icse2017}. 
The common solution to verify the apps correctness is Monkey\footnote{Monkey UI
Exerciser: \url{https://developer.android.com/studio/test/monkey}.}.  Monkey
generates pseudo-random events that interact with the GUI of the emulator or
the real device.
Monkey often obtains low code coverage because the events are completely
random~\cite{yerima_eurasip2019}, but it is quite efficient in terms of
execution time.
Other approaches try to exploit some information about the app to improve
coverage. For example, SmartDroid~\cite{zheng_spsm2012} exploits a combination
of static and dynamic techniques to trigger the APIs of interest.
DroidBot~\cite{li_icsec2017} is a test generator based on control-flow
graphs that can be extended to support custom exploration strategies.
Dynodroid~\cite{machiry_fse2013} monitors the app to guide the generation of
the next input event.
These approaches have a significant overhead on the execution of the app
because to generate useful events they require either to (i) rely on static
analysis of the code~\cite{zheng_spsm2012} or (ii) create a model at runtime
that helps the exploration~\cite{li_icsec2017}. In \our, we use Monkey as it is
lightweight and common among developers. 
}

\newcommand{\cclass}[1]{{\small\texttt{#1}}}
\newcommand{\cparam}[1]{{\small\texttt{#1}}}

\section{Crypto Library and Crypto Rules}\label{sec:pre}

{
A typical crypto library \textcolor{highlight}{(e.g., Java Cryptography
Architecture)} includes $7$ classes of tasks: (1) message digest, (2) symmetric
encryption, (3) asymmetric encryption, (4) key derivation/generation, (5)
random number generation, (6) key storage, and (7) SSL/TLS and certificates.
Fig.~\ref{fig:classes} shows the parameters used by \our. The parameters of
Fig.~\ref{fig:classes} are logged and used to check the rules. We do not claim
that this library is complete. We include the classes that are used by current
static tools and those that have a corresponding implementation in Java and
Android. These are the classes with the highest number of misuses in Android
and Java~\cite{\cryptoguard, \cryptolint, fahl_ccs2012}. Extensions are
possible, e.g., HKDF~\cite{krawczyk_2010} can be added to the key
derivation class.
}

\begin{enumerate}[(1), wide, labelwidth=!, labelindent=0pt]
\itemsep=0.25em

\item
\textcolor{highlight}{
\cclass{MessageDigest} implements crypto hash functions~\cite{katz_2014}.
}
These functions take as input an arbitrary amount of data and produce
fixed-length hash values, called digests. They are used to check data
integrity.  For this class, the most important parameter is the {algorithm}
(\cparam{alg}) that is used as hash function, for example, SHA$1$, SHA$256$.
Different libraries support different algorithms.

\item \cclass{SymmEncryption} contains block ciphers that are used for
symmetric encryption~\cite{katz_2014}. A block cipher takes as input a block of
data with fixed size (e.g., 128 bits) and a key (whose size is defined by the
algorithm) and it generates the corresponding output block (encrypted or
decrypted). A decrypted block of data is called plaintext, while an encrypted
block is the ciphertext. In addition to the algorithm  (\cparam{alg}), e.g.,
AES, used for encryption and decryption, we log the key (\cparam{key}) and some
other parameters. Block ciphers work on a fixed-size data block. Therefore, to
work on multiple blocks of data (\cparam{\#blocks}) they need to support some
operation modes (\cparam{mode}). For example, by using electronic code book
(ECB) each block is decrypted / encrypted independently from the other blocks.
With cipher block chaining (CBC), each block of plaintext is xored with the
previous block of ciphertext.  The initialization vector (IV) is a parameter
(\cparam{iv}) that defines the block that is xored with the very first block.
Other common operation modes are cipher feedback (CFB), output feedback (OFB),
and Galois/counter (GCM).  Another important parameter is the padding algorithm
(\cparam{pad}), which is the algorithm used to fill the last block of data if
the input  is not a multiple of the block size.  Example of padding algorithms
are \cparam{ZEROPADDING}, where the last block is filled with zeros,
\cparam{PKCS\#5}~\cite{rfc8018} and \cparam{PKCS\#7}~\cite{rfc5652}.

\item \cclass{AsymmEncryption} implements algorithms for public-key
cryptography~\cite{rivest_1990}. These algorithms use a key pair
(\cparam{key}): a public key and a private key. They can be used for (i)
encryption and decryption as well as (ii) signature and verification. For (i),
the message is encrypted with the public key of the receiver. It can be then
decrypted only with the private key of the receiver. For (ii), a message is
signed with the private key of the sender and verified with the corresponding
public key. The parameters of this class are the {algorithm} (\cparam{alg})
used for encryption, e.g., RSA, elliptic curves (EC) or digital signature
algorithm (DSA), and the padding (\cparam{pad}), e.g., \cparam{NOPADDING},
\cparam{PKCS1-v1.5} and \cparam{PSS}~\cite{jager_ccs2018}.

\item \cclass{KeyDerivation} implements algorithms to derive crypto
keys~\cite{katz_2014}. A key derivation function takes as input a password or a
passphrase (\cparam{pass}) and generates a key by using \textcolor{highlight}{a
salt (\cparam{salt}), i.e., a random value,} and by applying a function, e.g.,
hashing, for a fixed number of iterations (\cparam{iter}). The larger is the
number of iterations the harder is to implement brute-force
attacks~\cite{rfc8018}.

\begin{figure}[t]
\centering
\includegraphics[width=\linewidth]{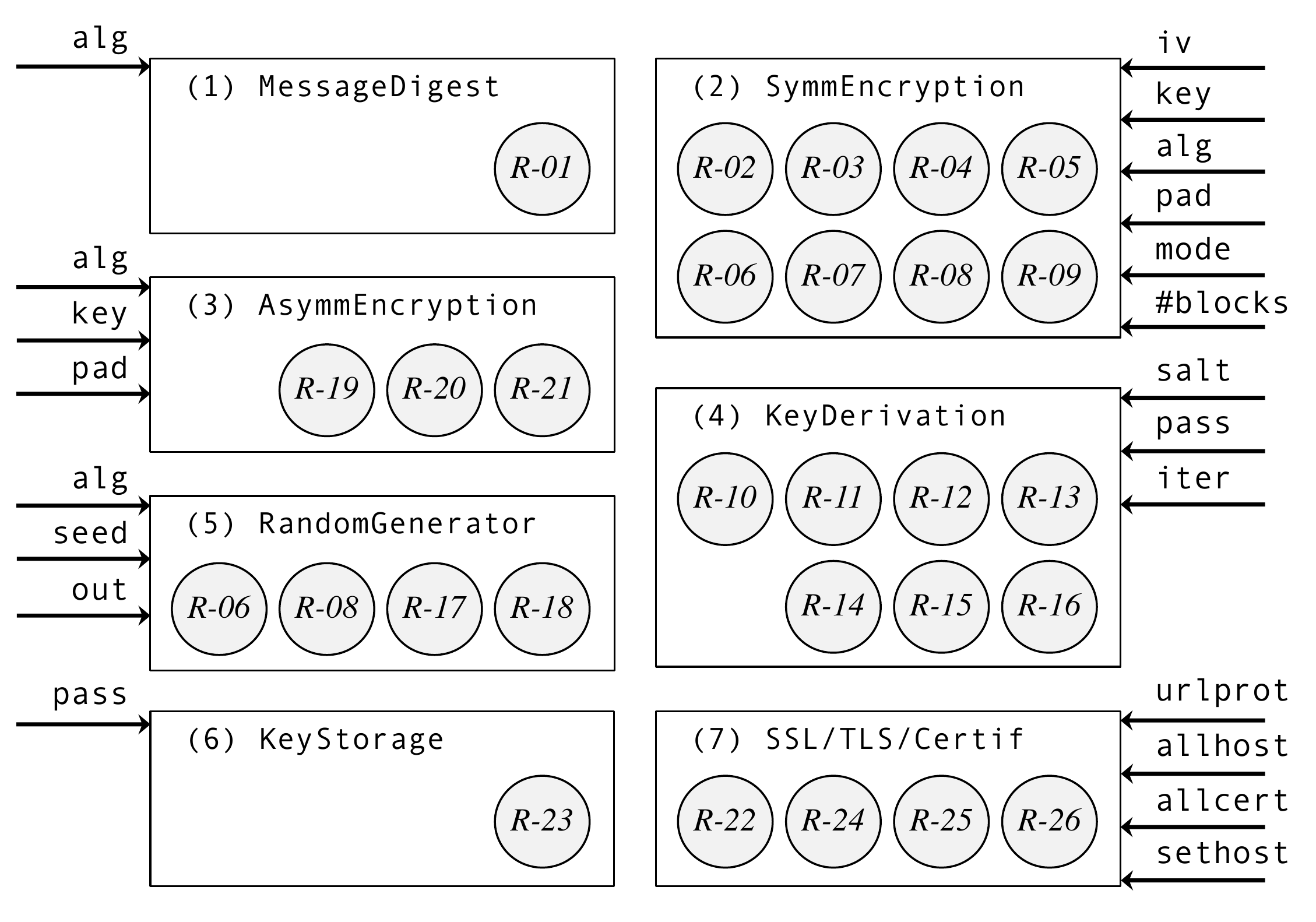}
\caption{Classes of a typical crypto library with their parameters (arrows
entering in the class). For each class we report the crypto rules of
\tablename~\ref{table:rules} that need parameters of that specific
class.}\label{fig:classes}
\vspace{-0.2cm}
\end{figure}

\item \cclass{RandomGenerator} implements algorithms for generating random
numbers.  The relevant parameters are the algorithm (\cparam{alg}) used for
generating the numbers, the bytes of the generated number (\cparam{out}), and
the seed (\cparam{seed}) for the generation.  In this paper we assume that
there are only two categories of algorithms: \cparam{Secure} and
\cparam{NotSecure}. The parameter \cparam{alg} is \cparam{Secure} if it
generates numbers suited for crypto, otherwise it is \cparam{NotSecure}. 

\item \cclass{KeyStorage} implements algorithms to store crypto keys,
certificates and other sensitive content. Usually, it takes as input a password
or a passphrase (\cparam{pass}) to store contents securely.

\item \cclass{SSL/TLS/Certif} is a class including multiple functions for
SSL/TLS and certificates: (1) connections that can be HTTP or HTTPS
(\cparam{urlprot}), (2) host name verification that can accept all the host
names or not (\cparam{allhost}), (3) certificate validation, which can trust
all certificates or not (\cparam{allcert}), and (4) host name verification for
SSL/TLS connections (\cparam{sethost})~\cite{fahl_ccs2012}.

\end{enumerate}


\newcounter{rulecounter}
\newcommand\symmnumber{\textit{R-\twodigits{\arabic{rulecounter}}}}
\newcolumntype{S}{>{\refstepcounter{rulecounter}\symmnumber}l}

\begin{table*}[!ht]
\renewcommand{\arraystretch}{1.1}
\centering
\begin{tabular}{cc}
\begin{tabular}{Sp{0.36\textwidth}c}
\toprule
\multicolumn{1}{c}{\textbf{ID}} & \textbf{Rule Description}& \textbf{Ref.} \\
\midrule
\label{rule:hash}
& Don't use broken hash functions (\texttt{SHA1}, \texttt{MD2}, \texttt{MD5}, ..)
& \cite{fischer_sp2017} \\ 
\label{rule:symmalg} 
& Don't use broken encryption alg. (\texttt{RC2}, \texttt{DES}, \texttt{IDEA} ..)
& \cite{fischer_sp2017}  \\
\label{rule:ecbmode}
& Don't use the operation mode \texttt{ECB} with $>$ 1 data block
& \cite{egele_ccs2013} \\
\label{rule:cbcmode} $\dag$
& Don't use the operation mode \texttt{CBC} (client/server scenarios)
& \cite{vaudenay_ecrypt2002} \\
\label{rule:constkey}
& Don't use a static (= constant) key for encryption 
& \cite{egele_ccs2013} \\
\label{rule:badkey} $\dag$
& Don't use a ``badly-derived'' key for encryption
& \cite{egele_ccs2013} \\
\label{rule:constiv}
& Don't use a static (= constant) initialization vector (IV)
& \cite{egele_ccs2013} \\
\label{rule:badiv} $\dag$ 
& Don't use a ``badly-derived'' initialization vector (IV)
& \cite{egele_ccs2013} \\
\label{rule:sameiv} $\dag$
&  Don't reuse the initialization vector (IV) and key pairs
& \cite{favre_web2018} \\
\label{rule:constsalt}
& Don't use a static (= constant) salt for key derivation
& \cite{egele_ccs2013} \\
\label{rule:shortsalt} $\dag$
& Don't use a short salt ($<$ 64 bits) for key derivation
& \cite{rfc8018}  \\
\label{rule:samesalt} $\dag$
& Don't use the same salt for different purposes
& \cite{favre_web2018} \\
\label{rule:iterat}
&  Don't use $<$ 1000 iterations for key derivation 
& \cite{rfc8018} \\
\bottomrule
\end{tabular}
&
\begin{tabular}{Sp{0.31\textwidth}c}
\toprule
\multicolumn{1}{c}{\textbf{ID}} & \textbf{Rule Description}& \textbf{Ref.} \\
\midrule
\label{rule:weakpass} $\dag$
& Don't use a weak password (score $<$ 3) 
& \cite{wheeler_usenix2016} \\
\label{rule:blackpass} $\dag$ 
& {Don't use a NIST-black-listed password}
& \cite{grassi_nist2017}  \\
\label{rule:reusepass}
& {Don't reuse a password multiple times}
& \cite{grassi_nist2017}  \\
\label{rule:statseed}
& {Don't use a static (= constant) seed for PRNG}
& \cite{nist_80022}  \\
\label{rule:unsaprng}
& {Don't use an unsafe PRNG (java.util.Random)}
& \cite{nist_80022}  \\
\label{rule:rsakeysize}
& Don't use a short key ($<$ 2048 bits) for {RSA}
& \cite{barker_nist2018} \\
\label{rule:rsatextbook} $\dag$
&  Don't use the textbook (raw) algorithm for {RSA}
& \cite{boneh_acrypt2000} \\
\label{rule:rsapadding} $\dag$
&  Don't use the padding \texttt{PKCS1-v1.5} for {RSA}
& \cite{bleich_crypt1998} \\ 
\label{rule:http}
& Don't use HTTP URL connections (use HTTPS)
 & \cite{fahl_ccs2012} \\
\label{rule:store}
& Don't use a static (= constant) password for store
& \cite{grassi_nist2017} \\
\label{rule:hostname}
& Don't verify host names in SSL in trivial ways
& \cite{fahl_ccs2012} \\ 
\label{rule:certif}
& Don't verify certificates in SSL in trivial ways
& \cite{fahl_ccs2012} \\
\label{rule:socket}
& Don't manually change the hostname verifier 
& \cite{fahl_ccs2012} \\
\bottomrule
\end{tabular} \\ \\
\end{tabular}
\caption{\normalfont Crypto rules that are considered in this paper. The symbol
$\dag$ indicates the rules that are not covered by other approaches (we
used \cite{\cryptoguard} as reference).
}\label{table:rules}
\vspace{-0.7cm}
\end{table*}

\vspace{-0.2cm}
\subsection{Threat Model and Crypto Rules}\label{sec:pre:rules}

{
\textcolor{highlight}{
\tablename~\ref{table:rules} reports the rules that are supported by \our. We
collected them from (i) papers ad (ii) documents published by NIST as well as
IETF.}  \figurename~\ref{fig:classes} shows how the rules relate to the crypto
classes. Some rules use parameters from more than one class
(e.g.,~\cryrule{rule:badkey} and \cryrule{rule:badiv}). We use the same threat
model of the current static tools. We briefly describe the crypto rules below.
The severity of most of these rules is discussed in~\cite{\cryptoguard}.
}

\smallskip

{
\textcolor{highlight}{
\underline{\cryrule{rule:hash}} does not let applications use broken hash
functions, e.g., those for which we can generate collisions, like
SHA1~\cite{stevens_2012}.}
\underline{\cryrule{rule:symmalg}} forbids the use of some 
broken algorithms for symmetric encryption, for example, Blowfish, DES, etc.
\underline{\cryrule{rule:ecbmode}} and \underline{\cryrule{rule:cbcmode}}  do
not allow applications to use the operation modes ECB and CBC, respectively.
ECB is well known to be vulnerable since identical blocks of plaintext are
encrypted to identical blocks of ciphertext. This breaks the property
of semantic security~\cite{gold_stoc1982}. CBC is instead vulnerable to 
padding oracle attacks in client-server scenarios~\cite{vaudenay_ecrypt2002}. 
\underline{\cryrule{rule:constkey}} and \underline{\cryrule{rule:badkey}} put
restrictions on how to generate keys. \cryrule{rule:constkey} requires that the
keys for symmetric encryption are randomly generated by the application instead
of being hard-coded in the app as constants. \cryrule{rule:badkey} requires
the keys to have enough randomness, i.e., they should be generated by using a
random generator that is considered secure for crypto.
\underline{\cryrule{rule:constiv}} and \underline{\cryrule{rule:badiv}} are
similar to \cryrule{rule:constkey} and \cryrule{rule:badkey}, but they consider
the IVs that are used in symmetric encryption instead of the keys. The IVs, in
fact, should always be random and non-constant to strengthen data confidentiality
when they are paired with some operation modes, e.g., GCM.
\underline{\cryrule{rule:sameiv}} requires that the same pair (key, IV) is
never reused to encrypt different messages. Reusing the same pair (key, IV)
makes the encryption predictable.
\underline{\cryrule{rule:constsalt}} is the same as \cryrule{rule:constkey}: it
is, however, applied to the salt used in key generation instead of the keys used
in symmetric encryption.
\underline{\cryrule{rule:shortsalt}} requires the salt to be large enough ($\ge$
$64$ bits) to protect the password used for key generation.
\underline{\cryrule{rule:samesalt}} prohibits the reuse of the same salt
because it defeats the purpose of adding randomness to the corresponding
password.
\underline{\cryrule{rule:iterat}} requires to use a sufficient number of
iterations to generate the key so that brute-force attacks become infeasible.
\underline{\cryrule{rule:weakpass}} and \underline{\cryrule{rule:blackpass}}
require to use a password that has not been black-listed and that is ``hard''
enough for password-based encryption, respectively.
\underline{\cryrule{rule:reusepass}} forbids using the same password multiple
times (e.g., constant passwords).
\underline{\cryrule{rule:statseed}} requires to use a random value as seed
instead of a constant value for pseudo-random number generation (PRNG). Using a
constant seed defeats the purpose of generating random number as the sequence
of numbers that is generated becomes predictable.
\underline{\cryrule{rule:unsaprng}} does not allow applications to use PRNGs
that are not approved for crypto operations, for example
{\small\texttt{java.util.Random}}~\cite{\cryptoguard}.
\underline{\cryrule{rule:rsakeysize}}, \underline{\cryrule{rule:rsatextbook}}
and \underline{\cryrule{rule:rsapadding}} forbid some configurations of the RSA
algorithm. In particular, the key should be  $\ge 2048$ bits and a padding
algorithm different from \cparam{NOPADDING} (\cryrule{rule:rsatextbook}) and
\cparam{PKCS1-v1.5} (\cryrule{rule:rsapadding}) must be used for encryption /
decryption.
\underline{\cryrule{rule:http}} forbids the use of HTTP and requires the use of
the more secure alternative HTTPS.
\underline{\cryrule{rule:store}} forbids the use of static passwords for key
storage.
\underline{\cryrule{rule:hostname}} and \underline{\cryrule{rule:certif}}
require to properly verify host names and certificates. For example, accepting
all host names or all certificates should not be allowed.
\underline{\cryrule{rule:socket}} forbids to modify the standard host name
verifier, which can lead to insecure communication over SSL/TLS.
}

\begin{figure*}[!ht]
\centering
\includegraphics[width=0.8\textwidth]{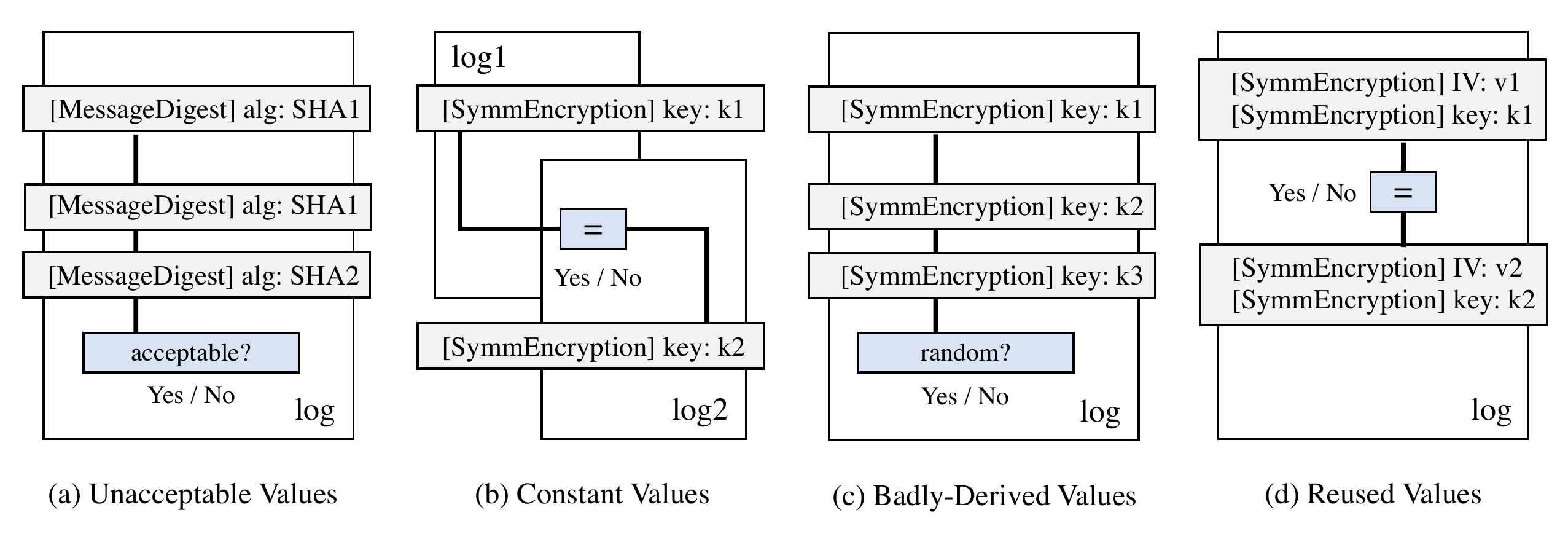}
\caption{We define four checking procedures to cover all the crypto rules of
\tablename~\ref{table:rules}. 
(a) We check if some unacceptable values are used to configure a parameter of a
crypto class (e.g., SHA1 for rule~\cryrule{rule:hash}).
(b) We check if a parameter is configured with constant values by verifying if
the same values are found in two different executions of an application
(e.g., same key for rule~\cryrule{rule:constkey}). 
(c) We check if the values of a parameter of a crypto class has enough
randomness (e.g., the keys for rule~\cryrule{rule:badkey}). 
(d) We check if some values of a parameter are reused multiple times during the
execution of an application (e.g., the pairs (key, IV) for \cryrule{rule:sameiv}).
}\label{fig:check}
\vspace{-0.4cm}
\end{figure*}

\section{Checking Crypto Rules Dynamically}\label{sec:check}

{
We define four {checking procedures} to cover the crypto rules reported in
TABLE~\ref{table:rules}. Each checking procedure covers multiple rules, while
each rule is verified by only one checking procedure. These checking procedures
are shown graphically in \figurename~\ref{fig:check} and explained in detail in
the next sections. These procedures are {generic}: they can be applied to new
crypto rules if needed. 
}

\subsection{Unacceptable Values}\label{sec:check:a}

{
The checking procedure of \figurename~\ref{fig:check} (a) extracts from the log
all the values of a parameter or a combination of parameters and verifies that
they can be used to configure the corresponding crypto class. All the values
that are collected from the log are sent to a rule-specific function that says
`yes' if the values are allowed by the rule or `no' otherwise.
For \cryrule{rule:hash}, for instance, we need to ensure that the parameter
{\cparam{alg}} of {\cparam{MessageDigest}} never takes one of the following
values: SHA$1$, MD$2$, MD$5$, etc.
This is the most basic checking procedure and it is used to check the highest
number of crypto rules. 
We describe how we check the crypto rules that fall under this type below. For
each rule, we report which property must be satisfied by all the values that
are collected for that rule.
}

\smallskip

\begin{lstlisting}[style=check]
(@\textit\normalfont{\cryrule{rule:hash}}@): MessageDigest.alg (@$\notin$@) {`SHA1', ..}
(@\textit\normalfont{\cryrule{rule:symmalg}}@): SymmEncryption.alg (@$\notin$@) {`DES', ..}
\end{lstlisting}

\noindent For rules \cryrule{rule:hash} and \cryrule{rule:symmalg} we simply
check that broken algorithms are not used for message digest and encryption,
respectively.

\begin{lstlisting}[style=check]
(@\textit\normalfont{\cryrule{rule:ecbmode}}@): SymmEncryption.mode (@$\neq$@) `ECB' or
        SymmEncryption.#blocks (@$=$@) 1 
(@\textit\normalfont{\cryrule{rule:cbcmode}}@): SymmEncryption.mode (@$\neq$@) `CBC'
\end{lstlisting}

\noindent For rules \cryrule{rule:ecbmode} and \cryrule{rule:cbcmode}, we check
that the operation modes ECB / CBC are not used. We accept the use of ECB for $1$
data block.

\begin{lstlisting}[style=check]
(@\textit\normalfont{\cryrule{rule:shortsalt}}@): KeyDerivation.salt (@$\ge$@) 64 bits
(@\textit\normalfont{\cryrule{rule:iterat}}@): KeyDerivation.iter (@$\ge$@) 1000
\end{lstlisting}

\noindent For key derivation we check that the lengths of the salts
in the log are always $\ge$ $64$ bits and the number of iterations 
is $\ge$ $1000$.

\begin{lstlisting}[style=check]
(@\textit\normalfont{\cryrule{rule:weakpass}}@): KeyDerivation.pass (@$\notin$@) BadPass
(@\textit\normalfont{\cryrule{rule:blackpass}}@): score(KeyDerivation.pass) (@$\ge$@) 3
\end{lstlisting}

\noindent For key derivation, we check if the password is broken (i.e., it
belongs to \texttt{BadPass}\footnote{We used a set of passwords
from: \url{https://github.com/cry/nbp}.}) or weak. To check if a password is
weak we use zxcvbn~\cite{wheeler_usenix2016} and consider it bad if it has a
score $<$ $3$.

\begin{lstlisting}[style=check]
(@\textit\normalfont{\cryrule{rule:unsaprng}}@): RandomGenerator.alg (@$=$@) `Secure'
\end{lstlisting}

\noindent We check that the algorithm to generate random numbers is \cparam{Secure},
i.e., it should generate truly-random numbers. For example in Java,
\cparam{java.secure.SecureRandom} must be used instead of \cparam{java.util.Random},
whose randomness is limited.

\begin{lstlisting}[style=check]
(@\textit\normalfont{\cryrule{rule:rsakeysize}}@): AsymmEncryption.alg (@$\neq$@) `RSA' or
        AsymmEncryption.key (@$\ge$@) 2048 bits
(@\textit\normalfont{\cryrule{rule:rsatextbook}}@): AsymmEncryption.alg (@$\neq$@) `RSA' or
        AsymmEncryption.pad (@$\neq$@) `NOPADDING'
(@\textit\normalfont{\cryrule{rule:rsapadding}}@): AsymmEncryption.alg (@$\neq$@) `RSA' or
        AsymmEncryption.pad (@$\neq$@) `PKCS1-v1.5'
\end{lstlisting}

\noindent These rules do not admit encryption keys that are $< 2048$ bits for RSA and
require some padding algorithm different from \cparam{NOPADDING} and \cparam{PKCS1-v1.5}
for encryption/decryption~\cite{bleich_crypt1998}.

\begin{lstlisting}[style=check]
(@\textit\normalfont{\cryrule{rule:http}}@): SSL/TLS/Cert.urlprot (@$\neq$@) `HTTP'
\end{lstlisting}

\noindent We check that HTTP is never used as a connection protocol.

\begin{lstlisting}[style=check]
(@\textit\normalfont{\cryrule{rule:hostname}}@): SSL/TLS/Cert.allhost (@$=$@) `False'
(@\textit\normalfont{\cryrule{rule:certif}}@): SSL/TLS/Cert.allcert  (@$=$@) `False'
(@\textit\normalfont{\cryrule{rule:socket}}@): SSL/TLS/Cert.sethost  not assigned
\end{lstlisting}

\noindent For rules \cryrule{rule:hostname} and \cryrule{rule:certif}, we
check that apps do not naively verify host names and certificates (e.g., they do
not verify the host name at all or they trust all certificates). For rule
\cryrule{rule:socket}, we check that the default host name verifier is not
replaced to avoid host name verification, e.g., in Java by creating
sockets\footnote{Android SSL:
\url{https://developer.android.com/training/articles/security-ssl}.}.

\subsection{Constant Values}\label{sec:check:b}

{
The checking procedure of \figurename~\ref{fig:check} (b) verifies if a
parameter of a crypto class is constant or not. For instance, for rule
\cryrule{rule:constkey} we need to ensure that applications do not use static
encryption keys that are hard-coded in the app. Ideally, the keys should be
generated with a proper random generator.
To verify the rules in this category, we examine the logs of two
executions of the same application and check that the values that are found in
one of the execution log is not present in the other and vice versa. 
For example, for rule \cryrule{rule:constkey} we check the following:
}

\begin{lstlisting}[style=check]
(@\textit\normalfont{\cryrule{rule:constkey}}@): { SymmEncryption.key }(@$_1$@) (@$\cap$@) 
        { SymmEncryption.key }(@$_2$@) (@$=$@) (@$\emptyset$@)
\end{lstlisting}

\noindent where we used \{\ \}$_1$ to indicate the values collected in the first
log and $\{\ \}_2$ the values collected in the second log. In a similar
way, we check the rules \cryrule{rule:constiv}, \cryrule{rule:constsalt},
\cryrule{rule:statseed}, and \cryrule{rule:store} with the {values of
{\cparam{SymmEncryption.iv}}, {\cparam{KeyDerivation.salt}}},
\mbox{{\cparam{RandomGenerator.seed}}, and {\cparam{KeyStorage.pass}}.}

\subsection{Badly-derived Values}\label{sec:check:c}

{
The checking procedure reported in \figurename~\ref{fig:check} (c) verifies if
a value is truly random or not. For rule \cryrule{rule:badkey}, for example,
we need to guarantee that the application uses encryption keys that have
enough randomness.
To verify the rules of this type, we collect all the values of the relevant
parameter and we make the following three checks sequentially (box
\textit{random?} of \figurename~\ref{fig:check} (c)):
\begin{enumerate}[1.]
\itemsep0.4em
\item if the value is obtained from \cclass{RandomGenerator} with \cparam{alg}
$=$ \cparam{`Secure'}, then we consider it a legit value;
\item if the value is obtained from \cclass{RandomGenerator} with \cparam{alg}
$\neq$ \cparam{`Secure'}, then we consider it a bad value;
\item otherwise we apply the NIST tests for randomness~\cite{nist_80022} and if
at least one test fails we consider it a bad value.
\end{enumerate} 
The first two checks try to determine the origin of the value, i.e., if
it has been generated by \cclass{RandomGenerator} (parameter \cparam{out}).
If the origin cannot be determined, e.g., the value is generated in some other
ways by the application, then we use the NIST tests.
\textcolor{highlight}{
For each NIST test we have three possible outcomes: (i) failure, (ii) success,
or (iii) skipped because there are not enough bits to apply the specific test.
We consider that an app violates a rule if at least one NIST test fails. This
policy can be easily changed by the user.
}
We apply this procedure to rules \cryrule{rule:badkey} and
\cryrule{rule:badiv}. Verifying the randomness of values is a challenging task.
While this test does not ensure that the values that pass the check are truly
random, it finds obvious sources of non-randomness.  Static approaches do not
typically check these types of rules.
}

\subsection{Reused Values}\label{sec:check:d}

{
The checking procedure of \figurename~\ref{fig:check} (d) checks if a value or
a combination of values of a parameter of a crypto class is reused across the
executions of an application.  For instance, for rule~\cryrule{rule:sameiv}, we
have to ensure that the same pair (key, IV) is never reused to encrypt
different messages. The checking procedure collects all the values from the
log and checks if there are duplicates:
\begin{lstlisting}[style=check]
(@\textit\normalfont{\cryrule{rule:sameiv}}@): containsDuplicates(
      { (SymmEncryption.key, 
          SymmEncryption.iv) }) = False
\end{lstlisting}
We used this checking procedure for the rules \cryrule{rule:sameiv} and 
\cryrule{rule:samesalt}. Static approaches do not typically check these
types of rules.
}

\newcommand{\javadoc}{\url{https://docs.oracle.com/javase/7/docs/technotes/guides/security/crypto/CryptoSpec.html}}

\section{Implementation of CRYLOGGER for Android}\label{sec:impl}

{
We implemented \our to detect crypto {misuses in Java} and Android apps by
instrumenting classes of the Java Cryptography Extension (JCE) and the Java
Cryptography Architecture (JCA), which are part of the Java standard
library\footnote{Documentation about JCA and JCE can be found here:
\javadoc~(Java 7).}. These classes provide a common interface for crypto algorithms to
all Java apps. This interface is then implemented by `providers', i.e.,
specific crypto libraries, e.g., SunJCE, BouncyCastle, etc. Thus, they are the
perfect place to detect crypto misuses in Android (as well as Java) apps.
\tablename~\ref{table:java} reports the mapping of the classes of
Section~\ref{sec:pre} (Crypto Classes in the table) to the Java classes that we
instrumented.
In some cases, a single crypto class, e.g., \cclass{RandomGenerator}, is mapped
to multiple Java classes, e.g., \cclass{Random} and \cclass{SecureRandom}.
In the appendices (\tablename~\ref{table:javafull}) we report for each class
the member methods that we instrumented and the parameters that we collected
for each Java class.
}

\begin{table}
\centering
\scriptsize
\begin{tabular}{ll}
\toprule
{\larger\textbf{Crypto Classes}} & {\larger\textbf{Java Classes}} \\
\midrule
\rowcolor{gray!20}
\texttt{MessageDigest} & 
\texttt{java.security.MessageDigest} \\
\rowcolor{white}
\texttt{SymmEncryption} & 
\texttt{javax.crypto.Cipher} \\ 
\rowcolor{gray!20}
\texttt{AsymmEncryption} & 
\texttt{javax.crypto.Cipher} \\ 
\rowcolor{gray!20}
& \texttt{java.security.Signature} \\
\rowcolor{white}
\texttt{KeyDerivation} & 
\texttt{javax.crypto.spec.PBEKeySpec} \\
& \texttt{javax.crypto.spec.PBEParameterSpec} \\
\rowcolor{gray!20}
\texttt{RandomGenerator} & 
\texttt{java.util.Random} \\
\rowcolor{gray!20}
& \texttt{java.security.SecureRandom} \\
\rowcolor{white}
\texttt{KeyStorage} & 
\texttt{java.security.KeyStore} \\
\rowcolor{gray!20}
\texttt{SSL/TLS/Certif.} & 
\texttt{java.net.URL} \\
\rowcolor{gray!20}
& \texttt{java.net.ssl.SSLContext} \\
\rowcolor{gray!20}
& \texttt{java.net.ssl.SocketFactory} \\
\rowcolor{gray!20}
& \texttt{java.net.ssl.HttpsURLConnection} \\
\rowcolor{white}
\bottomrule
\\
\end{tabular}
\caption{\normalfont Mapping from the crypto library of Section~\ref{sec:pre}
to the Java standard library.
}\label{table:java}
\vspace{-0.7cm}
\end{table}


\subsection{Automated Testing of Android Apps}\label{sec:impl:auto}

{
\textcolor{highlight}{We ran \our on $1780$ Android apps from the official Google
Play Store. These are the most popular free apps of $33$ different categories
(Section~\ref{sec:res2}).
In this section, we discuss how we automated the testing for such a large
number of apps.}
}

\smallskip

{
We implemented a Python script to perform the following nine steps. Step (S1)
starts an Android emulator, whose Java library has been instrumented with \our
(or we can use a real device). (S2) downloads the chosen app from the Google
Play Store market. (S3) configures the user interface (UI) of the emulator to
facilitate random testing (more details below). (S4) installs the app on the
emulator with the android debug bridge (ADB)\footnote{Android ADB:
\url{https://developer.android.com/studio/command-line/adb}.}. (S5) uses
Monkey to send random events
to the UI of the app (the number of UI events is configurable and Monkey can be
replaced with other tools). We call `events' the actions that can be performed
on the UI of an app, such as scrolling, touching, inserting text, etc. (S6)
collects the crypto log. (S7) uninstalls the app and deletes its data with ADB.
(S8) checks the crypto rules and reports the rules that have been violated.
(S9) tests another app starting from Step (S4), if it is necessary.
}

\smallskip

{
Android apps are UI driven~\cite{yerima_eurasip2019}. Therefore to verify
an app, there are two main alternatives: manual tests, where a user needs to
interact with the UI of the app, and automated tests, where the UI events are
generated by a tool~\cite{choudhary_ase2015}, e.g., Monkey.
Since the results of any dynamic tool, including \our, are as good as the
UI events used to exercise the app, it is critical to define how to test the
apps to detect crypto misuses.
Since we wanted to fully automate the testing process, we decided to exclude
the option of performing manual tests. We decided to use Monkey for the
experimental results in Sections~\ref{sec:res1} and \ref{sec:res2}. 
Monkey is the most popular tool for random-based testing and compared to other
tools for random-based generation is known to be the most
effective~\cite{choudhary_ase2015}.
The main advantage of Monkey is that it is fully automated. It is also fully
integrated in Android Studio, and thus supported on all the apps of the Google
Play Store and on different Android versions. In addition, it is fast because
to generate events it does not need to maintain any information (state) of the
app.
It has, however, two limitations: (1) random events generate unintended
behaviors, for instance, turning off Internet or closing the
app~\cite{yerima_eurasip2019}, and (2) poor app coverage since the events are
generated randomly, for example, Monkey cannot perform complex operations, such
as app registration or login.
}

\begin{enumerate}[(1), wide, labelwidth=!, labelindent=0pt]
\itemsep=0.5em

\item[(1) \textit{Unintended Behaviors}:]
{
To address this problem, we added Step (S3) mentioned above. This step (i)
activates the immersive mode\footnote{Immersive:
\url{https://developer.android.com/training/system-ui/immersive}.}, where an app
is fixed on the screen and there is no easy way to return to the home screen,
(ii) removes the quick settings, so that Monkey cannot interact with system
configurations, e.g., Wi-Fi, and (iii) disables physical buttons, e.g., power
and volume, to focus the attention of Monkey on the app. We observed that these
modifications eliminate most of the unintended behaviors.
}

\item[(2) \textit{Poor App Coverage:}]
{
To improve the coverage, we evaluated many tools for test generation, e.g.,
SmartDroid~\cite{zheng_spsm2012}, DroidBot~\cite{li_icsec2017}, and
Dynodroid~\cite{machiry_fse2013}.  Their main drawbacks are that the support is
limited (they work on specific versions of Android) and they are typically slower
than Monkey, as they need to keep some information about the state of the app
and update it to explore new behaviors (e.g., a control-flow
graph~\cite{li_icsec2017}).
Due to these limitations, we decided to use Monkey.
We noticed that Monkey is actually capable of triggering many of the crypto
misuses, even if the UI events are completely random. Most of the functions
that we instrumented (\tablename~\ref{table:javafull}) are, in fact, used to
initialize some basic, critical crypto classes, and therefore they are
relatively easy to trigger. We observed that Monkey achieves $\sim25\%$ of line
coverage on average, but it reports as many crypto misuses as \cryptoguardnref,
which employs static analysis (Section~\ref{sec:res1}).
This choice carries some limitations, i.e., the possibility of false negatives,
because some parts of the apps are hard to explore (e.g., login).
It is worth to mention, however, that \our can be configured to use any other
UI exercisers as well as manually-written sequences of UI events. For example, if
developers have sequences of events to stimulate their apps, it can exploit
those to obtain higher coverage.
In future, we plan to build our own UI event generator tool specialized for
crypto.
}

\end{enumerate}

\subsection{Details about Crypto Rules Checking}\label{sec:impl:details}

{
We used the checking procedures explained in Section~\ref{sec:check} to check
the crypto rules for the Android apps, but we made few adaptations.
The functions that we instrumented for rules~\cryrule{rule:hostname}
and~\cryrule{rule:certif} (\tablename~\ref{table:javafull}) take as input some
classes for which the developer of the application has to implement some
methods, e.g., the method \texttt{verify()} to verify the host name.  To obtain
the values of the parameters \cparam{allhost} and \cparam{allcert} that are
used by rules~\cryrule{rule:hostname} and~\cryrule{rule:certif}, during the
logging, we pass some erroneous values, such as \texttt{NULL} or empty strings, to
determine if those functions were implemented naively.
For the rules that require two executions (see Fig.~\ref{fig:check} (b)), we
obtain the two logs by running the application on two different instances of
the emulator. We also make sure that if we see a value that is in both logs,
then this is caused by constants hard-coded in the app.
}

%
%

\section{\textcolor{highlight}{Experimental Setup and Benchmarks}}\label{sec:res0}

{
We evaluated \our on two sets of benchmarks.
The first set consists of Android apps. We downloaded $2148$ free Android apps
from the Google Play Store. These cover the most popular free apps of $33$
different categories.
We discarded $110$ of these apps since they do not use any crypto APIs.
We discarded $258$ of these apps as they do not work on the Android emulator
either because they keep crashing or they require libraries that cannot be
installed in the emulator environment.
The results of running \our on the remaining $1780$ apps are discussed in
Section~\ref{sec:res2}. We used a random subset of these apps to compare \our
against \cryptoguardnref as described in Section~\ref{sec:res1}.
The second set of benchmarks is the \cryptoapinref, a set of Java applications
that include crypto misuses. The \cryptoapiname was originally proposed to
compare static approaches. We extended it and then used it to compare \our
against \cryptoguardname (see Section~\ref{sec:res1}).
}

\section{Results: Comparison with Cryptoguard}\label{sec:res1}

{
We compared \our against \cryptoguardnref, one of the most effective static
tools in detecting crypto misuses in Java-based applications. We could not
compare \our against a dynamic tool because, to the best of our knowledge, \our
is the only approach to detect misuses dynamically for a large number of rules
(Section~\ref{sec:rel}).
We chose \cryptoguardname among many available static tools, e.g., \cryptolintnref,
\cryslnref, because it has been recently shown that \cryptoguardname is the
tool with the lowest false positive and false negative rates among
them~\cite{afrose_secdev2019}. It is also the tool that supports the largest
number of crypto rules.
We compared \our and \cryptoguardname by using $2$ datasets. 
\textcolor{highlight}{The first consists of $150$ Android apps we randomly chose
from the set of $1780$ apps (Section~\ref{sec:res0}).}  For this dataset, we
evaluated the execution times and the number of crypto misuses found by the two
tools.
The second dataset is the \cryptoapinref, a set of Java benchmarks that include
crypto misuses. For this dataset, we determined the false positive and the
false negative rates of the two tools. We also extended the \cryptoapiname with
more benchmarks to cover cases relevant to dynamic approaches.
}

\begin{figure*}[!tbp]
\hspace{-1cm}
\centering\noindent
\begin{minipage}{\textwidth}
\begin{tabular}{c|@{ }c|@{ }c|@{ }c}
\begin{minipage}[t]{0.242\textwidth}
  {\hfil\includegraphics[width=\textwidth, clip, trim = {2.3cm 0.5cm 3.3cm 4.2cm}]{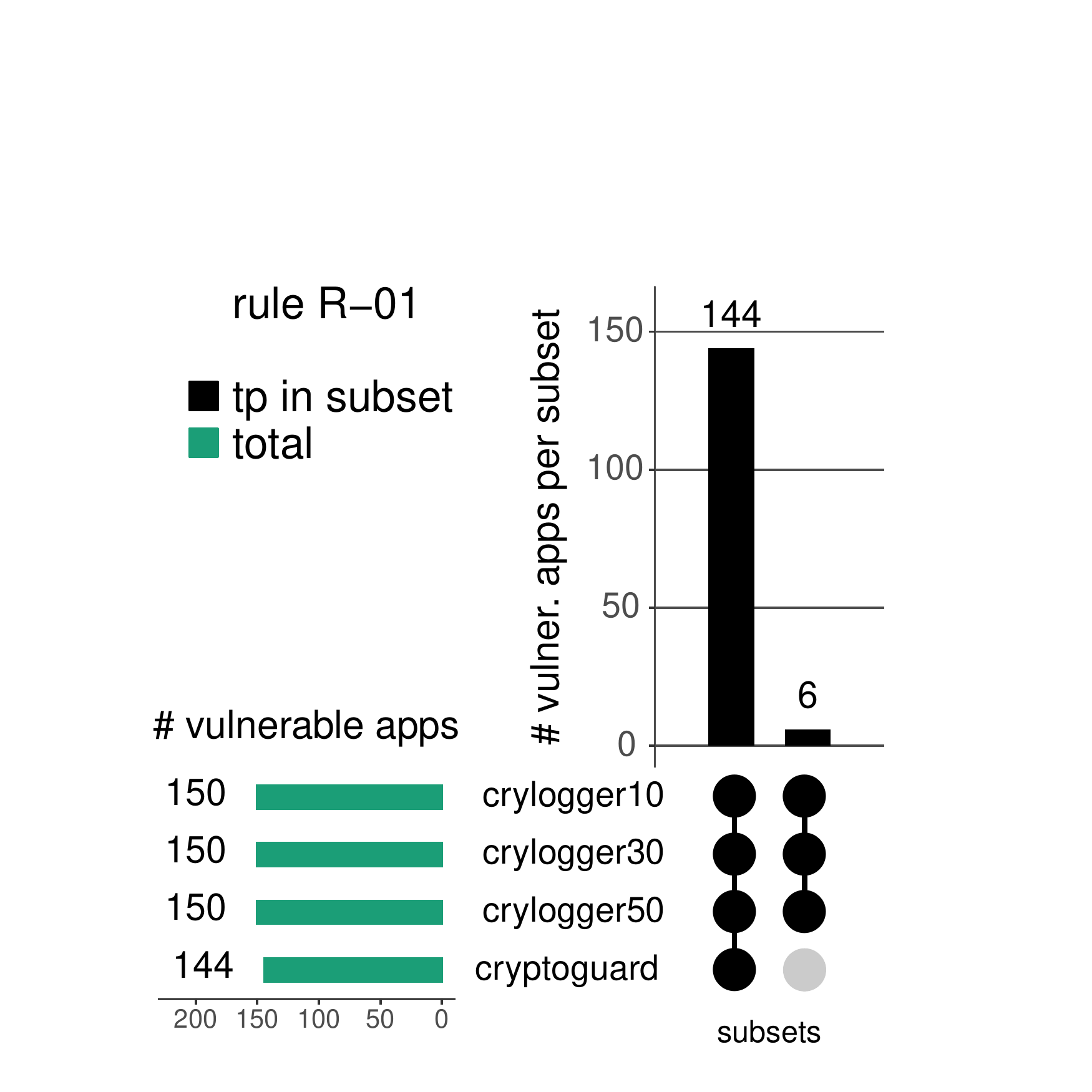}}
\end{minipage} &
\begin{minipage}[t]{0.242\textwidth}
  {\hfil\includegraphics[width=\textwidth, clip, trim = {2.3cm 0.5cm 3.3cm 4.2cm}]{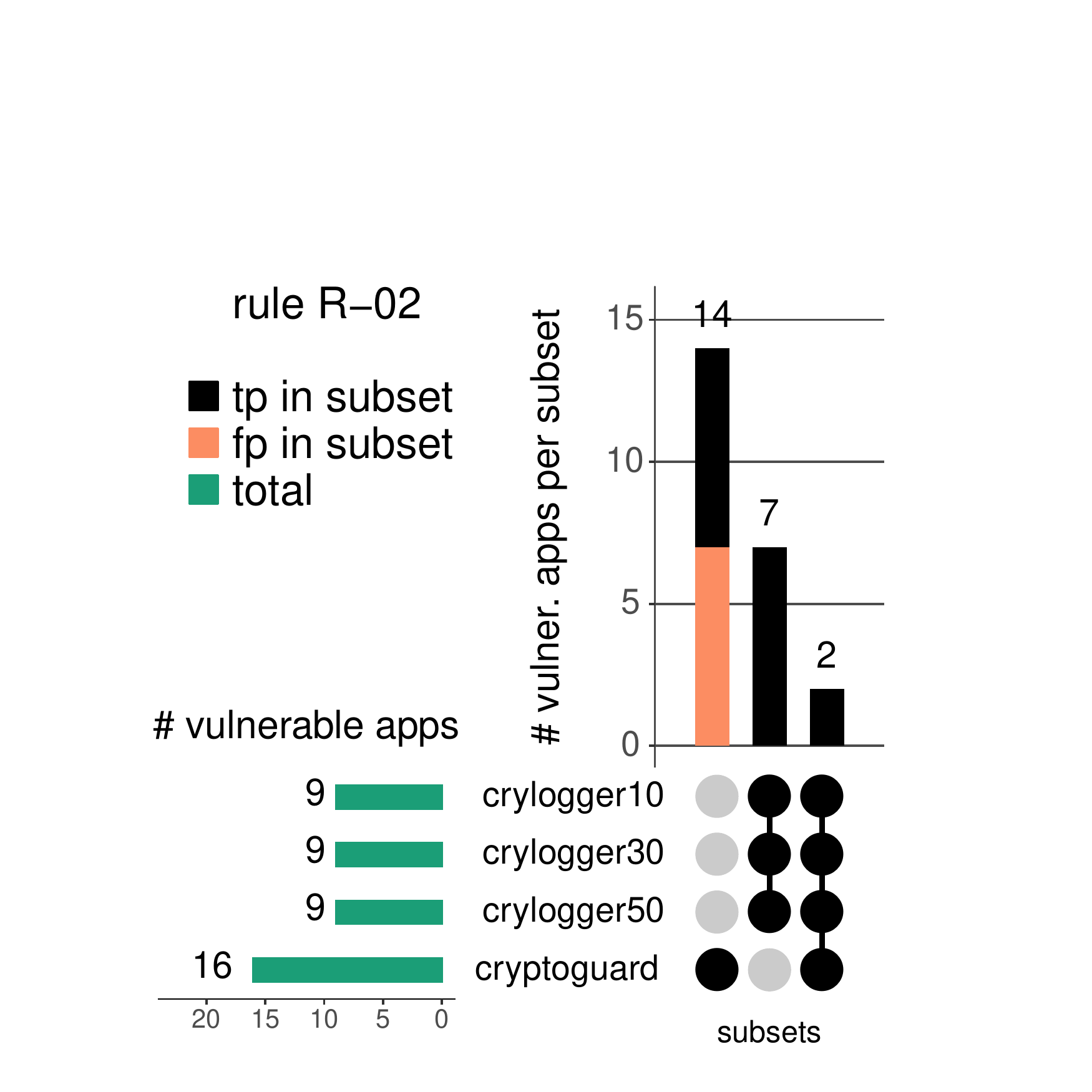}}
\end{minipage} &
\begin{minipage}[t]{0.242\textwidth}
  {\hfil\includegraphics[width=\textwidth, clip, trim = {2.3cm 0.5cm 3.3cm 4.2cm}]{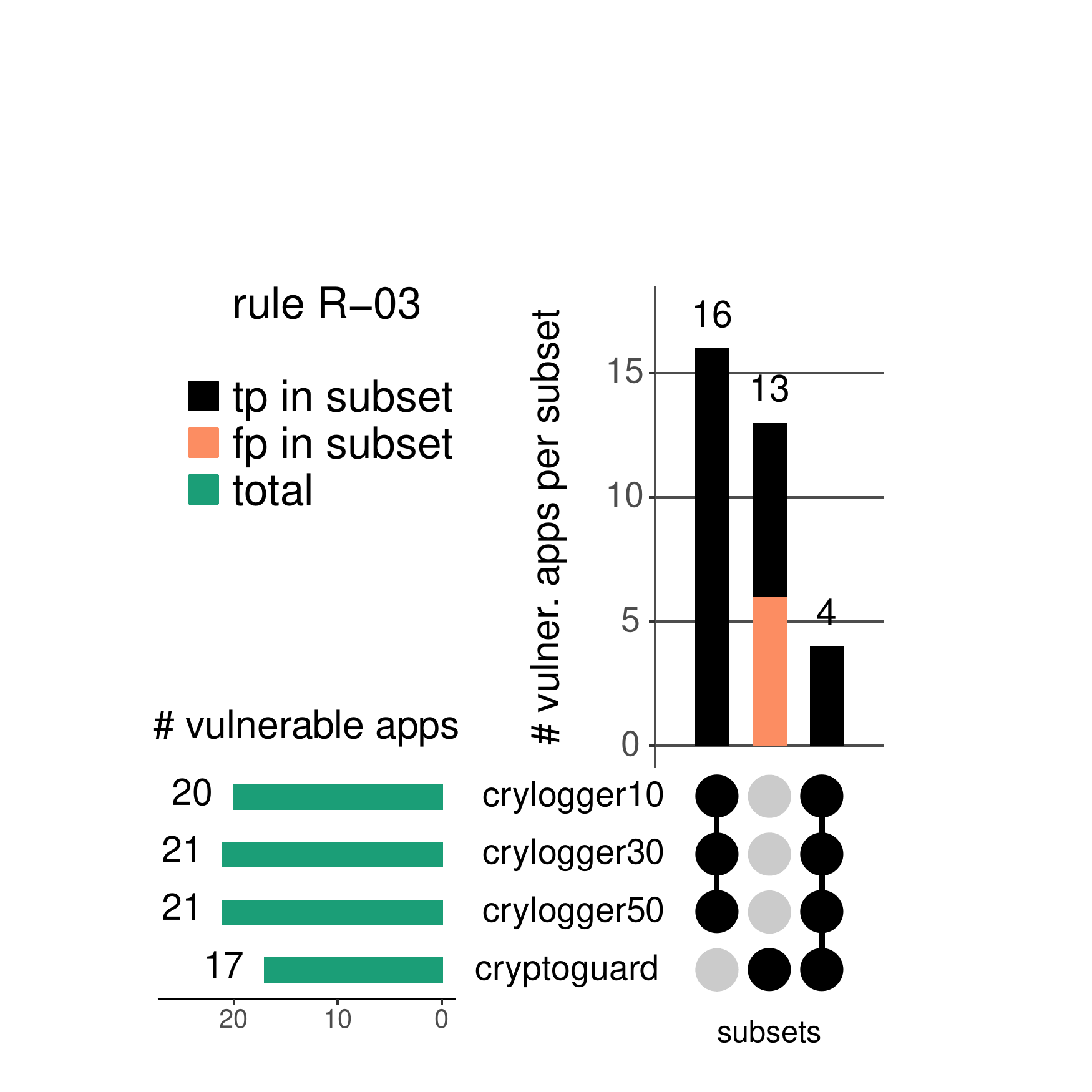}} 
\end{minipage} &
\begin{minipage}[t]{0.242\textwidth}
  {\hfil\includegraphics[width=\textwidth, clip, trim = {2.3cm 0.5cm 3.3cm 4.2cm}]{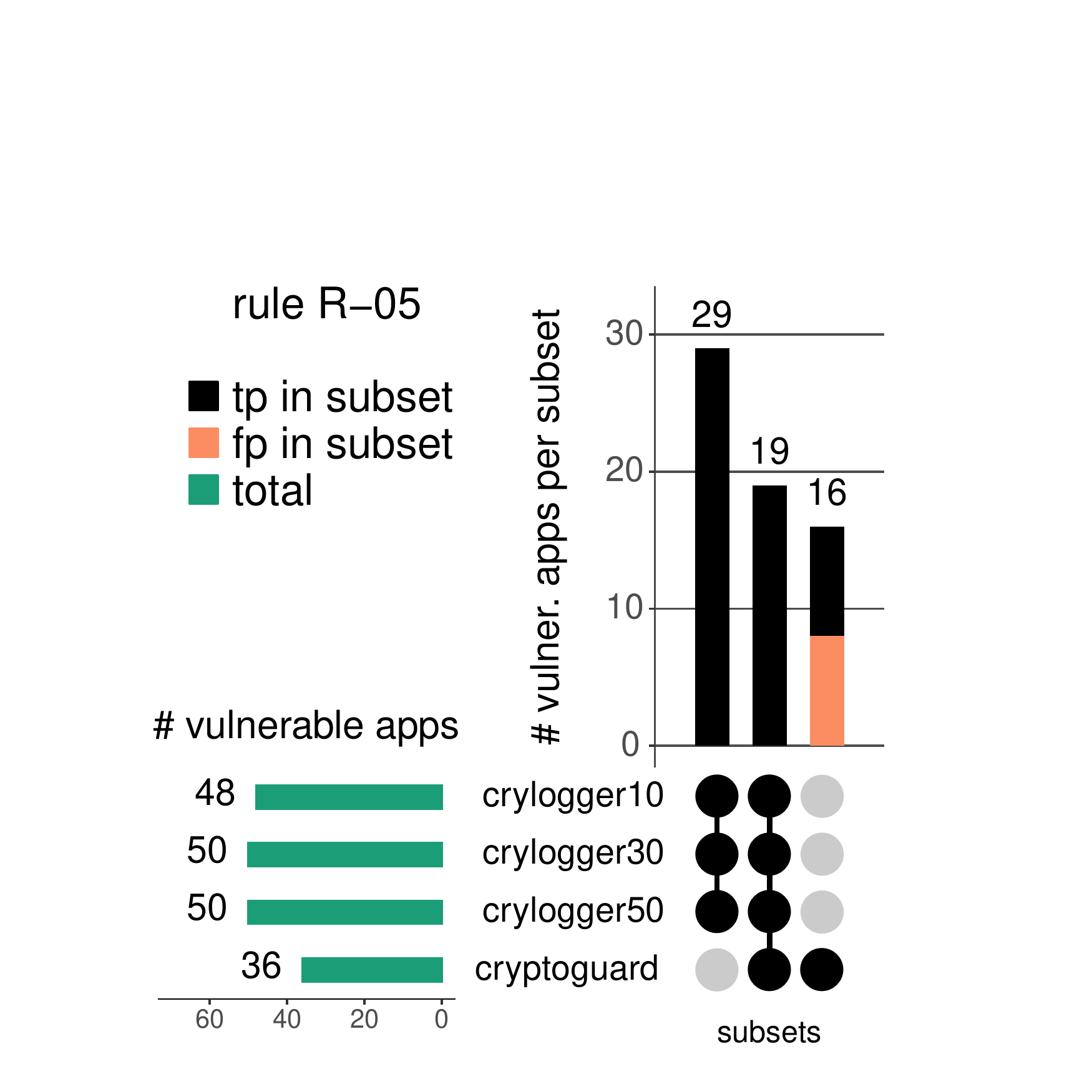}} 
\end{minipage} \\
\end{tabular}
\end{minipage}

\vspace{0.2cm}

\hspace{-1cm}
\centering\noindent
\begin{minipage}{\textwidth}
\begin{tabular}{c|@{ }c|@{ }c|@{ }c}
\begin{minipage}[t]{0.242\textwidth}
  {\hfil\includegraphics[width=\textwidth, clip, trim = {2.3cm 0.5cm 3.3cm 4.2cm}]{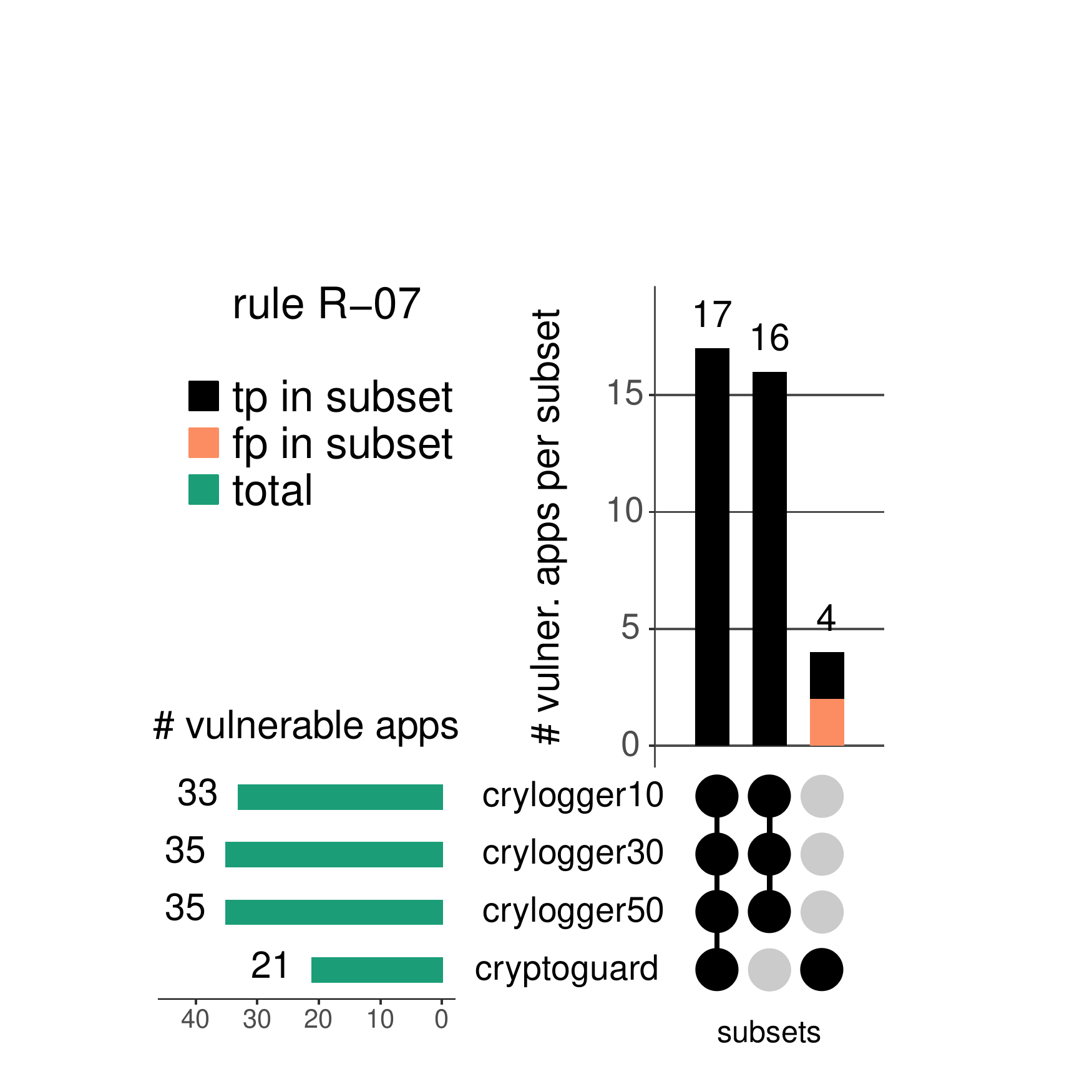}}
\end{minipage} &
\begin{minipage}[t]{0.242\textwidth}
  {\hfil\includegraphics[width=\textwidth, clip, trim = {2.3cm 0.5cm 3.3cm 4.2cm}]{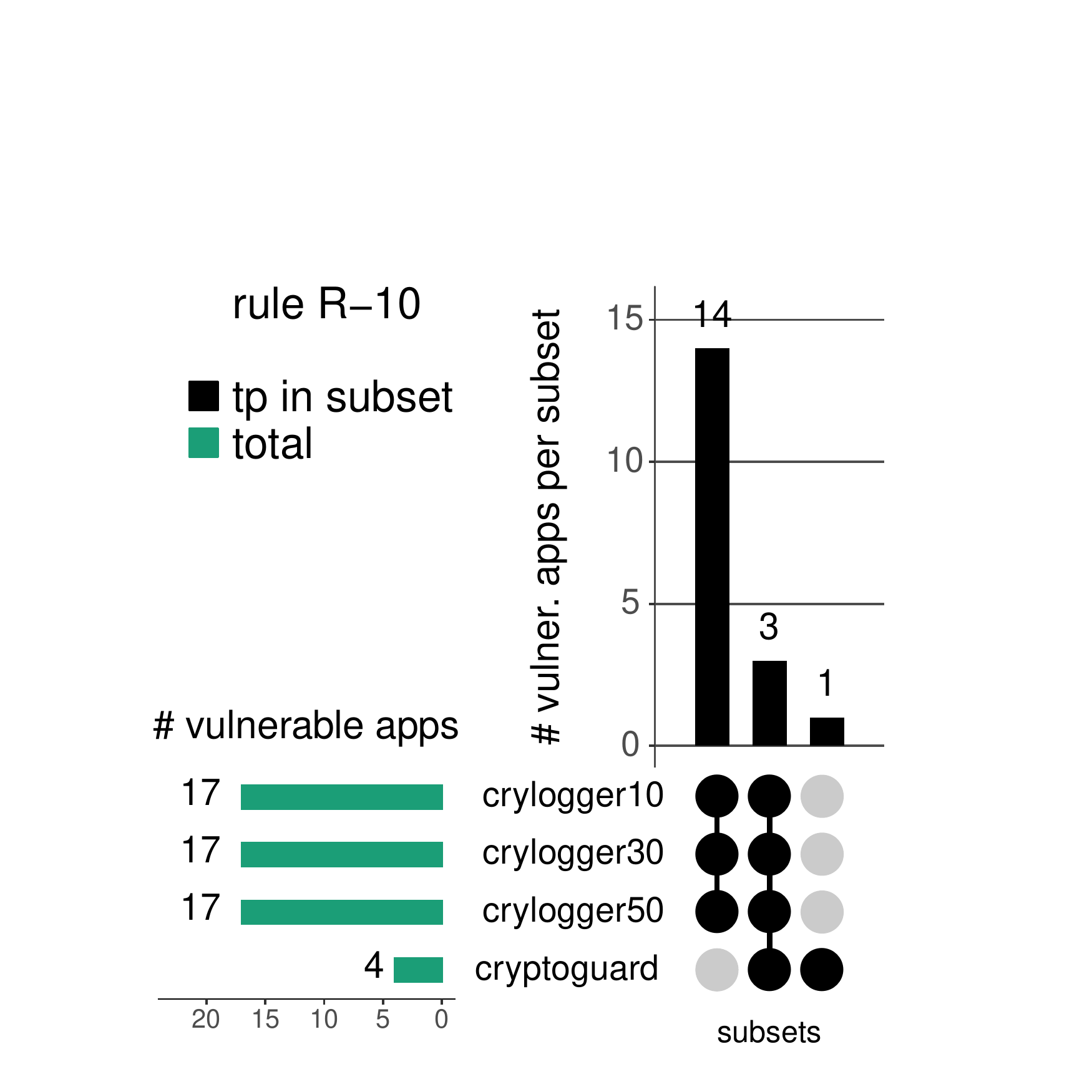}}
\end{minipage} &
\begin{minipage}[t]{0.242\textwidth}
  {\hfil\includegraphics[width=\textwidth, clip, trim = {2.3cm 0.5cm 3.3cm 4.2cm}]{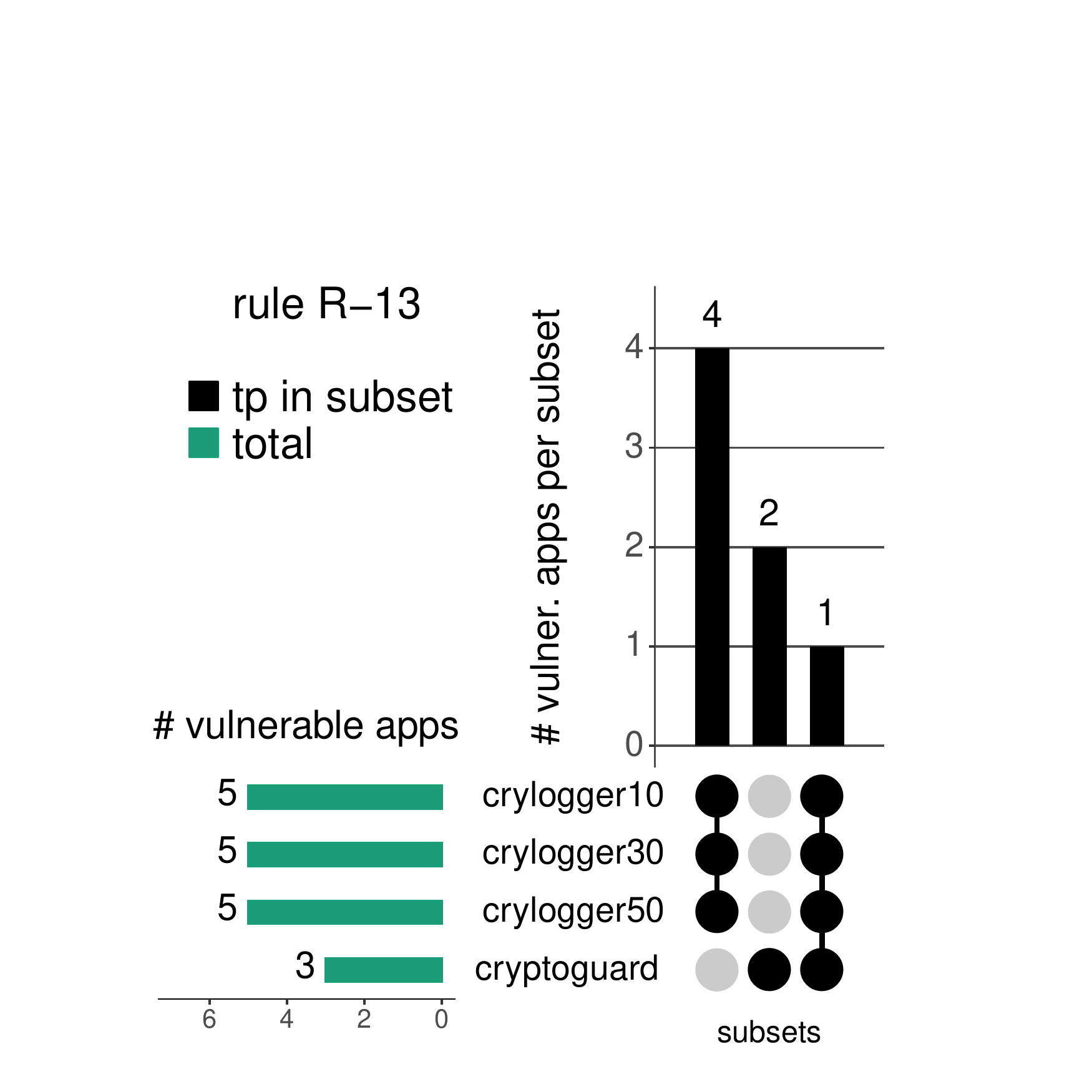}} 
\end{minipage} &
\begin{minipage}[t]{0.242\textwidth}
  {\hfil\includegraphics[width=\textwidth, clip, trim = {2.3cm 0.5cm 3.3cm 4.2cm}]{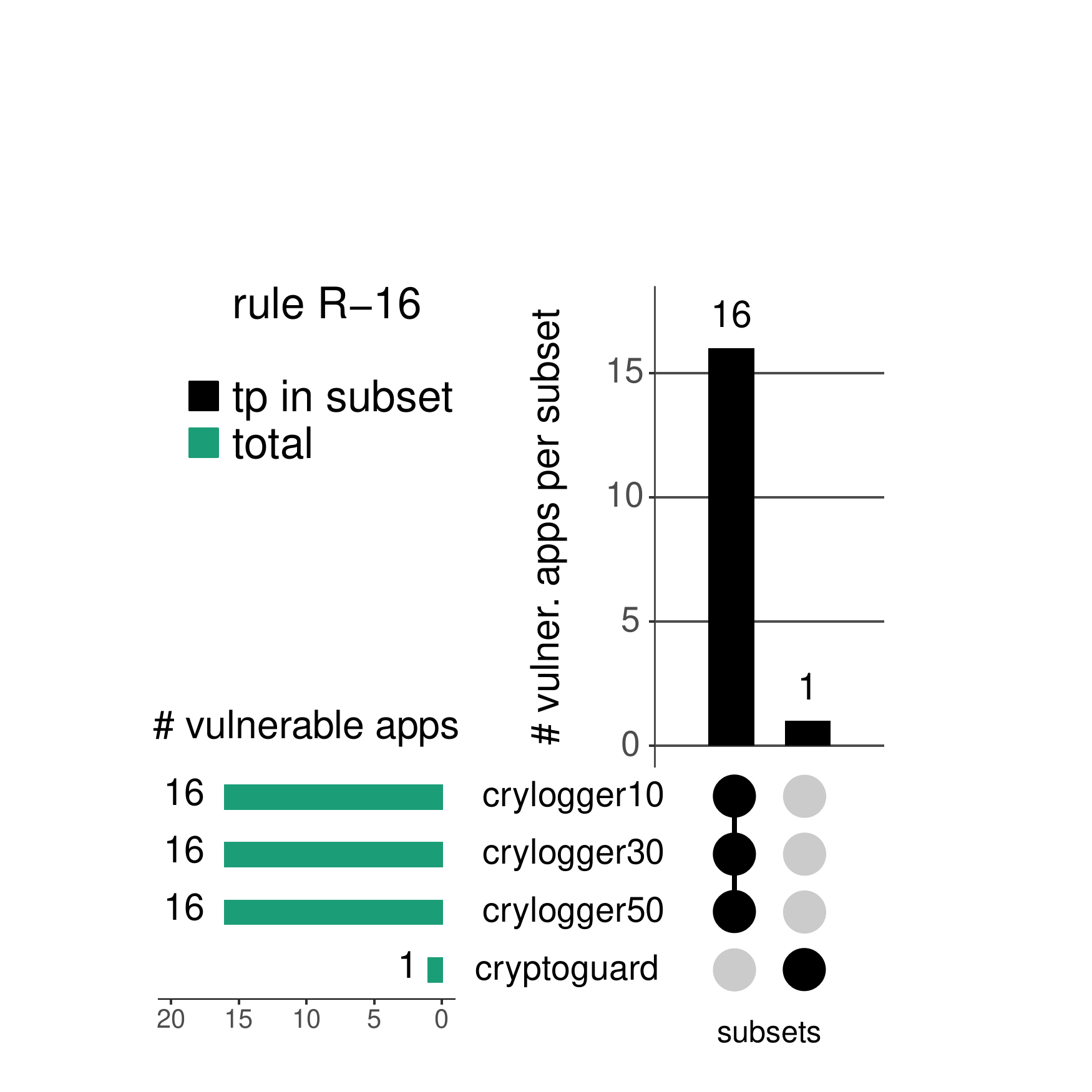}} 
\end{minipage} \\
\end{tabular}
\end{minipage}
\caption{
  \textbf{(Part 1)} Comparison of \ourff and \cryptoguardnref on 150 Android
  apps. Each graph is an \textit{upset plot}~\cite{lex_tvcg2014}. The
  \textbf{horizontal bars} indicate the number of apps flagged as vulnerable by
  \cryptoguardname and \ourff (that is run with 10k, 30k and 50k stimuli). The
  \textbf{vertical bars} indicate the number of apps flagged as vulnerable by a
  possible intersection of the four approaches (the three largest, non-empty
  intersections are reported). For example, for \cryrule{rule:symmalg}: 2 apps
  are considered vulnerable by all approaches, 14 apps are flagged as
  vulnerable by \cryptoguardname, but not by \ourff, and finally 7 apps are
  considered vulnerable by \ourff only.  The vertical bars distinguish the false
  positives (fp) obtained by reverse engineering and the true positives (tp)
  for \cryptoguardname.}\label{results:comparison1}
\vspace{-0.5cm}
\end{figure*}

\subsection{Android Apps: Results}\label{sec:res1:apps}

{
\textcolor{highlight}{We used $150$ free Android apps randomly chosen from the
dataset of $1780$ apps to compare \our and
\cryptoguardname\footnote{\url{https://github.com/franceme/cryptoguard}; vers:
03.07.03; commit: ba16c928.}. We could not use the entire dataset of $1780$ apps
of Section~\ref{sec:res0} because the false positives for \cryptoguardname must
be determined manually (see below).} For a fair comparison, we excluded the
rules that are supported by \our, but not by \cryptoguardname, and thus we
compared the two tools by checking $16$ crypto rules. For each rule, we
determined the number of apps that are marked as ``vulnerable'' by each tool
and analyzed the false positive and false negative rates.  We used $3$
configurations for \our where we varied the number of UI events that are
generated with Monkey: we used $10$k, $30$k and $50$k random events (same
random seed) to see how the number of input events impacts the number of
misuses that are identified.  In the following, we refer to the $3$
configurations as \our{10}, \our{30} and \our{50}, respectively.
}

\smallskip

{
The results of the comparison are reported
in~\figurename~\ref{results:comparison1} and \ref{results:comparison2}. Each
graph is an \textit{upset plot}~\cite{lex_tvcg2014, conway_bio2017} for a specific
rule. An upset plot is an alternative to the Venn diagrams to represent
sets and their intersections. In our context, the sets that are represented are
the sets of apps that are considered vulnerable by each approach (\our{10},
\our{30}, \our{50} and \cryptoguardname). 
The \underline{horizontal bars} are used to indicate the total number of {apps
that} are considered vulnerable by each approach. For instance, for rule
\cryrule{rule:ecbmode}, \cryptoguardname found $17$ vulnerable apps among the
$150$ apps that were analyzed, i.e., $17$ apps violate \cryrule{rule:ecbmode},
\our{50} and \our{30} flagged $21$ apps as vulnerable, and finally \our{10}
marked $20$ apps as vulnerable.
The \underline{vertical bars} are used to represent the intersections of the
sets of apps that are considered vulnerable by each approach. Specifically,
each vertical bar indicates the size of the intersection of the sets whose
circles below the bar are black.  For example, for rule \cryrule{rule:ecbmode}:
the $3$ configurations of \our identified $16$ crypto misuses that were not
found by \cryptoguardname; \cryptoguardname detected $13$ misuses that were not
found by the $3$ configurations of \our, and finally all the approaches agree
that $4$ apps are vulnerable. 
The vertical bars for \cryptoguardname distinguish the false positives (fp)
from the true positives (tp), because \cryptoguardname can produce false
positives. To make this distinction, we reverse engineered the apps by using
APKTool\footnote{\url{https://github.com/iBotPeaches/Apktool}; vers: 2.4.0; commit:
197d4687.} and verified if the API calls flagged as vulnerable by
\cryptoguardname could actually be called at runtime. We used a very
conservative approach to determine the false positives.
Starting from the flagged API call, we recursively built the sets of functions
that call that API until we obtained a fixed point. If a function that is part
of the package of the app is in the set, then we considered the API call a true
positive because there is the possibility that it could be called at runtime.
If none of the functions in the set is part of the package of the app, then we
considered the API call a false positive.
If the app was completely obfuscated with ProGuard\footnote{ProGuard:
\url{https://www.guardsquare.com/en/products/proguard}.}, thereby making it
impossible to determine its packages, then we assumed that the vulnerability
flagged by \cryptoguardname was a true positive. \textcolor{highlight}{In our case 6
apps were completely obfuscated.}
This process does not guarantee that all false positives are identified because
some paths in the code of the app could still be not executable (dead code),
but it helps to find the obvious sources of false positives.
}

\begin{figure*}[!tbp]
\hspace{-1cm}
\centering\noindent
\begin{minipage}{\textwidth}
\begin{tabular}{c|@{ }c|@{ }c|@{ }c}
\begin{minipage}[t]{0.242\textwidth}
  {\hfil\includegraphics[width=\textwidth, clip, trim = {2.3cm 0.5cm 3.3cm 4.2cm}]{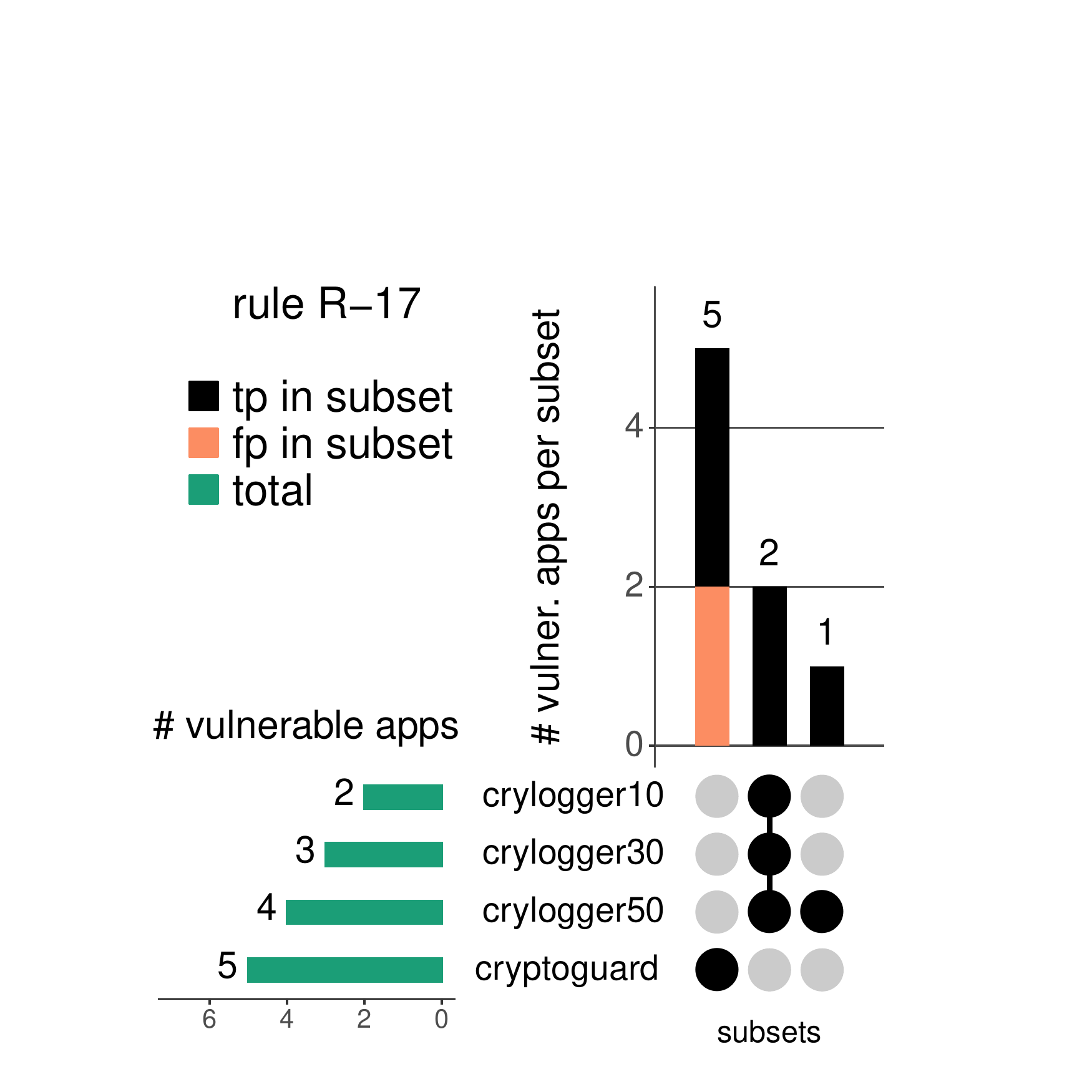}}
\end{minipage} &
\begin{minipage}[t]{0.242\textwidth}
  {\hfil\includegraphics[width=\textwidth, clip, trim = {2.3cm 0.5cm 3.3cm 4.2cm}]{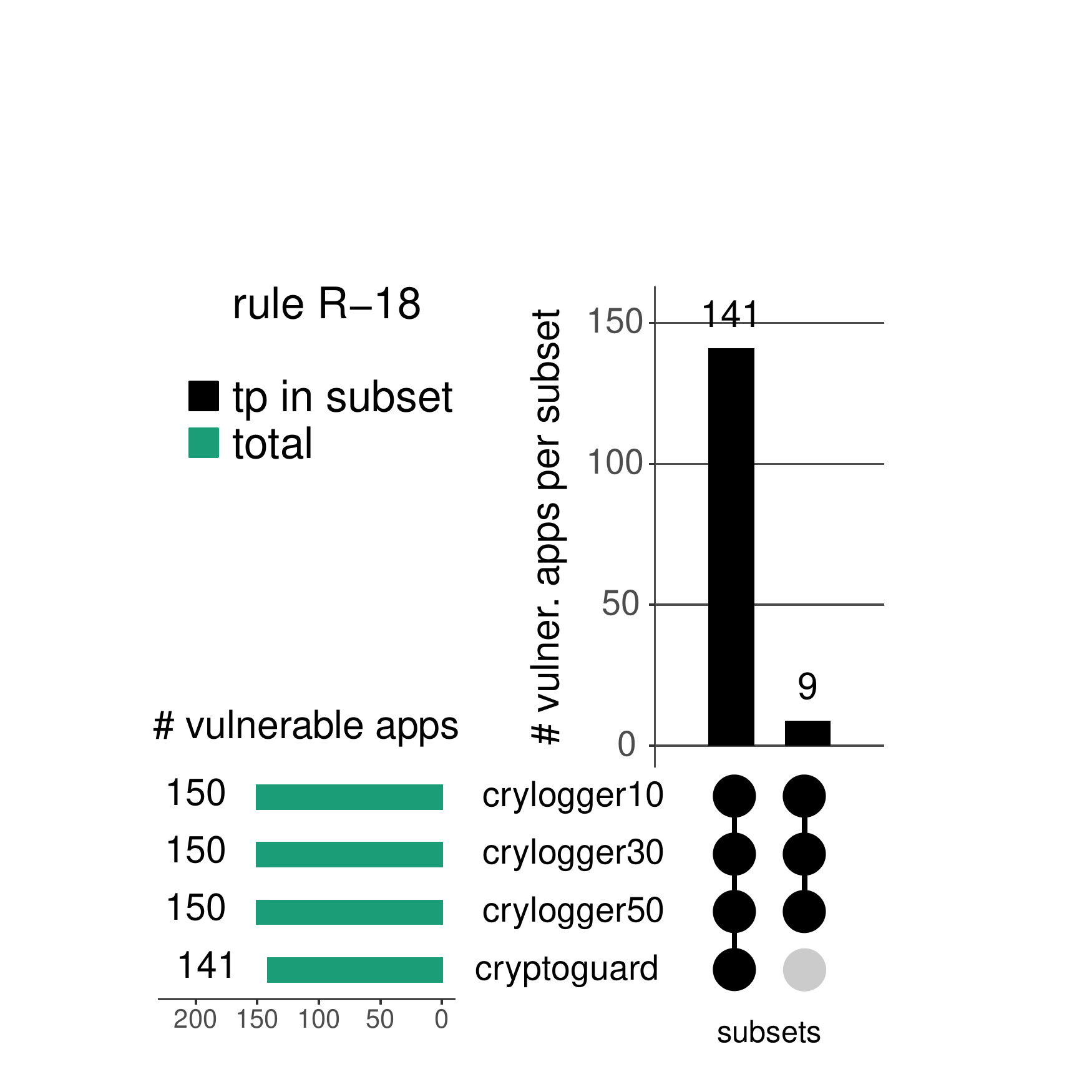}}
\end{minipage} &
\begin{minipage}[t]{0.242\textwidth}
  {\hfil\includegraphics[width=\textwidth, clip, trim = {2.3cm 0.5cm 3.3cm 4.2cm}]{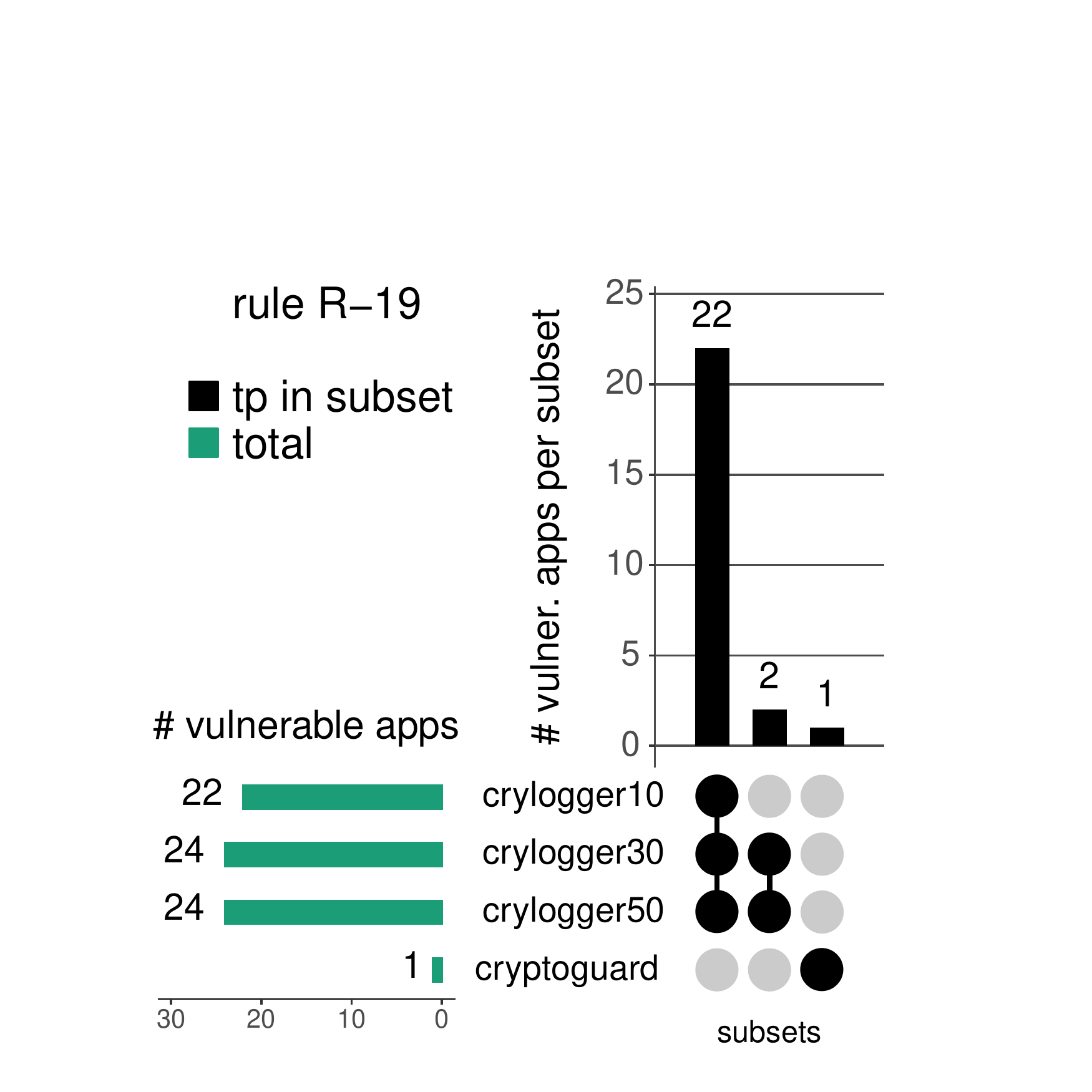}} 
\end{minipage} &
\begin{minipage}[t]{0.242\textwidth}
  {\hfil\includegraphics[width=\textwidth, clip, trim = {2.3cm 0.5cm 3.3cm 4.2cm}]{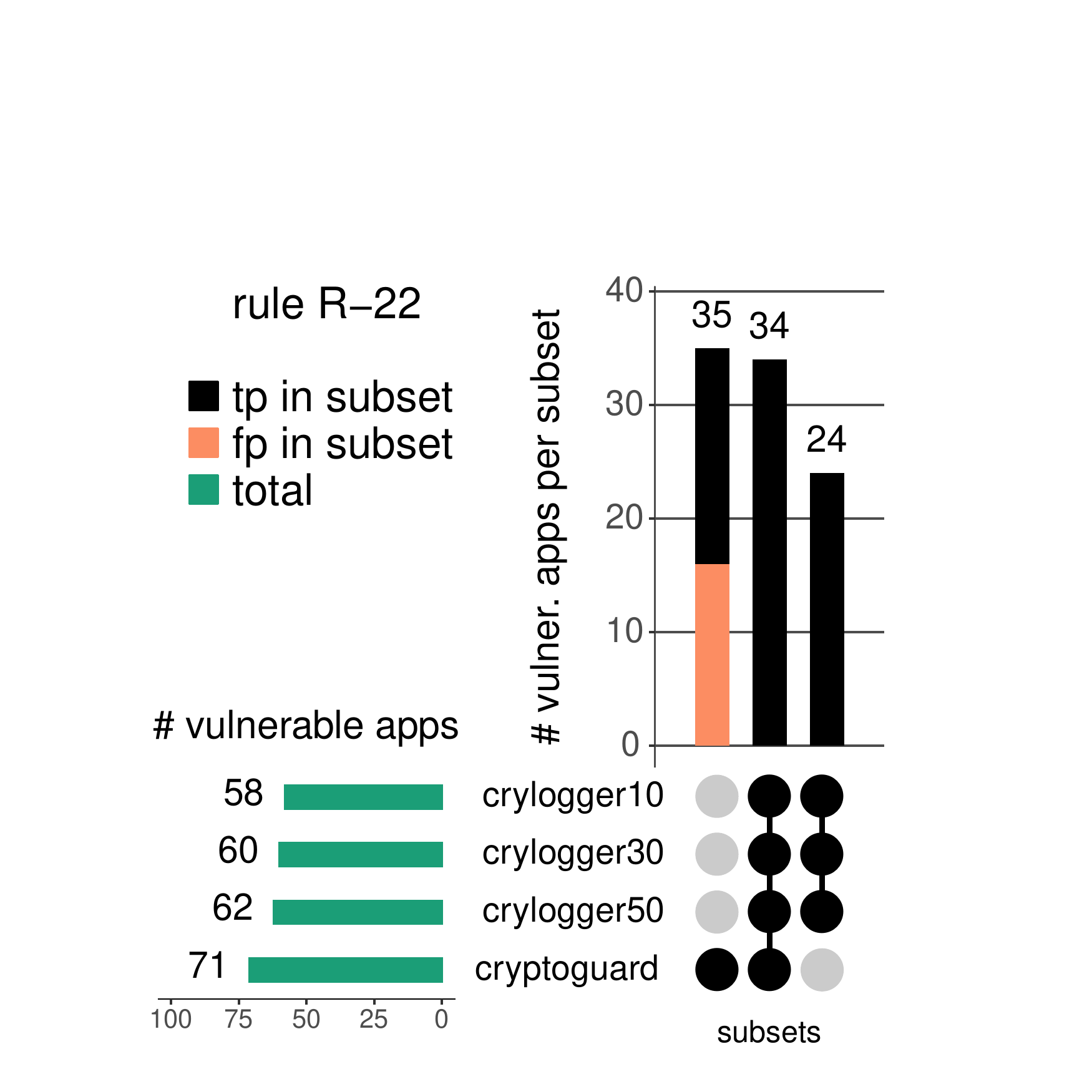}} 
\end{minipage} \\
\end{tabular}
\end{minipage}

\vspace{0.2cm}

\hspace{-1cm}
\centering\noindent
\begin{minipage}{\textwidth}
\begin{tabular}{c|@{ }c|@{ }c|@{ }c}
\begin{minipage}[t]{0.242\textwidth}
  {\hfil\includegraphics[width=\textwidth, clip, trim = {2.3cm 0.5cm 3.3cm 4.2cm}]{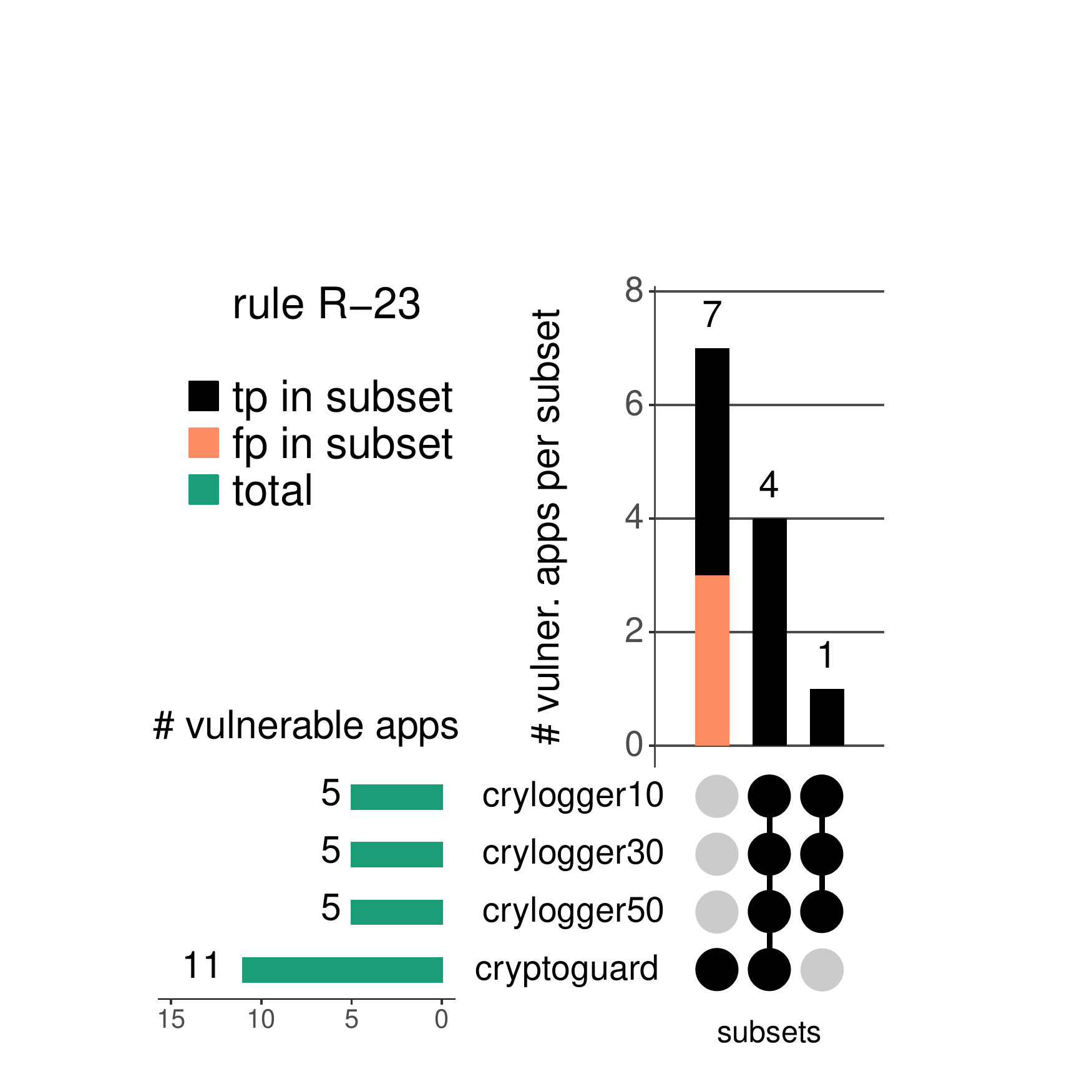}}
\end{minipage} &
\begin{minipage}[t]{0.242\textwidth}
  {\hfil\includegraphics[width=\textwidth, clip, trim = {2.3cm 0.5cm 3.3cm 4.2cm}]{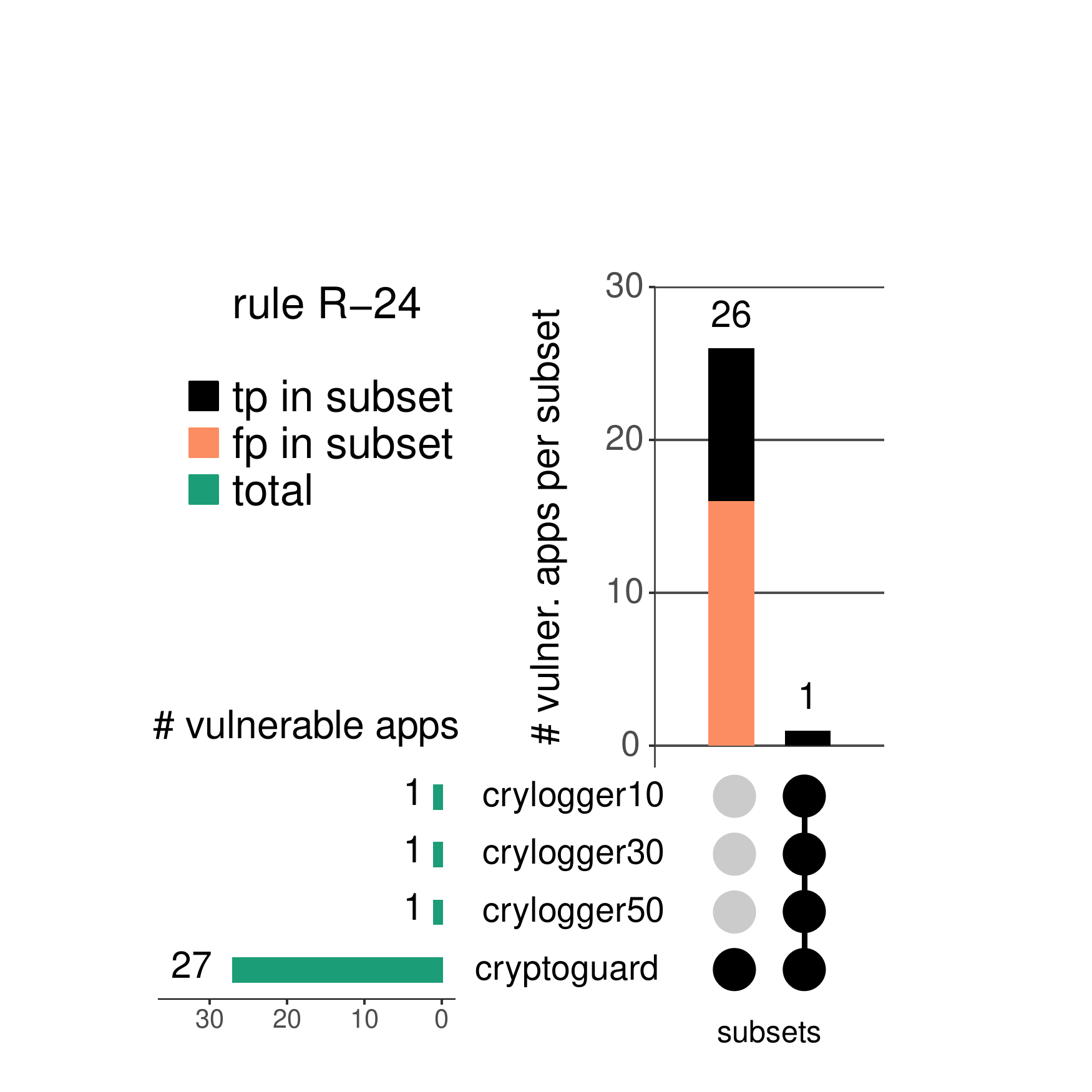}}
\end{minipage} &
\begin{minipage}[t]{0.242\textwidth}
  {\hfil\includegraphics[width=\textwidth, clip, trim = {2.3cm 0.5cm 3.3cm 4.2cm}]{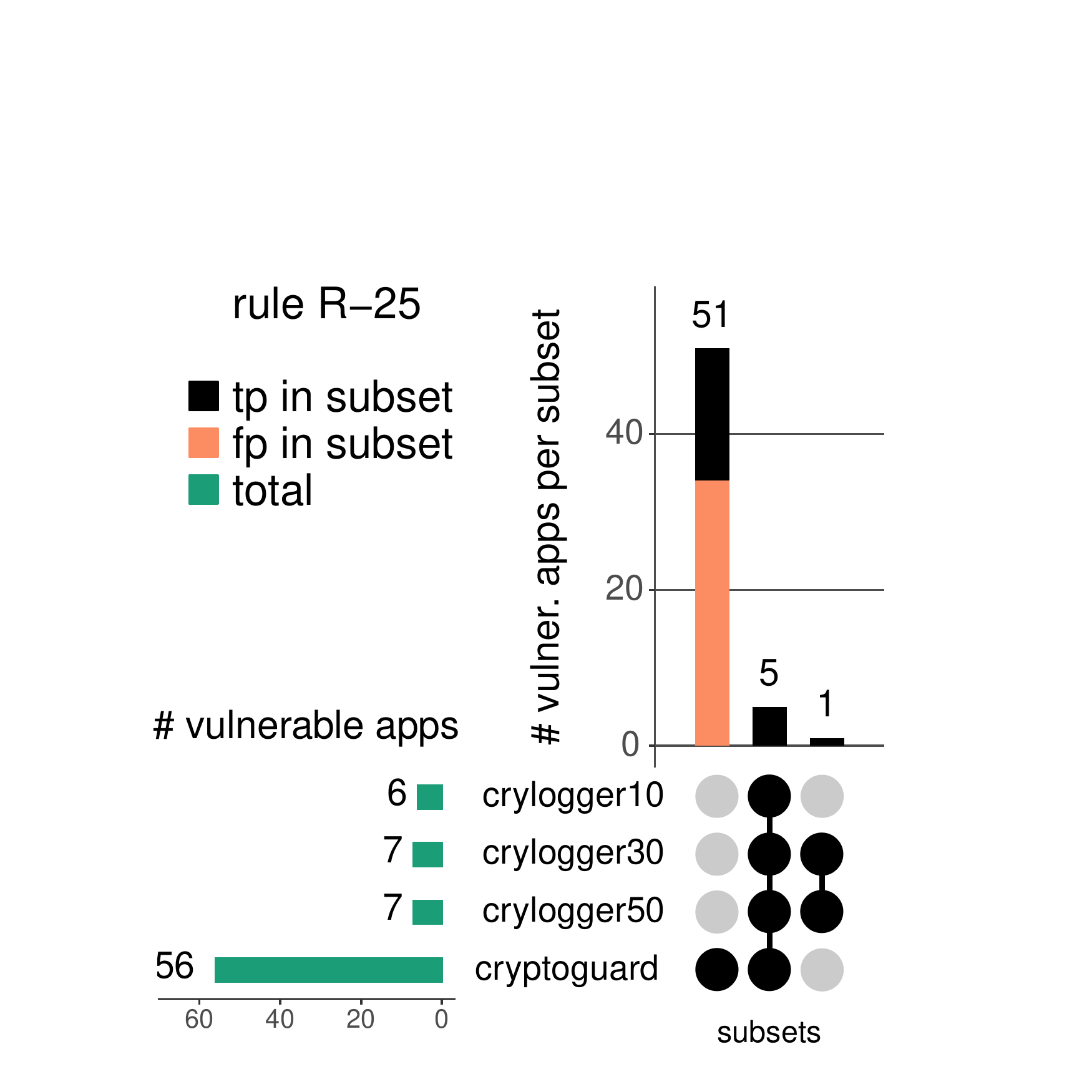}} 
\end{minipage} &
\begin{minipage}[t]{0.242\textwidth}
  {\hfil\includegraphics[width=\textwidth, clip, trim = {2.3cm 0.5cm 3.3cm 4.2cm}]{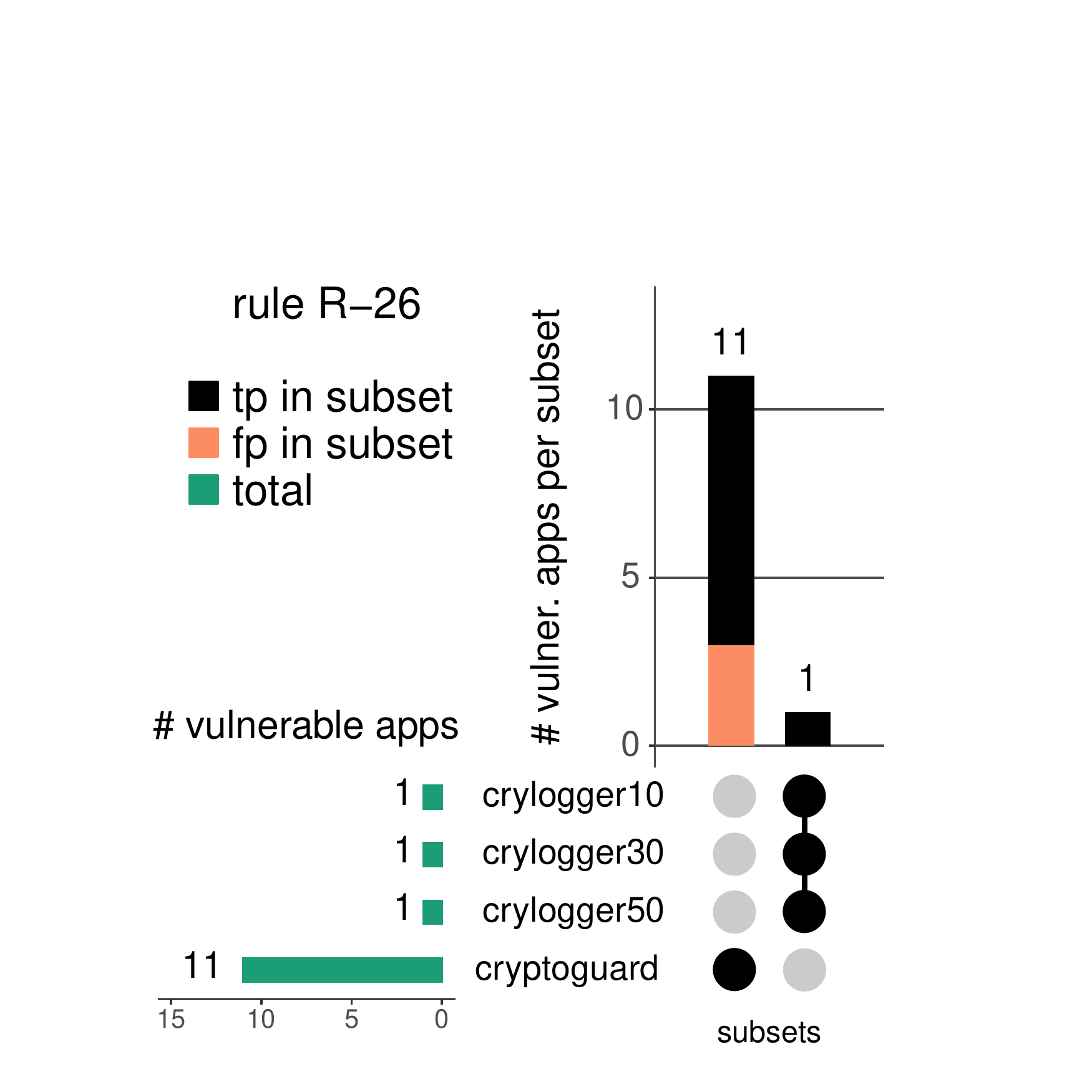}} 
\end{minipage} \\
\end{tabular}
\end{minipage}
\caption{
  \textbf{(Part 2)} Comparison of \ourff and \cryptoguardnref on 150 Android
  apps. Each graph is an \textit{upset plot}~\cite{lex_tvcg2014}. The
  \textbf{horizontal bars} indicate the number of apps flagged as vulnerable by
  \cryptoguardname and \ourff (that is run with 10k, 30k and 50k stimuli). The
  \textbf{vertical bars} indicate the number of apps flagged as vulnerable by a
  possible intersection of the four approaches (the three largest, non-empty
  intersections are reported). For example, for \cryrule{rule:http}: 35 apps
  are considered vulnerable by all approaches, 34 apps are flagged as
  vulnerable by \cryptoguardname, but not by \ourff, and finally 24 apps are
  considered vulnerable by \ourff only.  The vertical bars distinguish the false
  positives (fp) obtained by reverse engineering and the true positives (tp)
  for \cryptoguardname.}\label{results:comparison2}
\vspace{-0.5cm}
\end{figure*}

\smallskip

{
For most of the rules, excluding some cases (\cryrule{rule:hash},
\cryrule{rule:unsaprng}, \cryrule{rule:http}, \cryrule{rule:hostname},
\cryrule{rule:certif} and \cryrule{rule:socket}), we can observe the following:
(1) \cryptoguardname detected some crypto misuses that were not found by \our;
(2) \our detected some misuses that were not found by \cryptoguardname; (3) the
number of misuses detected by \our is higher than \cryptoguardname, considering
that the latter produces many false positives (we discuss some examples of
false positives in Section~\ref{sec:res1:fp}).
For some rules (\cryrule{rule:hash}, \cryrule{rule:unsaprng}) we can observe
that all the misuses detected by \cryptoguardname were also discovered by \our.
For other rules (\cryrule{rule:http}, \cryrule{rule:hostname},
\cryrule{rule:certif} and \cryrule{rule:socket}) we can observe that
\cryptoguardname found more crypto misuses compared to \our, but it produced a
significant number of false positives (in some cases the false positive rate is
$>50\%$). These rules are related to SSL/TLS and they require to evaluate the
security of the actual implementation of some Java functions, for example, the
function {\small\texttt{verify}} in the case of rule \cryrule{rule:hostname} or
the functions {\small\texttt{checkClientTrusted}},
{\small\texttt{checkServerTrusted}} and {\small\texttt{getAcceptedIssuers}} in
the case of rule \cryrule{rule:certif}. These tasks are better suited for
static analysis because it is necessary to prove that some parameters of the
functions are never used or the parameters of the functions do not influence
the return value~\cite{\cryptoguard}.
Overall, these results show that \our can complement the results that are
obtained through static analysis and it can be helpful in detecting misuses in
Android apps. By combining \our with powerful static tools such as
\cryptoguardname, it is possible to detect crypto misuses effectively.
We can also observe that it is sufficient to configure \our to use $30$k random
UI events to trigger most of the crypto misuses. We performed the same
experiments on the rules that are not supported by \cryptoguardname (see
\figurename~\ref{results:onlyour} in the appendices).
}

\subsection{Android Apps: Execution Time}\label{sec:res1:time}

{
We measured the average execution time required by the $3$ configurations of
\our and by \cryptoguardname to analyze the $150$ apps used for the comparison.
We obtained that \our{10} requires on average $146.4$ seconds per app, \our{30}
takes $287.4$ seconds, and \our{50} takes $751.7$ seconds to perform dynamic
analysis. \cryptoguardname requires $287.6$ seconds.
Other static tools are usually much slower. For example, the authors of
\cryptolintnref reported that $22.2\%$ of the apps they analyzed did not
terminate in $30$ minutes and $6.5\%$ ran out of memory.
This shows that the execution time of \our is comparable to the time required
by \cryptoguardname, confirming that both approaches are scalable.
}

\subsection{Android Apps: Coverage}\label{sec:res1:coverage}

{
We measured the line coverage, the method coverage and the class coverage of
the apps analyzed with the three configurations of \our.  We used
ACVTool~\cite{pilgun_ccs2018} to obtain this information. To calculate the
coverage, we considered only the files that are included in the main packages
of the apps, while excluding the files that belong to the third-party
libraries because they can contain code not callable from the apps.
The average line coverage for \our{10}, \our{30}, and \our{50} are $22.8\%$,
$25.3\%$, and $25.4\%$, respectively.
The average method coverage are $25.4\%$, $27.9\%$, and $27.9\%$, respectively.
The average class coverage are $32.8\%$, $35.4\%$, and $35.7\%$, respectively.
The coverage is relatively low and there are many lines of code that Monkey
could not explore ($\sim75\%$).
These results are not surprising because Monkey generates completely random UI
events~\cite{yerima_eurasip2019}. However, this shows that even if the
coverage is low, \our can detect misuses as the crypto APIs are easily
triggerable with random events.
}

\subsection{Android Apps: False Positives}\label{sec:res1:fp}

{
\figurename~\ref{results:comparison1} and \ref{results:comparison2} show that
\cryptoguardname can produce many false positives, especially for rules
\cryrule{rule:http} (false positives: 22.5\%), \cryrule{rule:hostname}
(59.3\%), \cryrule{rule:certif} (57.1\%) and \cryrule{rule:socket} (27.2\%). In
\figurename~\ref{fig:fp} we report two concrete examples of false positives
that we found. The first example is for rule \cryrule{rule:http}. We found that
many apps were flagged as vulnerable by \cryptoguardname because they include
the Java class {\small\texttt{HttpTesting}}.
While violating rule \cryrule{rule:http} due to the use of HTTP instead of
HTTPS, this class is meant to be used for testing and it is not instantiated at
runtime by any of the apps we analyzed. 
Similarly, for rule \cryrule{rule:hostname}, many apps were flagged because
they contain the Java class {\small\texttt{AdjustFactory}\footnote{The code is
available at \url{https://github.com/adjust/android_sdk}.}}.
The function reported in the second example of Fig.~\ref{fig:fp} is used only
for testing, as its name suggests, and it is never called at runtime by any of
the apps that we analyzed. This function was flagged as vulnerable by
\cryptoguardname.
}



\subsection{CryptoAPI-Bench: Results}

{
We compared \our against \cryptoguardname by using the
\cryptoapinref\footnote{\url{https://github.com/CryptoGuardOSS/cryptoapi-bench},
commit: ace0945.}, a set of Java benchmarks that include crypto misuses. The
\cryptoapiname has been proposed to compare \cryptoguardname and other static
approaches. Therefore, (1) the code is not directly executable, (2) it lacks
test cases that are useful for dynamic approaches, and (3) it misses test cases
for the rules that are not supported by \cryptoguardname.
We extended the \cryptoapiname such that (1) the code can be analyzed by static
approaches as well as executed by dynamic approaches, (2) we added new test
cases that are challenging for dynamic approaches, and (3) we included new test
cases for the rules supported by \our, but not by \cryptoguardname. In this
section, we discuss the result of the comparison on the modified \cryptoapiname
that we call \cryptoapiname{*}. For fairness, we consider the rules that are
supported by both \our and \cryptoguardname.  For fairness, we also report the
results on the original \cryptoapiname in Fig.~\ref{fig:benchcomplete}
(in the appendices).
}

\smallskip

{
\cryptoapiname contains six types of tests: (1) \textit{basic}: the
crypto misuse is in the function {\small\texttt{main}}; (2)
\textit{miscellaneous}: similar to basic, but the parameters for the API calls
are saved in data structures or go through data type conversions; (3)
\textit{interprocedural}: the misuse is in a function that is called by
{\small\texttt{main}} with $2$ or $3$ levels of indirection; (4) \textit{path
sensitive}: the crypto misuse is in a branch that is always evaluated to
{\small\texttt{true}} at runtime; (5) \textit{field sensitive}: the misuse is
in a member function and the relevant parameters are saved in the field of a
class; (6) \textit{multiple classes}: the relevant parameters of a misuse are
passed from a class to another class to reach the API call. We report an
example of each test in Fig.~\ref{fig:apitypes} (in the appendices).
Some of these tests are challenging for a static tool, but they are all the
same from a dynamic tool perspective. Therefore, we decided to add the following
type of test: (7) \textit{argument sensitive}: the misuse is triggered only if
a specific value is passed as input to {\small\texttt{main}}.
}

\begin{figure}
\vspace{-0.2cm}
{\hfil\includegraphics[width=0.45\textwidth, clip,
    trim = {0.2cm 5.4cm 0.2cm 5.4cm}]{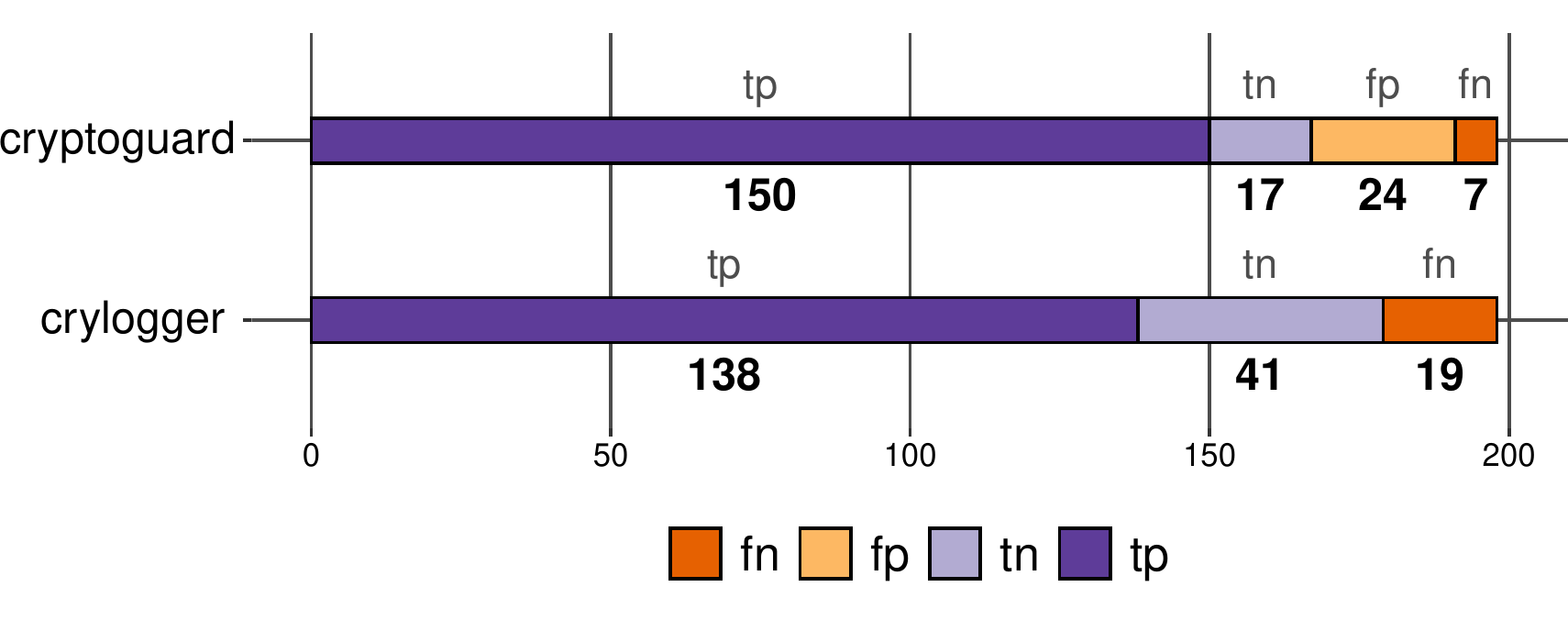}} 
\vspace{-0.2cm}
\caption{Comparison of \ourff and \cryptoguardnref on the \cryptoapiname{*}.
We report the number of false positives (fp), false negatives (fn),
true positives (tp) and true negatives (tn). ``True positive'': there is a
crypto misuse that is caught. ``True negative'': there is not a crypto
misuse and it is not caught.}\label{fig:compapi}
\vspace{-0.4cm}
\end{figure}

\smallskip

{
Fig.~\ref{fig:compapi} shows the results of the comparison of \our and
\cryptoguardname. The bars show the number of true positives (tp), true
negatives (tn), false positives (fp) and false negatives (fn).
In \cryptoapiname{*} there are $198$ tests in total, $157$ true positive tests,
i.e., tests in which there is a crypto misuse, and $41$ true negative tests,
i.e., tests in which there are no misuses. \our cannot produce any false
positives, but it produces $19$ false negatives, all for the tests that are
argument sensitive.
\cryptoguardname produces both false positives and false negatives. The false
positives are caused by tests that are path sensitive, and interprocedural
tests.
The false negatives are caused by the refinements that are applied by
\cryptoguardname~\cite{\cryptoguard}, interprocedural tests, and tests that are
path sensitive.
These results confirm that static tools can be complemented with \our to expose
more misuses as well as reduce the number of false positives.
}

\section{Results: Vulnerabilities in Android}\label{sec:res2}

{
\textcolor{highlight}{ We run \our on the $1780$ apps downloaded from the Google
Play Store (Section~\ref{sec:res0}).  We stimulated the apps with $30$k random
events as this was a good compromise between running time and number of
vulnerabilities found in a subset of these apps (Section~\ref{sec:res1}).}
The experiments took roughly $10$ days to run on an emulator running Android
9.0.0\_r36, to which we allocated $6$ cores (Intel Xeon E5-2650) and $16$ GB of
RAM.
}

\smallskip

{
Fig.~\ref{fig:andr} reports the results of the analysis. The graph reports the
total number of apps that violate the $26$ crypto rules checked by \our.
A very high number of apps use broken hash algorithms (\cryrule{rule:hash},
$99.1\%$) and unsafe random generator (\cryrule{rule:unsaprng}, $99.7\%$).
These results are more alarming than the ones that were obtained statically
in~\cite{\cryptoguard}, $85.3\%$ and $84.0\%$, respectively.
\textcolor{highlight}{\our, similarly to static tools, cannot determine
exactly how hash functions or random numbers are used in the apps by
using rules \cryrule{rule:hash} and \cryrule{rule:unsaprng} only. 
While for \cryrule{rule:hash} it is challenging to determine how hash functions
are actually used, for \cryrule{rule:unsaprng} we can check if non-truly random
numbers are used as values for keys and initialization vectors with
\cryrule{rule:badkey} and \cryrule{rule:badiv}.
These rules are not supported by static tools and they give more precise
information about the use of non-truly random numbers.  We decided to keep rule
\cryrule{rule:unsaprng} to compare \our against other static tools, but we
suggest using rules \cryrule{rule:badkey} and rule \cryrule{rule:badiv} for a
more precise analysis. Other more subtle uses of hash functions can produce
false positives, e.g., when broken hash functions are used with non-sensitive
data or when the property of collision resistant is not
required. }
For other rules, e.g., \cryrule{rule:ecbmode}, \cryrule{rule:iterat}, and
\cryrule{rule:http}, we obtained results more similar to~\cite{\cryptoguard}.
A surprising number of apps reuse the same (key, IV) pairs
(\cryrule{rule:sameiv}, $31.3\%$), which was never reported before. 
Many apps also use badly-generated keys (\cryrule{rule:badkey}, $36.1\%$),
badly-generated IVs (\cryrule{rule:badiv}, $6.6\%$), and reuse salts for
different purposes (\cryrule{rule:samesalt}, $6.6\%$), which are rules that
were not checked by other tools before.
For rule \cryrule{rule:hash} we found that $99.0\%$ of the apps that violate
\cryrule{rule:hash} use SHA1 and $99.7$\% use MD5 as message digest algorithm.
For \cryrule{rule:symmalg}, we found that $81.0\%$ of the apps that use broken
symmetric algorithms use DES, while $16.7$\% still use Blowfish.
We found that $82.8\%$ of the apps that violate \cryrule{rule:iterat} use $\le
3$ iterations for key derivation, which is much lower compared to the suggested
value ($1000$).
For \cryrule{rule:weakpass} and \cryrule{rule:blackpass} we found that $27.1\%$
of the apps use ``changeit'' as password, while $8.5\%$ use ``dontcare''.
For RSA, we saw that $97.7\%$ use $1024$ bits as key size ($2048$ is the
suggested value).  These results confirm what was obtained in previous works by
using static analysis~\cite{\cryptoguard, \cryptolint} and show that \our can
analyze a large number of apps automatically.
}

\begin{figure}
\vspace{-0.5cm}
\begin{center}
   \includegraphics[width=0.5\textwidth, clip,
    trim = {5.8cm 7cm 5.8cm 6.4cm}]{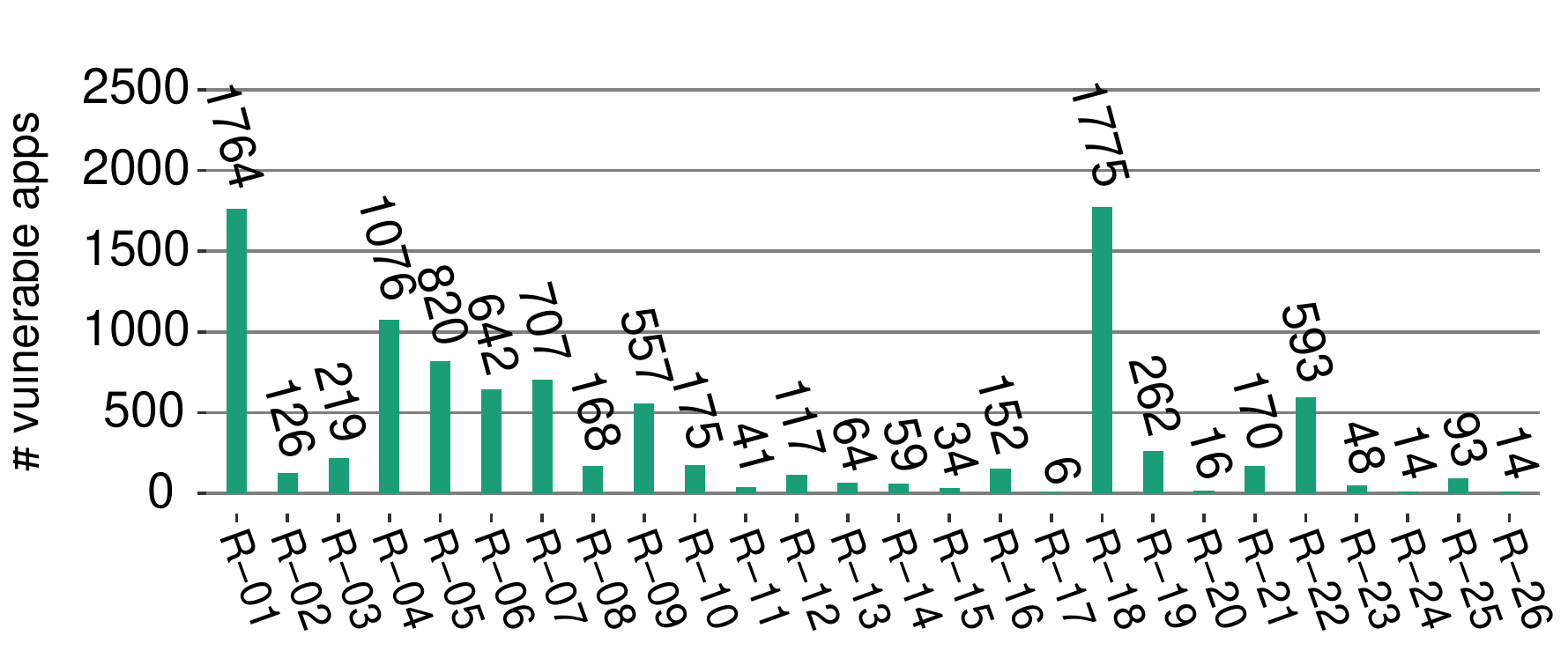} 
\end{center}
\vspace{-0.4cm}
\caption{Number of vulnerable Android apps for each crypto rule. We analyzed 
$1780$ Android apps with \ourff configured to generate $30$k random events with 
Monkey. We downloaded the apps from the official 
Google Play Store. The dataset of apps was collected between September and
October 2019.}\label{fig:andr}
\vspace{-0.4cm}
\end{figure}

\newcommand\appv[1]{\textit{A-#1}}
\newcommand\libv[1]{\textit{L-#1}}

\begin{figure*}
\centering
\scriptsize
\setlength{\tabcolsep}{2pt}
\begin{tabular}[t]{c@{\hspace{14pt}}cc}
\textsc{Developers Feedback} &
\multicolumn{2}{c}{\textsc{Reverse Engineering}} \\
\begin{tabular}[t]{clll}
\toprule
\textbf{ID} & \multicolumn{1}{l}{\textbf{Type (\#Downloads)}} & \textbf{Analyzed Violations} \\
\midrule
\rowcolor{gray!20}
\textit{A-01} & File Manager      (100M+) & 
  \textit{R-02},
  \textit{R-03},
  \textit{R-05},
  \textit{R-07}, \\
& &
  \textit{R-08},
  \textit{R-09},
  \textit{R-10},
  \textit{R-12}, \\
& &
  \textit{R-19} \\
\rowcolor{white}
\textit{A-02} & Data Transfer     (10M+)  &
  \textit{R-16},
  \textit{R-23} \\
\rowcolor{gray!20}
\textit{A-03} & Video Streaming   (10M+)  &
  \textit{R-09},
  \textit{R-20},
  \textit{R-22} \\
\rowcolor{white}
\textit{A-04} & Newspaper App     (5M+)   &
  \textit{R-01},
  \textit{R-19},
  \textit{R-20},
  \textit{R-23} \\
\rowcolor{gray!20}
\textit{A-05} & Social \& News    (5M+)   &
  \textit{R-05},
  \textit{R-06},
  \textit{R-07},
  \textit{R-08} \\
& &
  \textit{R-10},
  \textit{R-16},
  \textit{R-19} \\
\rowcolor{white}
\textit{A-06} & Language Learning (1M+)   &
  \textit{R-16} \\
\rowcolor{gray!20}
\textit{A-07} & Music Streaming   (1M+)   &
  \textit{R-01},
  \textit{R-05},
  \textit{R-06},
  \textit{R-09} \\
\rowcolor{white}
\textit{A-08} & Video Streaming   (1M+)   &
  \textit{R-16},
  \textit{R-23} \\
\rowcolor{gray!20}
\textit{L-01} & Advertisement (N.A.) &
  \textit{R-09} \\
\rowcolor{white} 
\textit{L-02} & Advertisement (N.A.) &
  \textit{R-07},
  \textit{R-08},
  \textit{R-10} \\
\rowcolor{white}
\bottomrule \\
\end{tabular} &
\begin{tabular}[t]{clll}
\toprule
\textbf{ID} & \multicolumn{1}{l}{\textbf{Type (\#Downloads)}} & \textbf{Analyzed Violations} \\
\midrule
\rowcolor{gray!20}
\textit{A-09} & Messaging      (100M+) & 
  \textit{R-01} \\
\rowcolor{white}
\textit{A-10} & Entertainment    (100M+)  &
  \textit{R-18},
  \textit{R-22} \\
\rowcolor{gray!20}
\textit{A-11} & Movie Reviews   (100M+)  &
  \textit{R-18},
  \textit{R-19},
  \textit{R-21} \\
\rowcolor{white}
\textit{A-12} & Book Reading     (50M+)   &
  \textit{R-02},
  \textit{R-03},
  \textit{R-05},
  \textit{R-06} \\
\rowcolor{gray!20}
\textit{A-13} & Passw. Manager    (50M+)   &
  \textit{R-02},
  \textit{R-03},
  \textit{R-04},
  \textit{R-05} \\
& &
  \textit{R-06},
  \textit{R-07},
  \textit{R-08} \\
\rowcolor{white}
\textit{A-14} & Passw. Manager (50M+)   &
  \textit{R-17} \\
\rowcolor{gray!20}
\textit{A-15} & Screen Utils   (10M+)   &
  \textit{R-01} \\
\rowcolor{white}
\textit{A-16} & File Manager   (10M+)   &
  \textit{R-01} \\
\rowcolor{gray!20}
\textit{A-17} & Video Streaming   (10M+)   &
  \textit{R-04} \\

\rowcolor{white}
\textit{A-18} & Video Streaming   (10M+)   &
  \textit{R-04},
  \textit{R-07},
  \textit{R-08},
  \textit{R-21}, \\
& &
  \textit{R-23} \\
\rowcolor{gray!20}
\textit{A-19} & Video Streaming   (10M+)   &
  \textit{R-09},
  \textit{R-20},
  \textit{R-22} \\

\rowcolor{white}
\textit{A-20} & Live Events Info  (10M+)   &
  \textit{R-11},
  \textit{R-16} \\

\rowcolor{gray!20}
\textit{A-21} & Video Streaming   (10M+)   &
  \textit{R-11},
  \textit{R-13} \\

\rowcolor{white}
\textit{A-22} & Video Streaming   (10M+)   &
  \textit{R-14},
  \textit{R-15},
  \textit{R-16} \\
\rowcolor{gray!20}
\textit{A-23} & Newspaper App   (5M+)   &
  \textit{R-01},
  \textit{R-19},
  \textit{R-20},
  \textit{R-21} \\
\rowcolor{white}
\bottomrule \\
\end{tabular} &
\begin{tabular}[t]{clll}
\toprule
\textbf{ID} & \multicolumn{1}{l}{\textbf{Type (\#Downloads)}} & \textbf{Analyzed Violations} \\
\midrule
\rowcolor{gray!20}
\textit{A-24} & Mail Manager      (5M+) & 
  \textit{R-04},
  \textit{R-05},
  \textit{R-06},
  \textit{R-10} \\
& &
  \textit{R-12},
  \textit{R-13},
  \textit{R-16} \\
\rowcolor{white}
\textit{A-25} & Video Streaming   (5M+)  &
  \textit{R-19},
  \textit{R-21},
  \textit{R-24},
  \textit{R-25} \\
& &
  \textit{R-26} \\
\rowcolor{gray!20}
\textit{A-26} & Stocks Manager   (5M+)  &
  \textit{R-22} \\
\rowcolor{white}
\textit{A-27} & Authentication     (5M+)   &
  \textit{R-23} \\
\rowcolor{gray!20}
\textit{A-28} & Video Streaming   (1M+)   &
  \textit{R-10},
  \textit{R-16} \\
\rowcolor{white}
\textit{A-29} & Blog Reading    (1M+)   &
  \textit{R-14},
  \textit{R-15},
  \textit{R-16} \\
\rowcolor{gray!20}
\textit{A-30} & Book Reading    (1M+)   &
  \textit{R-14},
  \textit{R-15},
  \textit{R-16} \\
\rowcolor{white}
\textit{A-31} & Healthcare Info    (1M+)   &
  \textit{R-24},
  \textit{R-25},
  \textit{R-26} \\
\rowcolor{gray!20}
\textit{A-32} & Music Streaming    (1M+)   &
  \textit{R-24},
  \textit{R-25},
  \textit{R-26} \\
\rowcolor{white}
\textit{A-33} & Newspaper App   (500K+)   &
  \textit{R-03},
  \textit{R-05},
  \textit{R-06},
  \textit{R-10} \\
& &
  \textit{R-13},
  \textit{R-16},
  \textit{R-24},
  \textit{R-25} \\
& &
  \textit{R-26} \\
\rowcolor{gray!20}
\textit{A-34} & Entertainment    (100K+)   &
  \textit{R-10},
  \textit{R-11},
  \textit{R-13},
  \textit{R-16} \\
\rowcolor{white}
\textit{A-35} & Passw. Manager    (100k+)   &
  \textit{R-13} \\
\rowcolor{gray!20}
\textit{A-36} & Video Streaming    (100K+)   &
  \textit{R-22} \\
\rowcolor{white}
\bottomrule \\
\end{tabular} \\
\end{tabular}
\caption{\normalfont The first table from the left reports the characteristics
of the Android apps for which we received feedback from their developers.  The
other tables report the characteristics of the apps that we reverse
engineered. The rules reported in the last column of each table are those
that were analyzed by the developers or by us.
}\label{figure:disclosure}
\vspace{-0.4cm}
\end{figure*}

\subsection{\textcolor{highlight}{Disclosure of Vulnerabilities}}

{
\textcolor{highlight}{
We contacted $306$ developers of Android apps and libraries to disclose the
vulnerabilities reported in Fig.~\ref{fig:andr}. We respected the disclosure
policies of the companies we contacted. 
Starting from the apps that violate $18$ rules (the highest number of
violations in our dataset), we contacted all the apps with $\ge$ $9$ rule
violations.
All the apps are popular: they have from hundreds of thousands of downloads to
more than $100$ millions.
Unfortunately, only $18$ developers answered our first email of request and
only $8$ of them followed back with us multiple times providing useful feedback
on our findings.  
We also contacted $6$ developers of popular Android libraries and received
answers from $2$ of them. The characteristics of the $8$ apps and $2$ libraries
for which we received feedback are reported in the first table from the
left of Fig.~\ref{figure:disclosure}.  We preferred to anonymize the apps and
libraries because (i) we do not want to associate the feedback we received to
the company of the app or its employers, and (ii) we consider some of the
attacks possible although developers considered them out-of-scope because they
require privilege escalation.
}
}

\smallskip

{
\textcolor{highlight}{
Apps \appv{01}, \appv{04}, and \appv{07} violate rule \cryrule{rule:hash}.
Their developers told us that {MD$5$} or {SHA$1$} are used for hashing
non-sensitive values. 
App \appv{01} violates also rules \cryrule{rule:symmalg} (DES) and
\cryrule{rule:ecbmode}: the developers justified the use of broken algorithms
saying that they do not pose concrete risks to their users.
\appv{01}, \appv{05}, and \appv{07} violate rules related to poor encryption
parameters, such as constant keys (\cryrule{rule:constkey},
\cryrule{rule:badkey}), IVs (\cryrule{rule:constiv}, \cryrule{rule:badiv}) and
salts (\cryrule{rule:constsalt}). The developers adopted poor encryption
practices to encrypt data that are stored locally on the smartphone. They
consider these issues outside of their threat model since privilege escalation
attacks are required to exploit them.
\appv{03} uses repeating (key, IV) pairs (\cryrule{rule:sameiv}): the
developers agreed that it is a real issue and they plan to fix it. They reused
the same pairs because they experienced app crashing when using fresh pairs.
\appv{02}, \appv{05}, \appv{06}, and \appv{08} use constant passwords
(\cryrule{rule:reusepass}, \cryrule{rule:store}) to encrypt data. The
developers do not plan to fix these problems because a privilege escalation
attack is necessary to access the data.
The developers of \appv{01}, \appv{04} and \appv{05} told us that using a short
RSA key (\cryrule{rule:rsakeysize}) does not pose concrete risks.
\libv{09} is a popular library for advertisements. The library uses the same
(key, IV) pairs to store data locally. The same (key, IV) pairs are reused
across different apps, i.e., all the apps using this library end up using the
same sequence of (key, IV) pairs.
About $30\%$ of the apps in our dataset share the same sequence of pairs which
are used to encrypt data in the private folder of each app. The library
developers confirmed this issue, but they classified it as out-of-scope. Note
that this experiments cannot be replicated by static tools and it is an example
of how \our can perform inter-app analysis.
\libv{10} is a common library for advertisements. The library employs weak
encryption practices to store data locally. We talked with the library
developers. They were aware of the issue and said that the data are not
security critical.
}
}

\smallskip

{
\textcolor{highlight}{
This analysis reveals that the threat model of \our and all the other static
tools is not aligned with the developers' threat model. Developers claim that
sensitive data can be encrypted poorly if they are stored only locally because
privilege escalation is required to access them.
Unfortunately, side-channel attacks can also access the
data~\cite{tang_usenix17}.
While we recommend to always adopt safe crypto practices, one way to to avoid
such types of warnings in \our is to log when data are stored on the local
storage (e.g., in classes such as \cparam{File} or \cparam{KeyStore}) and
discard the corresponding violations.
Developers are also more interested to rules that, if violated, pose concrete
security threats as also reported in~\cite{\cryptoguard}. For example, while
setting a minimum size for keys (\cryrule{rule:rsakeysize}) is important, the
effects of its violation are hard to assess.
Since the feedback we received from developers is limited to a few apps, we
decided to analyze some apps manually to determine if the vulnerabilities of
Fig.~\ref{fig:andr} are exploitable. 
}
}

\subsection{\textcolor{highlight}{Analysis of Vulnerabilities}}

{
\textcolor{highlight}{
We reverse engineered $28$ apps with APKTool and
JADX\footnote{\url{https://github.com/skylot/jadx}; vers: 1.1.0, commit:
cc29da8.}.
We chose half of the apps among the most popular apps of our dataset
(Section~\ref{sec:res0}) with the highest number of violations.  We chose the
remaining half randomly. The apps characteristics are shown in
Fig.~\ref{figure:disclosure}.
We performed the following steps for reverse engineering: (i) we used APKTool
and JADX to obtain the Java code from the binary (apk) of the app, (ii) we
analyzed the app with \our, which we extended to log the stack trace for
each rule violation, and (iii) we manually analyzed the code starting from the
flagged API call to understand its purpose in the app. We spent on average $6$
hours per app for code analysis.
}
}

\smallskip

{
\textcolor{highlight}{
A significant number of these apps ($14/28$) are vulnerable to attacks, even
though some may be considered out-of-scope by developers. Most of the rules
($22/26$) are effective in detecting at least one vulnerable app.
App \appv{13} violates many rules related to encryption. This app uses
encryption to manage subscriptions to premium features and users data.  The
subscription and the users data are stored locally on the app and attacker can
read the data as well as fake subscriptions.
Similarly, apps \appv{18}, \appv{20}, \appv{24}, \appv{25}, \appv{33}, and
\appv{34} store critical users data (emails, answers to security questions,
etc.) by using weak encryption algorithms.
\appv{22}, \appv{29}, and \appv{30} store SSL/TLS certificates with weak
password-based encryption.
\appv{14} uses a constant seed (\cryrule{rule:statseed}) to randomly generate 
keys used for encryption of users data, so the keys can be easily obtained.
Apps \appv{31}, \appv{32}, and \appv{33} are vulnerable to man-in-the-middle
attacks because they violate \cryrule{rule:hostname}, \cryrule{rule:certif},
and \cryrule{rule:socket}.  These apps download copyrighted videos/music as
well as ads, which can be intercepted by attackers.
The other violations can be considered false positives. Some are caused by
`imprecise' rules. For example, on $3$ apps each, rules \cryrule{rule:hash} and
\cryrule{rule:unsaprng} flag secure uses of hash algorithms and random number
generators for non-sensitive data. 
Similarly, \cryrule{rule:cbcmode} flags $3$ apps that use CBC encryption for
scenarios different from client/server.
Other violations come from (i) employing weak encryption schemes to obfuscate
non-sensitive data and (ii) legacy practices such as using \cparam{PCKS\#1} as padding
scheme in SSL/TLS instead of more secure alternatives such as \cparam{OAEP}.
}
}

\smallskip

{
\textcolor{highlight}{
This analysis confirms that the threat model of \our and all the other static
tools does not completely align with the developers' threat model and some
rules produce false positives. 
}
}

\vspace{-0.15cm}
\section{Discussions and Limitations}\label{sec:lim}

{
\textcolor{highlight}{
In this section, we discuss the advantages of dynamic approaches over static
approaches and our current limitations. 
}
}

\smallskip

{
\textit{Why a Dynamic Approach?} 
}
\textcolor{highlight}{
{
To date, most of the approaches to detect crypto misuses are based on static
analysis, which provides many benefits.
Static analysis can analyze the code without executing it, and this is
especially important for Android apps since UI test generators are not
required.  Static analysis can scale up to a large number of applications and,
thanks to recent improvements~\cite{\cryptoguard}, it can analyze massive code
bases.
Static analysis has, however, some limitations. It can produce false positives,
i.e., alarms can be raised on correct calls to crypto APIs due to imprecise
slicing algorithms.
These alarms add up to those raised on parts of the applications that
are not security critical (see Section~\ref{sec:res2}).  This makes it hard
to analyze a large number of applications.
Some static approaches~\cite{\cryptoguard} also incur in many false negatives.
Some misuses escape detection because the exploration is pruned prematurely
to improve scalability.
In addition, static analysis misses some crypto misuses in the code that is
loaded dynamically. This prevents analyses on critical
code~\cite{poeplau_ndss2014}.
Also, static analysis can be inherently done on a single application only. It
is not possible to perform inter-application analysis, as the one we did with
\our on an Android library (Section~\ref{sec:res2}).
On the other hand, dynamic analysis is not a perfect antidote. Dynamic analysis
is as good as the test generator that is used to run the applications. We
discuss the main limitations of dynamic analysis in the next paragraphs. 
}
}

\smallskip

{
\textit{False Positives}. 
}
\textcolor{highlight}{
{
Although dynamic analysis, theoretically, should avoid false
positives, these are possible when detecting crypto misuses
(Section~\ref{sec:res2}).
It is hard to distinguish critical parts of the application, which should obey
to the rules, from less critical parts where the data are not sensitive.
In addition, the threat model adopted by app developers can differ from the one
adopted in the research community. This requires complex manual analyses.
One possible solution is to log additional information in other classes (e.g.,
\cparam{File}) to determine if rule violations can be discarded.  This would
greatly reduce the false positives, but it is hard to implement with general
solutions.
}
}

\smallskip

{
\textit{False Negatives}. 
}
{
\textcolor{highlight}{Crypto misuses escape detection if they are not exercised
during the execution. 
In Section~\ref{sec:res1}, we showed that for many Android apps, \our
confirmed the results reported by \cryptoguardname and found misuses missed by
\cryptoguardname. In other contexts, it might be harder to trigger
the crypto APIs depending on the specific application.
One possible solution is to complement \our with a static tool in order to
expose the misuses that cannot be triggered at runtime.}
}

\section{Concluding Remarks}\label{sec:conc}

\textcolor{highlight}{
{
We presented \our, the first tool that detects crypto misuses dynamically, while
supporting a large number of rules.
We released \our open-source to allow the community to use a dynamic
tool alongside static analysis.
We hope that application developers will adopt it to check their applications
as well as the third-party libraries that they use.
}
}

\section*{Acknowledgments}\label{sec:ack}
{
This work was supported in part by the NSF (A\#: 1527821 and 1764000), a gift
from Bloomberg, DARPA HR0011-18-C-0017, and N00014-17-1-2010.
}

{
\bibliographystyle{IEEEtran}
\bibliography{ref}
}

\begin{table*}[b]
\vspace{-6.5cm}
\renewcommand{\arraystretch}{1.05}
\footnotesize
\begin{center}
\begin{tabular}{llll}
\\ \\ \\ \\
\toprule
\textbf{Package} & 
\textbf{Class} &
\textbf{Function} &
\textbf{Logged Data} \\
\midrule
\rowcolor{gray!20}
java.security
  & {MessageDigest} & \textit{byte[]} digest (\textit{void})                
    & \texttt{alg} \\
\rowcolor{gray!20}
  &                      & \textit{int} digest (\textit{byte[]}, \textit{int}, \textit{int})   
    & \\
\rowcolor{white}
javax.crypto
  & {Cipher}& \textit{void} init (\textit{int}, \textit{Key}, \textit{SecureRandom}) 
    & \texttt{alg, mode,} \\
  &         & \textit{void} init (\textit{int}, \textit{Key}, \textit{AlgorithmParameters}, \textit{SecureRandom}) 
    & \texttt{pad, key,} \\
  &         & \textit{void} init (\textit{int}, \textit{Key}, \textit{AlgorithmParameterSpec}, \textit{SecureRandom}) 
    & \texttt{iv} \\
  &         & \textit{void} init (\textit{int}, \textit{Certificate}, \textit{SecureRandom}) 
    & \textit{} \\
\rowcolor{gray!20}
  & {Cipher}& \textit{byte[]} doFinal (\textit{void})
    & \texttt{out} \\
\rowcolor{gray!20}
  &         & \textit{int} doFinal (\textit{byte[]}, \textit{int})
    & \textit{} \\
\rowcolor{gray!20}
  &         & \textit{byte[]} doFinal (\textit{byte[]})
    & \textit{} \\
\rowcolor{gray!20}
  &         & \textit{byte[]} doFinal (\textit{byte[]}, \textit{int}, \textit{int})
    & \textit{} \\
\rowcolor{gray!20}
  &         & \textit{int} doFinal (\textit{byte[]}, \textit{int}, \textit{int}, \textit{byte[]})
    & \textit{} \\
\rowcolor{gray!20}
  &         & \textit{int} doFinal (\textit{byte[]}, \textit{int}, \textit{int}, \textit{byte[]}, \textit{int})
    & \textit{} \\
\rowcolor{gray!20}
  &         & \textit{int} doFinal (\textit{ByteBuffer}, \textit{ByteBuffer})
    & \textit{} \\
\rowcolor{white}
java.security
  & {Signature} & \textit{void} initVerify (\textit{PublicKey})                
    & \texttt{alg, key} \\
  &             & \textit{void} initVerify (\textit{Certificate})   
    & \\
  &             & \textit{void} initSign (\textit{PrivateKey})   
    & \\
  &             & \textit{void} initSign (\textit{PrivateKey, SecureRandom})   
    & \\
\rowcolor{gray!20}
javax.crypto.spec
  & {PBEKeySpec} & PBEKeySpec (\textit{char[]}) 
    & \texttt{pass, salt,} \\
\rowcolor{gray!20}
  &              & PBEKeySpec (\textit{char[]}, \textit{byte[]}, \textit{int}) 
    & \texttt{iter} \\
\rowcolor{gray!20}
  &              & PBEKeySpec (\textit{char[]}, \textit{byte[]}, \textit{int}, \textit{int}) 
    & \textit{} \\
\rowcolor{white}
javax.crypto.spec
  & {PBEParameterSpec} & PBEParameterSpec (\textit{byte[]}, \textit{int}) 
    & \texttt{salt, iter} \\
  &                    & PBEParameterSpec (\textit{byte[]}, \textit{int}, \textit{AlgorithmParameterSpec}) 
    & \textit{} \\
\rowcolor{gray!20}
java.security
    & SecureRandom & {SecureRandom} (\textit{void})
    & \texttt{seed, out} \\
\rowcolor{gray!20}
    &              & {SecureRandom} (\textit{byte[]})
    & \\
\rowcolor{gray!20}
    &              & \textit{void} setSeed (\textit{byte[]})
    & \\
\rowcolor{white}
    & SecureRandom & \textit{void} nextBytes (\textit{byte[]}) \\
    &              & \textit{void} setSeed (\textit{byte[]})
    & \\
\rowcolor{gray!20}
java.util
     & Random& {Random} (\textit{void})
     & \textit{constructor} \\
\rowcolor{white}
     & Random & \textit{int} next (\textit{int})
     & \texttt{out} \\
     &        & \textit{void} nextBytes (\textit{byte[]})
     & \textit{} \\
\rowcolor{gray!20}
java.security
  & {KeyStore} & \textit{Key} getKey (\textit{String}, \textit{char[]})                
    & \texttt{pass} \\
\rowcolor{gray!20}
  &            & \textit{void} load (\textit{InputStream}, \textit{char[]})   
    & \textit{} \\
\rowcolor{gray!20}
  &            & \textit{void} load (\textit{LoadStoreParameter})   
    & \textit{} \\
\rowcolor{gray!20}
  &            & \textit{void} store (\textit{OutputStream}, \textit{char[]})   
    & \textit{} \\
\rowcolor{gray!20}
  &            & \textit{void} store (\textit{LoadStoreParameter})   
    & \\
\rowcolor{white}
java.net
    & URL & URL (\textit{String}, \textit{String}, \textit{int}, \textit{String})
    & \texttt{urlprotl} \\
    &     & URL (\textit{URL}, \textit{String}, \textit{URLStreamHandler})
    & \texttt{} \\
\rowcolor{gray!20}
javax.net.ssl
    & HttpsURLConnection & \textit{void} setHostnameVerifier (\textit{HostnameVerifier})
    & \texttt{allhost} \\
\rowcolor{gray!20}
    &                    & \textit{void} setDefaultHostnameVerifier (\textit{HostnameVerifier})
    & \texttt{sethost} \\
\rowcolor{white}
javax.net.ssl
    & SSLContext & \textit{void} init (\textit{KeyManger[]}, \textit{TrustManager[]}, \textit{SecureRandom})
    & \texttt{allcert} \\
\rowcolor{gray!20}
javax.net.ssl
    & SocketFactory & \textit{SocketFactory} getDefault (\textit{void})
    & \texttt{sethost} \\
\rowcolor{white}
\bottomrule \\
\end{tabular}
\captionof{table}{\normalfont Java functions that have been instrumented and the parameters
 that are logged as defined in~Fig.~\ref{fig:classes}.}\label{table:javafull}
\end{center}
\end{table*}

\begin{figure*}[!th]
\hspace{-1cm}
\centering\noindent
\begin{minipage}{\textwidth}\centering
\begin{tabular}{c|@{ }c|@{ }c|@{ }c}
\begin{minipage}[t]{0.246\textwidth}
  {\hfil\includegraphics[width=\textwidth, clip, trim = {2.3cm 0.5cm 3.3cm 4.2cm}]{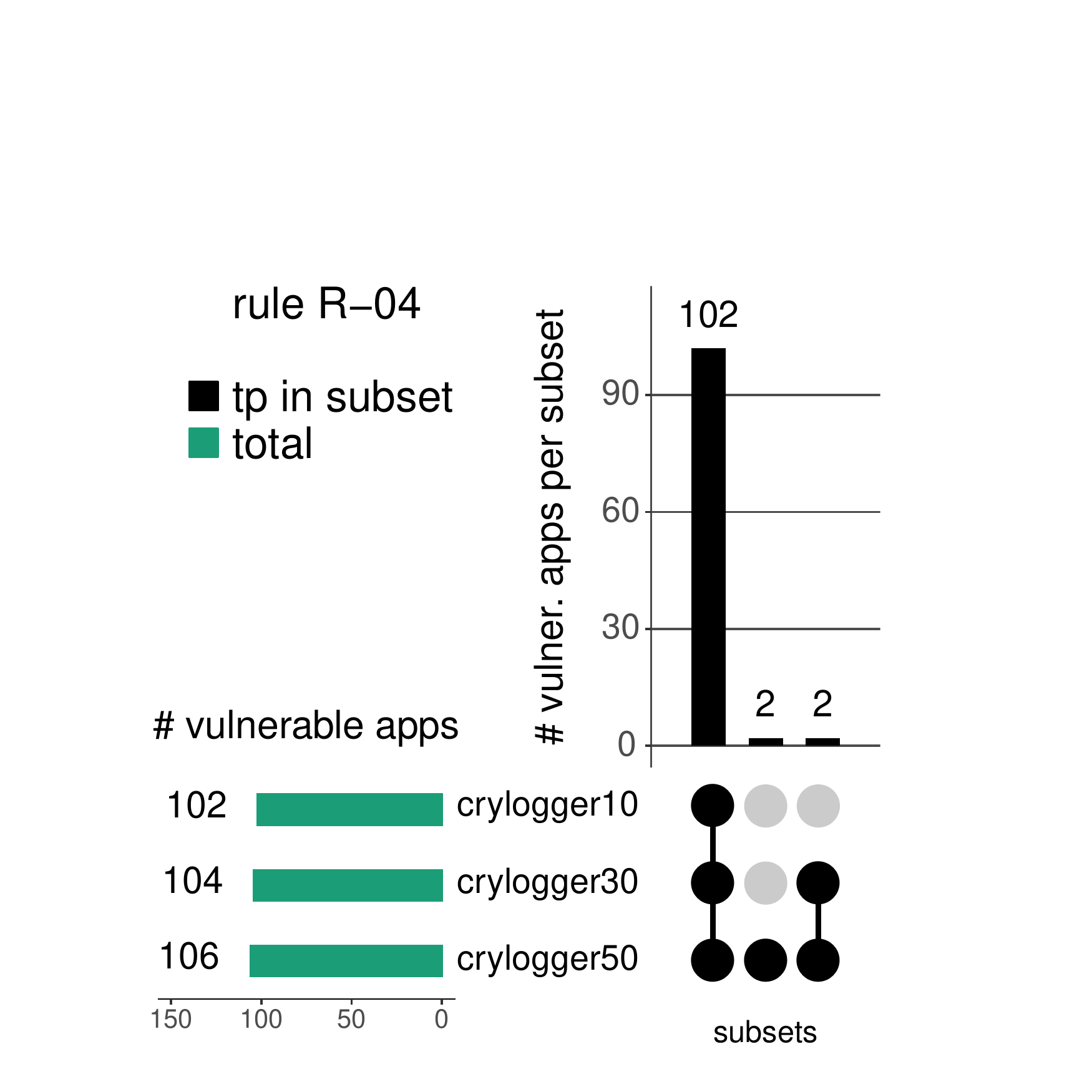}}
\end{minipage} &
\begin{minipage}[t]{0.246\textwidth}
  {\hfil\includegraphics[width=\textwidth, clip, trim = {2.3cm 0.5cm 3.3cm 4.2cm}]{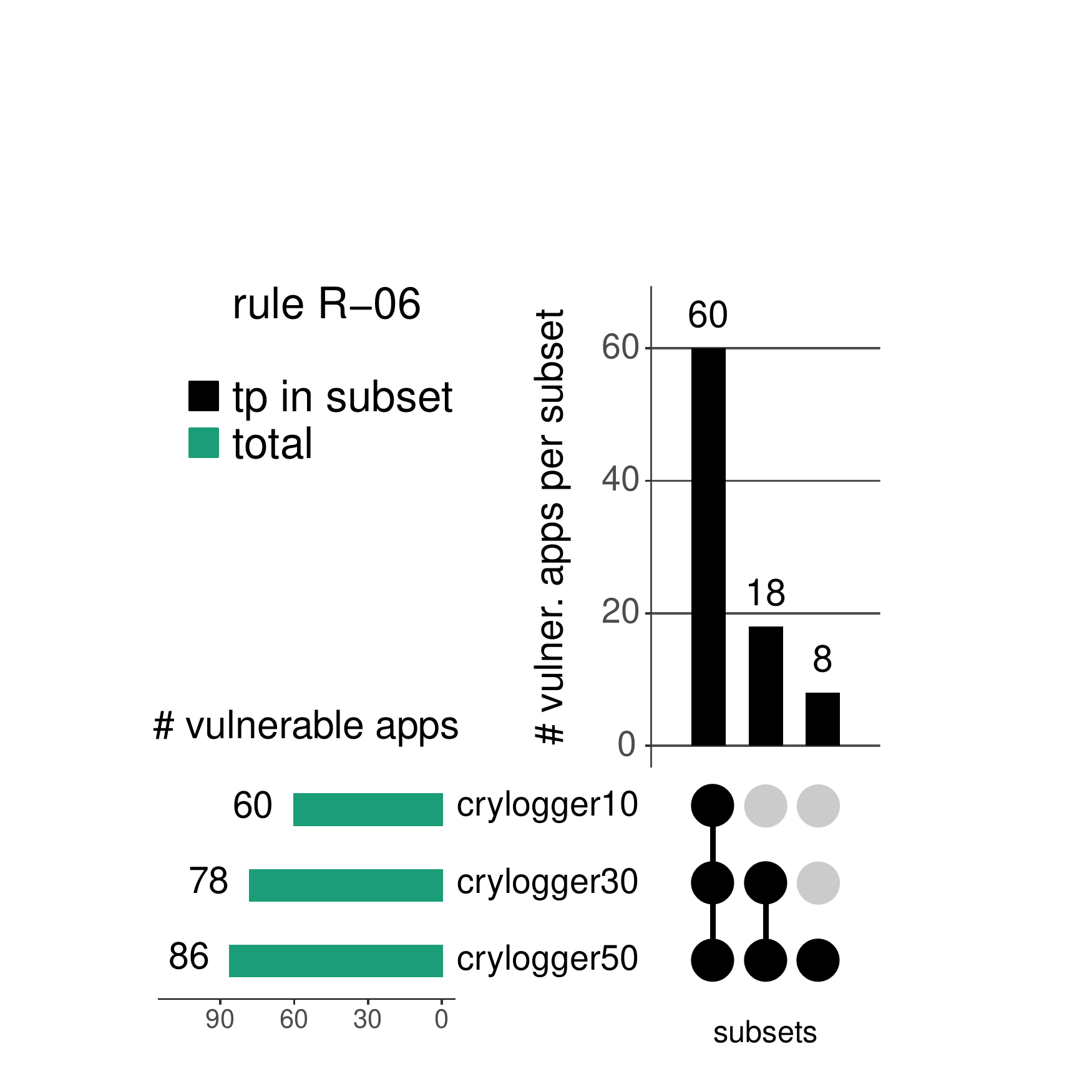}}
\end{minipage} &
\begin{minipage}[t]{0.246\textwidth}
  {\hfil\includegraphics[width=\textwidth, clip, trim = {2.3cm 0.5cm 3.3cm 4.2cm}]{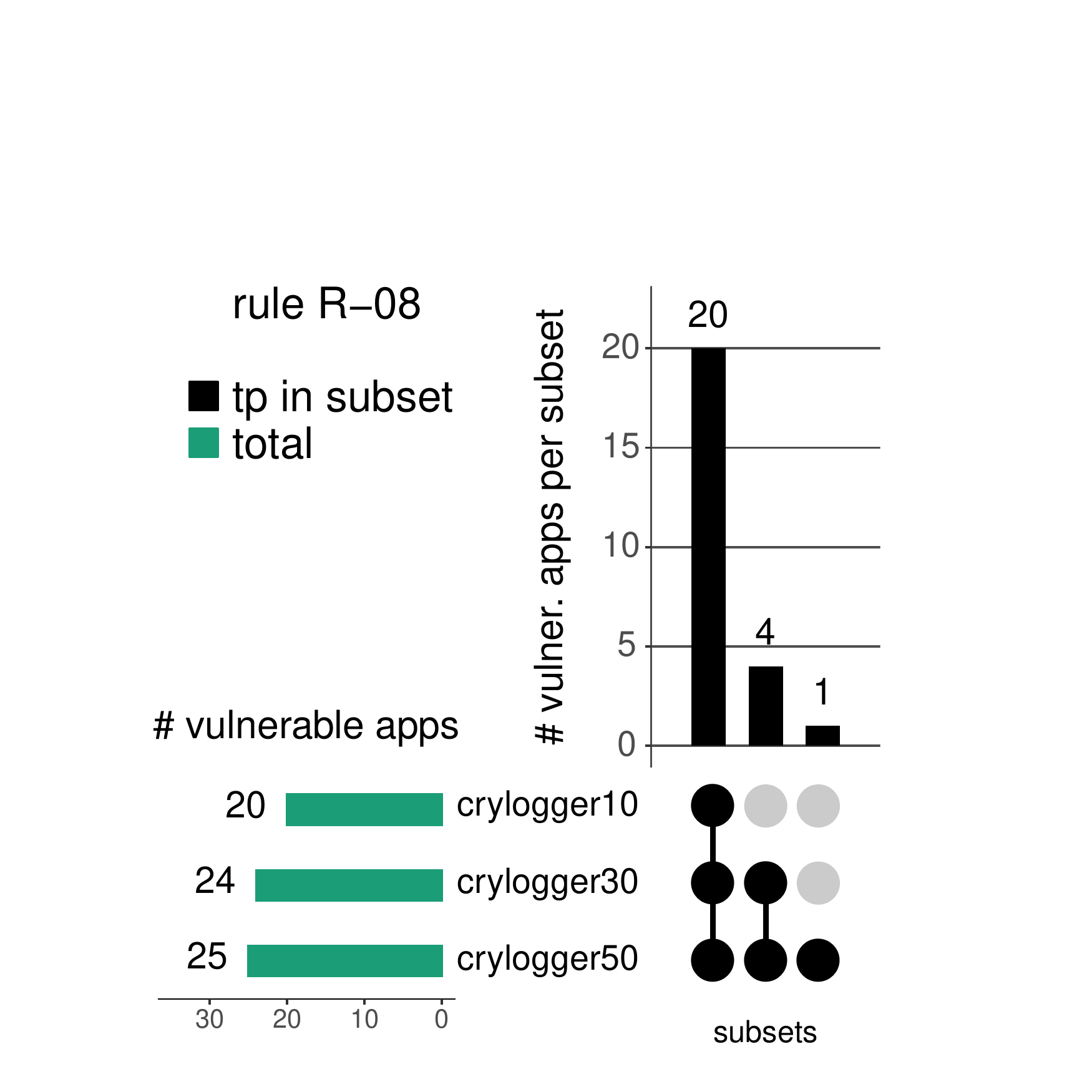}} 
\end{minipage} &
\begin{minipage}[t]{0.246\textwidth}
  {\hfil\includegraphics[width=\textwidth, clip, trim = {2.3cm 0.5cm 3.3cm 4.2cm}]{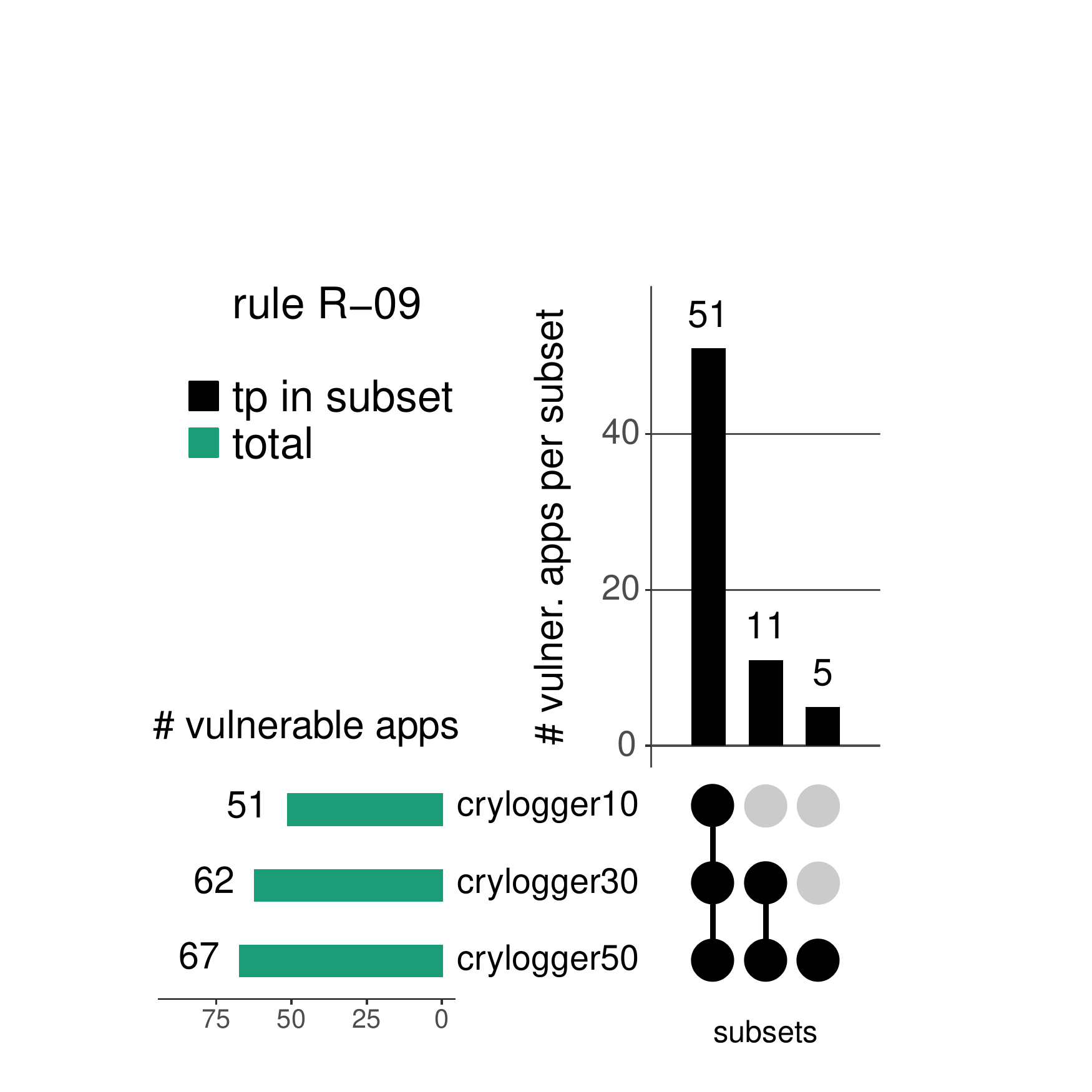}} 
\end{minipage} \\
\end{tabular}
\end{minipage}

\vspace{0.2cm}

\hspace{-1cm}
\centering\noindent
\begin{minipage}{\textwidth}\centering
\begin{tabular}{c|@{ }c|@{ }c|@{ }c}
\begin{minipage}[t]{0.246\textwidth}
  {\hfil\includegraphics[width=\textwidth, clip, trim = {2.3cm 0.5cm 3.3cm 4.2cm}]{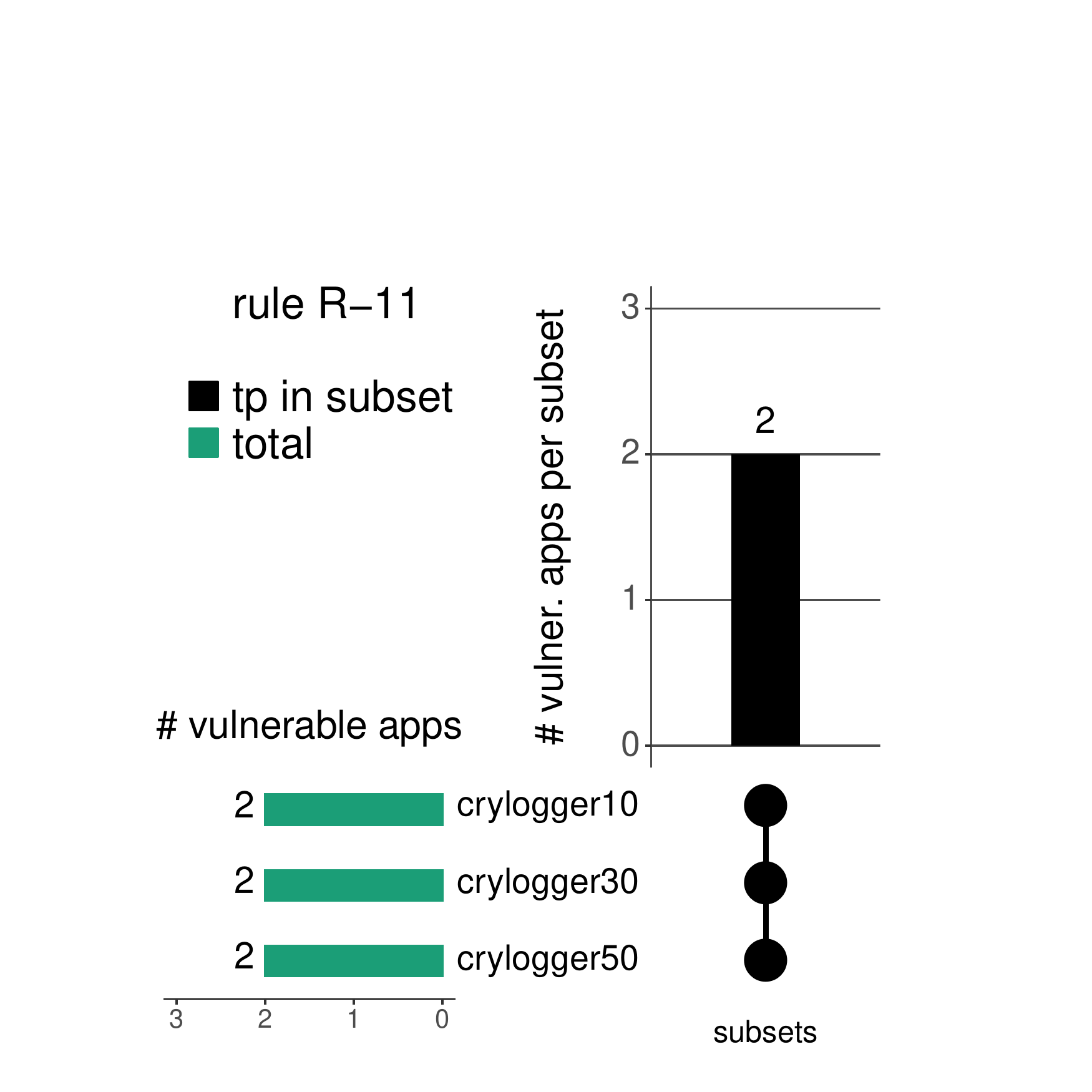}}
\end{minipage} &
\begin{minipage}[t]{0.246\textwidth}
  {\hfil\includegraphics[width=\textwidth, clip, trim = {2.3cm 0.5cm 3.3cm 4.2cm}]{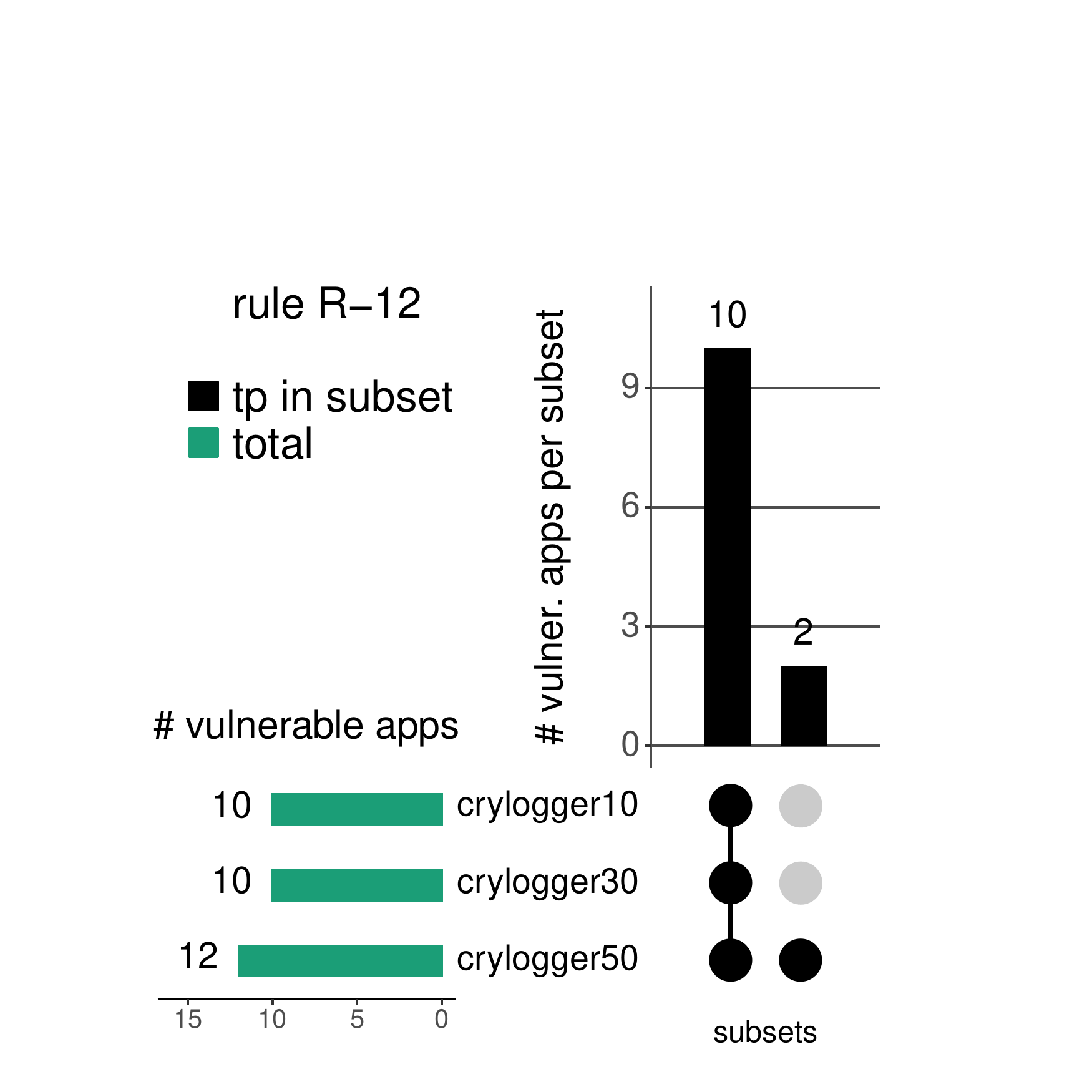}}
\end{minipage} &
\begin{minipage}[t]{0.246\textwidth}
  {\hfil\includegraphics[width=\textwidth, clip, trim = {2.3cm 0.5cm 3.3cm 4.2cm}]{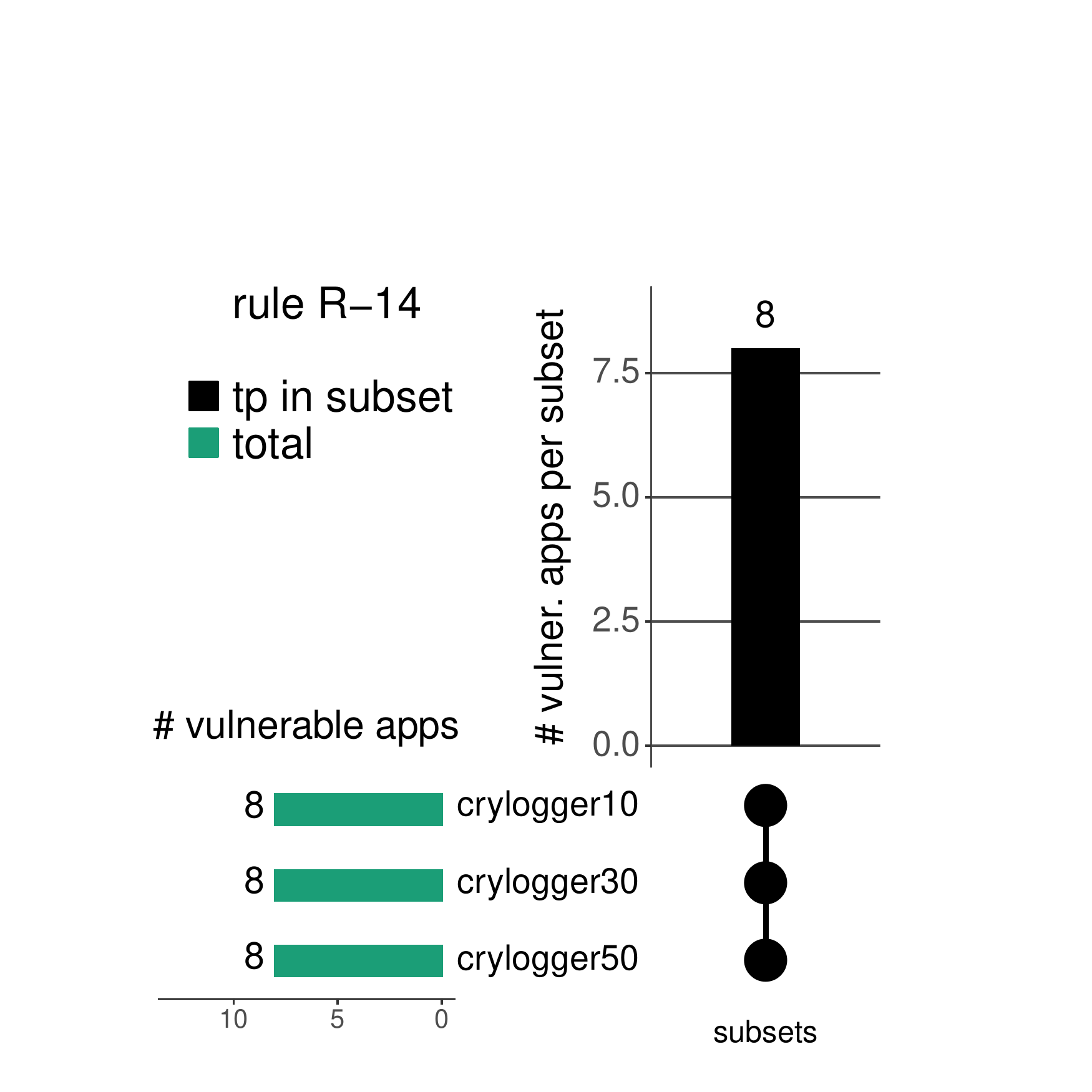}} 
\end{minipage} &
\begin{minipage}[t]{0.246\textwidth}
  {\hfil\includegraphics[width=\textwidth, clip, trim = {2.3cm 0.5cm 3.3cm 4.2cm}]{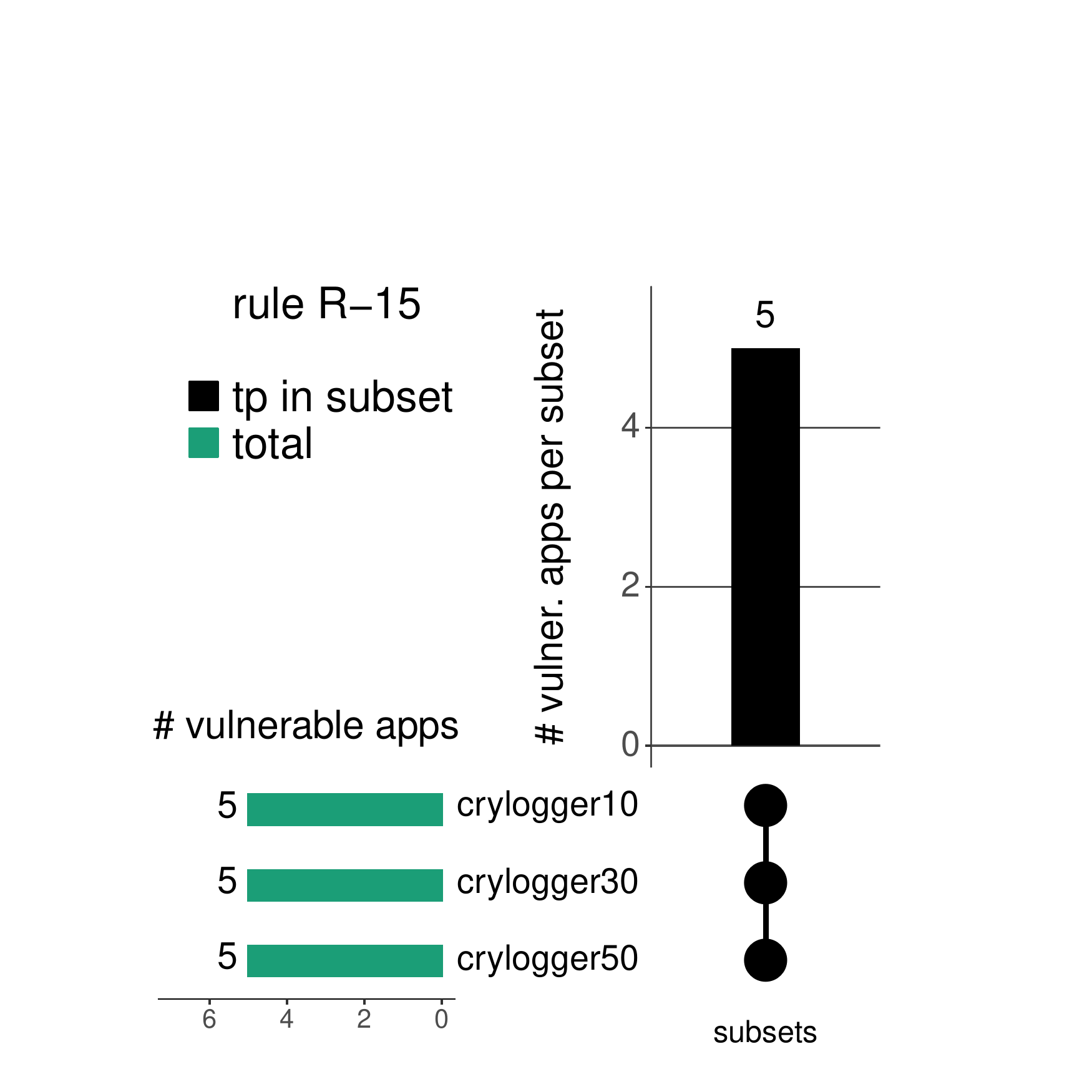}}
\end{minipage} \\
\end{tabular}
\end{minipage}

\vspace{0.2cm}

\centering\noindent
\begin{minipage}{\textwidth}\centering
\begin{tabular}{cc|@{ }c}
\hspace{-0.3cm} &
\begin{minipage}[t]{0.246\textwidth}
  {\hfil\includegraphics[width=\textwidth, clip, trim = {2.3cm 0.5cm 3.3cm 4.2cm}]{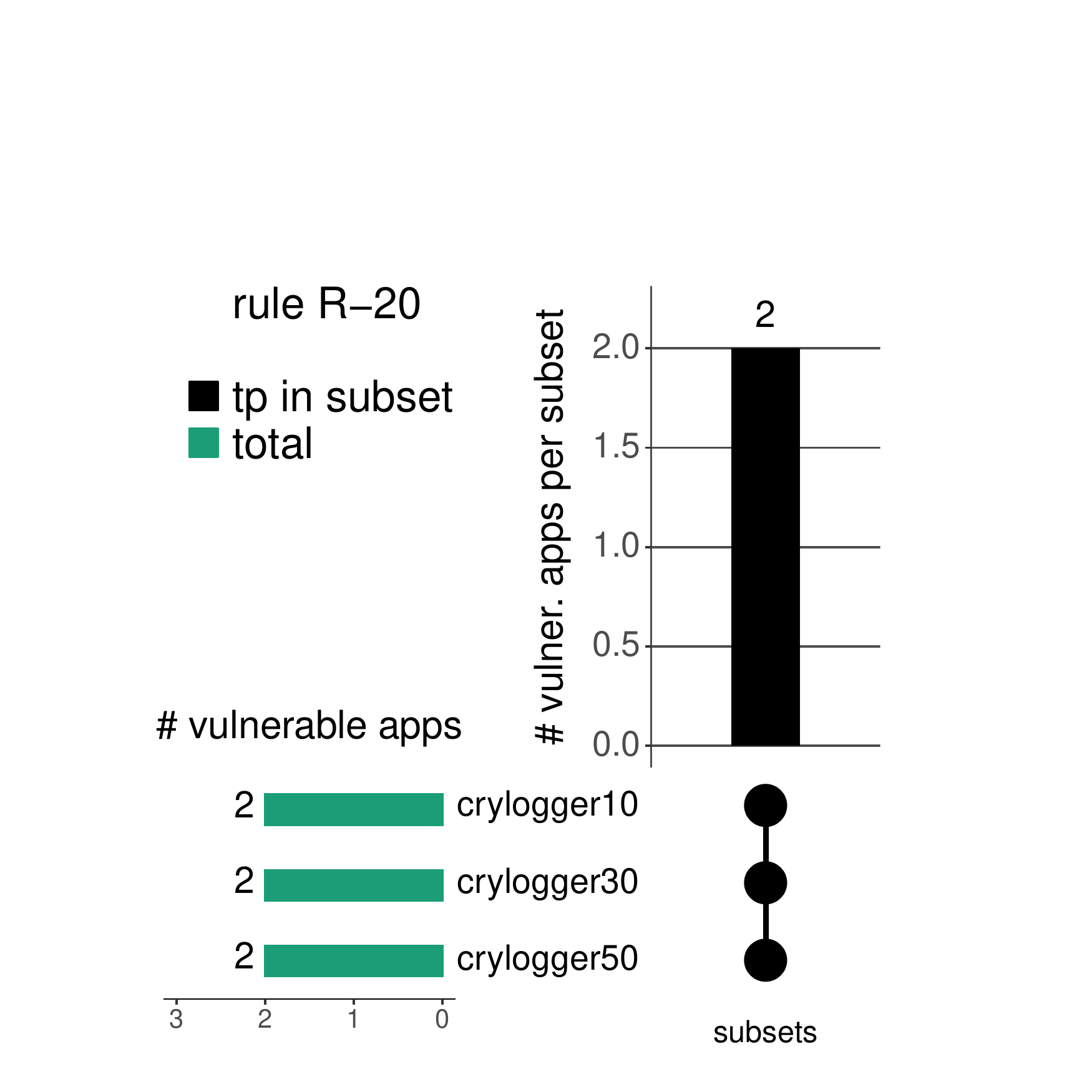}}
\end{minipage} &
\begin{minipage}[t]{0.246\textwidth}
  {\hfil\includegraphics[width=\textwidth, clip, trim = {2.3cm 0.5cm 3.3cm 4.2cm}]{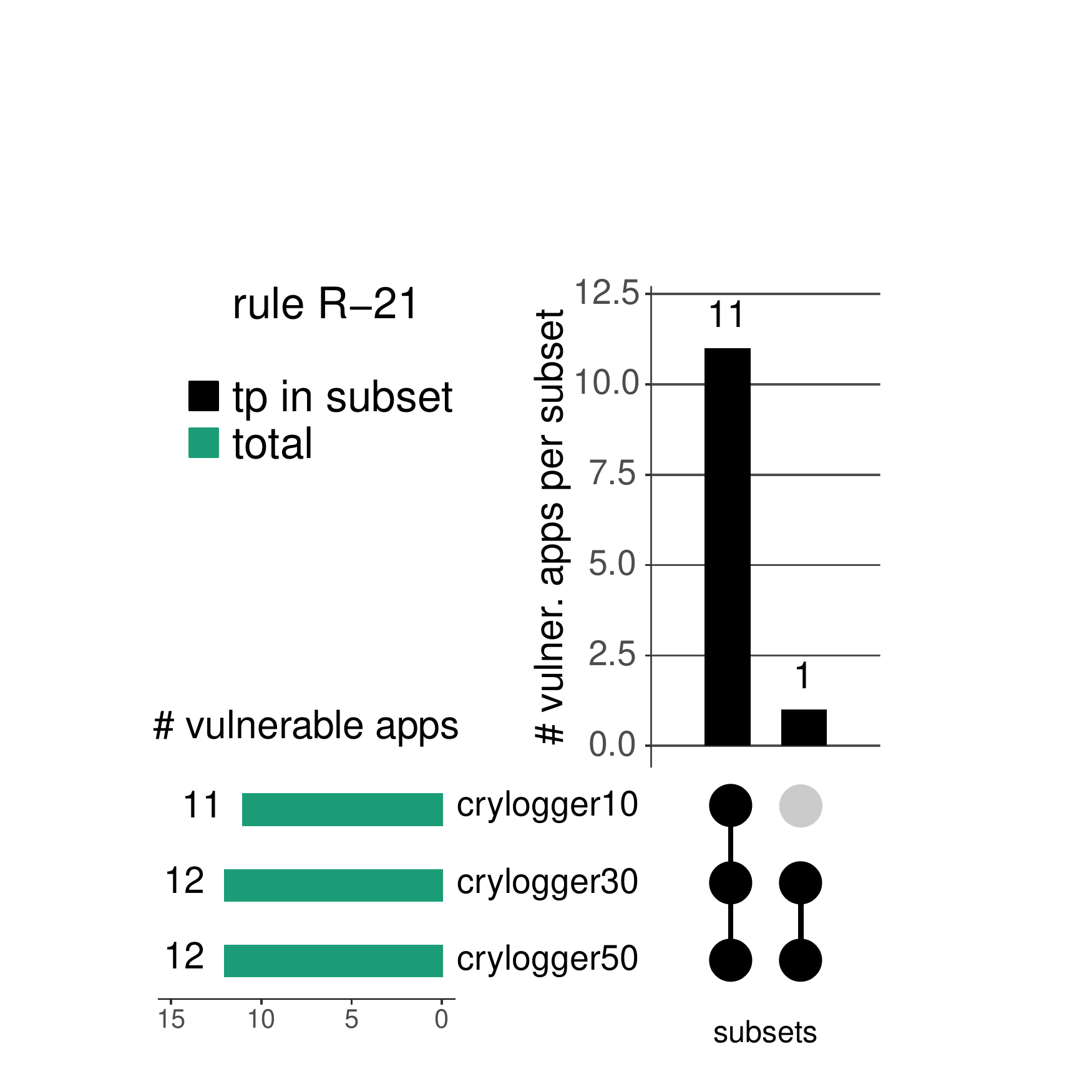}} 
\end{minipage} \\
\end{tabular}
\end{minipage}

\caption{
  Comparison of \ourff with 10k, 30k and 50k random stimuli on $150$ Android apps.
  Each graph is an \textit{upset plot}~\cite{lex_tvcg2014, conway_bio2017}. The
  \textbf{horizontal bars} indicate the number of apps flagged as vulnerable by
  \ourff with 10k, 30k and 50k stimuli; the \textbf{vertical bars} indicate the
  number of apps flagged as vulnerable by a possible intersection of the
  approaches (the $3$ largest, non-empty intersections are reported). For
  example, for \cryrule{rule:badiv}: $20$ apps are considered vulnerable by all
  the approaches, $4$ apps are flagged as vulnerable by using 30k and
  50k stimuli only, and $1$ app is considered vulnerable by using 50k stimuli
  only.}\label{results:onlyour}
\end{figure*}

\colorlet{high}{gray!20}

\begin{figure*}[p]
\begin{center}
\begin{tabular}{cc}
\begin{minipage}{0.48\textwidth}
\centering
\begin{minipage}{0.9\textwidth}
\lstlistingname{\ 1. Basic}
\begin{lstlisting}[linebackgroundcolor={}]
public class Test_X {
   public static void main(String[] args) {
      String algorithm = "AES/ECB/PKCS5PADDING";
      Cipher c = Cipher.getInstance(algorithm);
   }
}
\end{lstlisting}
\end{minipage}

\begin{minipage}{0.9\textwidth}
\lstlistingname{\ 3. Interprocedural}
\begin{lstlisting}[linebackgroundcolor={}]
public class Test_X {
   public static void main(String[] args) {
      String algorithm = "AES/ECB/PKCS5PADDING";
      method1(algorithm);
   }
   public static void method1(String algorithm) {
      method2(algorithm);
   }
   public static void method2(String algorithm) {
      Cipher c = Cipher.getInstance(algorithm);
   }
}
\end{lstlisting}
\end{minipage}

\begin{minipage}{0.9\textwidth}
\lstlistingname{\ 5. Field Sensitive}
\begin{lstlisting}[linebackgroundcolor={
    \ifnum \value{lstnumber}=10 \color{high}\fi
    \ifnum \value{lstnumber}=11 \color{high}\fi
    \ifnum \value{lstnumber}=12 \color{high}\fi
    \ifnum \value{lstnumber}=13 \color{high}\fi}]
public class Test_X {
   String algorithm;
   public Test_X(String alg) {
      algorithm = alg;
   }
   public method(String alg) {
      alg = algorithm;
      Cipher c = Cipher.getInstance(alg);
   }
   public static void main(String[] args) {
      Test_X x = new Test_X("AES/ECB/PKCS5PADDING");
      x.method("AES/CBC/PKCS5PADDING");
   }
}
\end{lstlisting}
\end{minipage} 

\begin{minipage}{0.9\textwidth}
\lstlistingname{\ 7. Argument Sensitive}
\begin{lstlisting}[linebackgroundcolor={\color{high}}]
public class Test_X {
   public static void main(String[] args) {
      if (condition(args)) {
        algorithm = "AES/CBC/PKCS5PADDING";
        Cipher c = Cipher.getInstance(algorithm);
      }
   }
}
\end{lstlisting}
\end{minipage}

\end{minipage} &
\begin{minipage}{0.48\textwidth}
\centering
\begin{minipage}{0.9\textwidth}
\lstlistingname{\ 2. Miscellaneous}
\begin{lstlisting}[linebackgroundcolor={}]
public class Test_X {
   public static void main(String[] args) {
      String alg = "AES/ECB/PKCS5PADDING";
      // Use of a simple data structure
      DataStructure data = new DataStructure(alg);
      Cipher c = Cipher.getInstance(data.get());
   }
}
\end{lstlisting}
\begin{lstlisting}[linebackgroundcolor={}]
public class Test_X {
   public static void main(String[] args) {
      String alg = "AES/ECB/PKCS5PADDING";
      // Conversion to another type
      Othertype type = ConvertOtherType(alg);
      Cipher c = Cipher.getInstance(data.get());
   }
}
\end{lstlisting}
\end{minipage}

\begin{minipage}{0.9\textwidth}
\lstlistingname{\ 4. Path Sensitive}
\begin{lstlisting}[linebackgroundcolor={}]
public class Test_X {
   public static void main(String[] args) {
      int choice = 2;
      String algorithm = "AES/ECB/PKCS5PADDING";
      if (choice > 1)
         algorithm = "AES/CBC/PKCS5PADDING";
      Cipher c = Cipher.getInstance(algorithm);
   }
}
\end{lstlisting}
\end{minipage}

\begin{minipage}{0.9\textwidth}
\lstlistingname{\ 6. Multiple Classes}
\begin{lstlisting}[linebackgroundcolor={
    \ifnum \value{lstnumber}=2 \color{high}\fi
    \ifnum \value{lstnumber}=3 \color{high}\fi
    \ifnum \value{lstnumber}=4 \color{high}\fi}]
public class Test_X {
   public static void main(String[] args) {
      method1("AES/ECB/PKCS5PADDING");
   }
   public static void method1(String algorithm) {
      Test_Y y = new Test_Y();
      y.method(algorithm);
   }
}
public class Test_X {
   public void method2(String algorithm) {
      Cipher c = Cipher.getInstance(algorithm);
   }
}
\end{lstlisting}
\end{minipage} 

\end{minipage} \\
\end{tabular}
\caption{The types of benchmarks that are present in the \cryptoapinref. We
highlighted our modifications to make the benchmarks executable
(Section~\ref{sec:res1}). The first 6 types of benchmarks (\textit{basic,
miscellaneous, interprocedural, path sensitive, field sensitive, multiple
classes}) were originally proposed in~\cite{\cryptoapi}.  We added
\textit{argument-sensitive} tests so that the \cryptoapiname can be used to
evaluate dynamic approaches. 
}\label{fig:apitypes}
\end{center}
\end{figure*}

\begin{figure*}[!h]
\begin{center}
\vspace{-1.5cm}%
\begin{tabular}{c@{\hspace{1cm}}c}
\begin{minipage}{0.44\textwidth}
\begin{lstlisting}[]
package com.google.api.client.testing.http;
class HttpTesting {
  static String SIMPLE_URL = "http://google.com"
  public HttpTesting() {
    GenericUrl url = new GenericUrl(SIMPLE_URL);
  } ...
\end{lstlisting}
\end{minipage} 
&

\begin{minipage}{0.44\textwidth}
\begin{lstlisting}[]
package com.adjust.sdk;
class AdjustFactory {
  public static void useTestConnectionOptions() {
    con.setHostnameVerifier(new HostnameVerifier() {
      public boolean verify(String h, SSLSession s)
      { return true; } ...
\end{lstlisting}
\end{minipage}
\\
\end{tabular}
\end{center}
\caption{Examples of false positives for rules \cryrule{rule:http}
and \cryrule{rule:hostname} for \cryptoguardnref.}\label{fig:fp}
\end{figure*}

\begin{figure*}[b]
\renewcommand*{\arraystretch}{1.05}
\small
\begin{center}
\begin{tabular}{ccc}
\textbf{(a) Original \cryptoapinref} & 
\textbf{(b) Modified \cryptoapiname} & 
\textbf{(c) New Tests} \\
\begin{tabular}[t]{c
p{0.2cm}p{0.2cm}p{0.2cm}p{0.2cm}
p{0.2cm}p{0.2cm}p{0.2cm}}
\toprule
&
\multicolumn{4}{c}{{\cryptoguardnref}} & 
\multicolumn{3}{c}{{\ourff}} \\
\cmidrule(lr){2-5}
\cmidrule(lr){6-8}
\textbf{Rule ID}
& \textbf{TP} & \textbf{TN} & \textbf{FP} & \textbf{FN} 
& \textbf{TP} & \textbf{TN} & \textbf{FN} \\
\midrule
\cryrule{rule:hash}       & 24 & 1 & 4 & 0 
                        & 24 & 5 & 0 \\
\cryrule{rule:symmalg}    & 30 & 1 & 5 & 0 
                        & 30 & 6 & 0 \\
\cryrule{rule:ecbmode}    & 6 & 1 & 1 & 0 
                        & 6 & 2 & 0 \\
\cryrule{rule:constkey}   & 5 & 2 & 1 & 2 
                        & 7 & 3 & 0 \\
\cryrule{rule:constiv}    & 8 & 1 & 1 & 0 
                        & 8 & 2 & 0 \\
\cryrule{rule:constsalt}  & 7 & 1 & 1 & 0 
                        & 7 & 2 & 0 \\
\cryrule{rule:iterat}     & 5 & 1 & 1 & 2 
                        & 7 & 2 & 0 \\
\cryrule{rule:reusepass}  & 7 & 2 & 1 & 1 
                        & 8 & 3 & 0 \\
\cryrule{rule:statseed}   & 13 & 1 & 2 & 1 
                        & 14 & 3 & 0 \\
\cryrule{rule:unsaprng}   & 1 & 1 & 0 & 0 
                        & 1 & 1 & 0 \\
\cryrule{rule:rsakeysize} & 4 & 0 & 1 & 1 
                        & 5 & 1 & 0 \\
\cryrule{rule:http}       & 6 & 2 & 1 & 0 
                        & 6 & 3 & 0 \\
\cryrule{rule:store}      & 7 & 2 & 1 & 0 
                        & 7 & 3 & 0 \\
\cryrule{rule:hostname}   & 1 & 1 & 0 & 0 
                        & 1 & 1 & 0 \\
\cryrule{rule:certif}     & 3 & 0 & 0 & 0 
                        & 3 & 0 & 0 \\
\cryrule{rule:socket}     & 4 & 0 & 0 & 0 
                        & 4 & 0 & 0 \\
\midrule
\textbf{Total}          & 131 & 17 & 20 & 7
                        & 138 & 37 & 0 \\
\bottomrule
\end{tabular}
& \begin{tabular}[t]{c
p{0.2cm}p{0.2cm}p{0.2cm}p{0.2cm}
p{0.2cm}p{0.2cm}p{0.2cm}}
\toprule
&
\multicolumn{4}{c}{{\cryptoguardnref}} & 
\multicolumn{3}{c}{{\ourff}} \\
\cmidrule(lr){2-5}
\cmidrule(lr){6-8}
\textbf{Rule ID}
& \textbf{TP} & \textbf{TN} & \textbf{FP} & \textbf{FN} 
& \textbf{TP} & \textbf{TN} & \textbf{FN} \\
\midrule
\cryrule{rule:hash}       & \textbf{\textcolor{blue}{28}} & 1 & 4 & 0 
                          & 24 & 5 & \textbf{\textcolor{blue}{4}} \\
\cryrule{rule:symmalg}    & \textbf{\textcolor{blue}{35}} & 1 & 5 & 0
                          & 30 & 6 & \textbf{\textcolor{blue}{5}} \\
\cryrule{rule:ecbmode}    & \textbf{\textcolor{blue}{7}} & 1 & \textbf{\textcolor{blue}{5}} & 0
                          & 6 & \textbf{\textcolor{blue}{6}} & \textbf{\textcolor{blue}{1}}\\
\cryrule{rule:constkey}   & \textbf{\textcolor{blue}{6}} & 2 & 1 & 2 
                          & 7 & 3 & \textbf{\textcolor{blue}{1}} \\
\cryrule{rule:constiv}    & \textbf{\textcolor{blue}{9}} & 1 & 1 & 0
                          & 8 & 2 & \textbf{\textcolor{blue}{1}} \\
\cryrule{rule:constsalt}  & \textbf{\textcolor{blue}{8}} & 1 & 1 & 0
                          & 7 & 2 & \textbf{\textcolor{blue}{1}} \\
\cryrule{rule:iterat}     & \textbf{\textcolor{blue}{6}} & 1 & 1 & 2
                          & 7 & 2 & \textbf{\textcolor{blue}{1}} \\
\cryrule{rule:reusepass}  & \textbf{\textcolor{blue}{8}} & 2 & 1 & 1
                          & 8 & 3 & \textbf{\textcolor{blue}{1}} \\
\cryrule{rule:statseed}   & \textbf{\textcolor{blue}{14}} & 1 & 2 & 1
                          & 14 & 3 & \textbf{\textcolor{blue}{1}} \\
\cryrule{rule:unsaprng}   & {1} & 1 & 0 & 0 
                          & 1 & 1 & 0 \\
\cryrule{rule:rsakeysize} & \textbf{\textcolor{blue}{5}} & 0 & 1 & 1
                          & 5 & 1 & \textbf{\textcolor{blue}{1}} \\
\cryrule{rule:http}       & \textbf{\textcolor{blue}{7}} & 2 & 1 & 0
                          & 6 & 3 & \textbf{\textcolor{blue}{1}} \\
\cryrule{rule:store}      & \textbf{\textcolor{blue}{8}} & 2 & 1 & 0
                          & 7 & 3 & \textbf{\textcolor{blue}{1}} \\
\cryrule{rule:hostname}   & 1 & 1 & 0 & 0 
                          & 1 & 1 & 0 \\
\cryrule{rule:certif}     & 3 & 0 & 0 & 0 
                          & 3 & 0 & 0 \\
\cryrule{rule:socket}     & 4 & 0 & 0 & 0 
                          & 4 & 0 & 0 \\
\midrule
\textbf{Total}          & \textbf{\textcolor{blue}{150}} & 17 & \textbf{\textcolor{blue}{24}} &
                                7 
                        & 138 & 41 & \textbf{\textcolor{blue}{19}} \\
\bottomrule
\end{tabular}
& \begin{tabular}[t]{c
p{0.2cm}p{0.2cm}p{0.2cm}p{0.2cm}
p{0.2cm}p{0.2cm}p{0.2cm}}
\toprule
&
\multicolumn{3}{c}{{\ourff}} \\
\cmidrule(lr){2-4}
\textbf{Rule ID}
& \textbf{TP} & \textbf{TN} & \textbf{FN} \\
\midrule
\cryrule{rule:cbcmode}     & 4 & 2 & 1 \\
\cryrule{rule:badkey}      & 6 & 2 & 1 \\
\cryrule{rule:badiv}       & 6 & 2 & 1 \\
\cryrule{rule:sameiv}      & 6 & 2 & 1 \\
\cryrule{rule:shortsalt}   & 7 & 2 & 1 \\
\cryrule{rule:samesalt}    & 1 & 1 & 1 \\
\cryrule{rule:weakpass}    & 7 & 2 & 1 \\
\cryrule{rule:blackpass}   & 7 & 2 & 1 \\
\cryrule{rule:rsatextbook} & 5 & 1 & 1 \\
\cryrule{rule:rsapadding}  & 5 & 1 & 1 \\
\midrule
\textbf{Total}             & 54 & 17 & 10 \\
\bottomrule
\end{tabular}
\\
\end{tabular}
\end{center}
\caption{Results for the \cryptoapinref. (a) Comparison of \cryptoguardnref and
\ourff on the original \cryptoapiname. In this case, we made the benchmarks
executable with a dynamic tool by adding a \texttt{main} to all benchmarks. (b)
Comparison of \cryptoguardname and \ourff on our modified version of the
\cryptoapiname. We added tests cases to (i) highlight the problem of false
positives (Section~\ref{sec:res2}) and (ii) show the limitations of dynamic
approaches in activating paths that are rarely executed. (c) Benchmarks that we
added for the rules supported only by \ourff on the modified
\cryptoapiname.}\label{fig:benchcomplete}
\end{figure*}

\end{document}